\documentclass[aip,amsmath,amssymb,reprint]{revtex4-1}

\usepackage{graphicx}
\usepackage{dcolumn}
\usepackage{bm}
\usepackage[utf8]{inputenc}
\usepackage[T1]{fontenc}
\usepackage{mathptmx}
\usepackage{mathrsfs}
\usepackage{etoolbox}

\usepackage{xcolor}

\newcommand{\vect}[1]{\mathbf{#1}}
\newcommand{\vers}[1]{\hat{\mathbf{#1}}}

\draft % marks overfull lines with a black rule on the right

\begin{document}

% Use the \preprint command to place your local institutional report number 
% on the title page in preprint mode.
% Multiple \preprint commands are allowed.
%\preprint{}

\title{Kinetics of isotropic to string-like phase switching in electrorheological fluids of
nanocubes} %Title of paper

% repeat the \author .. \affiliation  etc. as needed
% \email, \thanks, \homepage, \altaffiliation all apply to the current author.
% Explanatory text should go in the []'s, 
% actual e-mail address or url should go in the {}'s for \email and \homepage.
% Please use the appropriate macro for the type of information

% \affiliation command applies to all authors since the last \affiliation command. 
% The \affiliation command should follow the other information.

\author{L. Tonti}
 \affiliation{Department of Chemical Engineering, The University of Manchester, Manchester, M13 9PL, UK}
 
\author{F. A. García Daza}%
\affiliation{Department of Chemical Engineering, The University of Manchester, Manchester, M13 9PL, UK}

\author{A. Patti}
\email{a.patti@ugr.es}
\affiliation{Department of Chemical Engineering, The University of Manchester, Manchester, M13 9PL, UK}
\affiliation{Department of Applied Physics, University of Granada, Fuente Nueva s/n, Granada, 18071, Spain}

% Collaboration name, if desired (requires use of superscriptaddress option in \documentclass). 
% \noaffiliation is required (may also be used with the \author command).
%\collaboration{}
%\noaffiliation

\date{\today}

\begin{abstract}
Applying an electric field to polarisable colloidal particles, whose permittivity differs from that of the dispersing medium, generates induced dipoles that promote the formation of string-like clusters and ultimately alter the fluid mechanical and rheological properties. Complex systems of this kind, whose electric-field-induced rheology can be manipulated between that of viscous and elastic materials, are referred to as electrorheological fluids. By dynamic Monte Carlo simulations, we investigate the dynamics of self-assembly of dielectric nanocubes upon application of an electric field. Switching the field on induces in-particle dipoles and, at sufficiently large field intensity, leads to string-like clusters of variable length across a spectrum of volume fractions. The kinetics of switching from the isotropic to the string-like state suggests the existence of two mechanisms, the first related to the nucleation of chains and the second to the competition between further merging and separation. We characterise the transient unsteady state by following the chain length distribution and analysing the probability of transition of nanocubes from one chain to another over time. Additionally, we employ passive microrheology to gain an insight into the effect of the electric field on the viscoelastic response of our model fluid. Not only do we observe that it becomes more viscoelastic in the presence of the field, but also that its viscoelasticity assumes an anisotropic signature, with both viscous and elastic moduli in planes perpendicular to the external field being larger than those along it.
\end{abstract}

% insert suggested PACS numbers in braces on next line

\maketitle %\maketitle must follow title, authors, abstract and \pacs

% Body of paper goes here. Use proper sectioning commands. 
% References should be done using the \cite, \ref, and \label commands
\section{Introduction}
In his 1949 seminal paper, Winslow demonstrated that the application of an electric field to high-dielectric-constant particles dispersed in low-viscosity oils induces their self-assembly in fibrous filaments, with dramatic changes in the rheological properties (\textit{e.g.} shear modulus, viscosity, yield stress) of the suspension  \cite{Winslow1949}. These systems, generally referred to as electrorheological (ER) fluids, are a class of smart soft materials that can adapt their viscoelasticity in response to an electric field of a given intensity. Due to their relatively simple manufacturing, significant responsiveness to external stimuli and reversible recovery, ER fluids are especially suitable for industry-relevant applications, such as damping systems, microfluidics, tactile displays, where an abrupt change in material rheology is required  \cite{hao2001,zhang2012,dong2019}. 

As originally predicted by Clausius  \cite{Clausius1879} and Mossotti  \cite{Mossotti1850}, the application of an external electric field generates an induced dipole in non-conducting suspended particles when their dielectric constant differs from that of the dispersing medium. Under these circumstances, particles polarise and tend to align with each other and with the direction of the electric field \cite{Toor1993}. The effect of external fields on suspensions of spherical particles has been extensively investigated by experiments \cite{Martin1992,Martin1998,Cao2006,Horvat2012,Belijar2016} and simulations \cite{Baxter1996,Klingenberg1989,Whittle1990,Hass1993,Fertig2021,Fertig2021a}. Thanks to recent advances in chemical \cite{sun2002}, physical \cite{Manoharan2003} and biosynthetic \cite{Shankar2004} techniques for the synthesis of anisotropic particles \cite{xiang2006,sacanna2010,okuno2010,cortie2012,sacanna2013,Khlebtsov2015,rossi2015}, which were key to discover new phases either driven by simple excluded-volume effects \cite{Glotzer2007,Xi2007,Torquato2010,Damasceno2012,Donaldson2017,Dennison2017,Joe2017, Patti2018, Chiappini2019,Cuetos2019,Casquilho2021} or induced by external stimuli \cite{Yan2013,Okada2018,Anke2014,Smallenburg2012,Hynniyen2005}, electrorheology has been extended to more exotic colloidal suspensions \cite{Rex1998,Qi2002,Kai2017}. Polarisation has then been employed to explore the fabrication of an intriguing family of novel materials, termed colloidal polymers, where a field-induced alignment of particles in polymer-like chains of tunable flexibility is subsequently made permanent even when the field is switched off \cite{Vutukuri2012,Vutukuri2017}. Most of the work on colloidal polymers and, more generally, on ER fluids has so far explored suspensions of spherical particles, with very few exceptions. One of these exceptions is the very recent experimental work by Cai and co-workers who prepared colloidal polymers of micron-sized $\alpha$-Fe$_2$O$_3$ cubes combining dipolar-directed assembly and in situ hydrolysis-condensation of tetraethylorthosilicate \cite{Huang2022}. Despite the widespread interest in ER fluids of colloidal spheres, particle shape definitely plays a crucial role in the kinetics of clustering as well as in the resulting rheology of the suspension, thus opening the path to novel materials with tunable properties.

We have recently developed a stochastic method to mimic the dynamics of Brownian particles \textit{via} standard Markov Chain Monte Carlo (MC) simulations with specific settings on the particle elementary moves. This method, referred to as dynamic Monte Carlo (DMC), allows one to implement discontinuous interaction potentials and generate time trajectories in equilibrium or over transitory unsteady states \cite{Patti2012, Cuetos2015, Corbett2018, Daza2020, chiappini2020, patti2021, Daza2022}. In this work, we apply a novel DMC technique to assess the response of a suspension of perfect hard dielectric nanocubes under the application of an electric field. At sufficiently low volume fraction and moderate field strength, colloidal hard nanocubes are expected to assemble into string-like clusters, as predicted by equilibrium MC simulations and observed experimentally \cite{Vutukuri2014}. While the phase behaviour of these ER fluids has been well documented, very little attention has been given to the kinetics of such an isotropic to string-like phase switching and to its impact on the viscoelastic properties of the material. To bridge this gap, here we investigate the responsiveness of nanocubes to the application of an electric field and compute the time scales associated to the formation of these chains and their length distribution. Additionally, we apply passive microrheology (MR) to infer, from the computation of the mean square displacement of a free-diffusing tracer, the elastic and viscous moduli of this model ER fluid for both the field-off and field-on cases. While dilute suspensions only exhibit negligible changes in their viscoelastic response to an electric field, denser suspensions, where the presence of fibrous structures is more significant, show an evident increase of the viscous and elastic moduli and the occurrence of a strong space-dependent rheology.

%%%%%%%%%%%%%%%%%%%%%%%%%%%%%%%%%%%%%%%%%%%%%%%%%
%%%%%%%%%%%%%%% METHODOLOGY %%%%%%%%%%%%%%%%%%%%%
%%%%%%%%%%%%%%%%%%%%%%%%%%%%%%%%%%%%%%%%%%%%%%%%%

\section{Computational Methodology}

%%%%%% model

\subsection{Model}
We set the length of a cube edge $\sigma$, the reciprocal of the thermal energy $\beta \equiv 1/k_\text{B}T$, with $k_\text{B}$ the Boltzmann constant and $T$ the absolute temperature, and the solvent viscosity $\mu$ as system units. It follows that time has units $\tau \equiv \beta \mu \sigma^3$. Polarised cubes are modeled \textit{via} hard core interactions, and the induced dipole-dipole interaction between their centers of mass is considered only in the presence of an electric field. We apply an algorithm based on the separating axis theorem to detect overlaps between two nanocubes \cite{Gottschalk1996}. Hard core models neglect more complex phenomena of anisotropic interactions, observed in simulations and experiments of dielectric nanocubes with electric double layers, where the type of the solvent and the salt concentration proved to have a role in the preferential relative orientation as they get closer to each other \cite{Rosenberg2020}.

We modelled the induced polarisation of the cubic particles using a point-dipole approximation, \textit{i.e.} $\vect{p}_{i} = \alpha E_0 \vers{E} = p \vers{E}$, where $p$ is the magnitude of the dipole moment, $\alpha$ the particle polarizability and $E_0\vers{E}$ the external field.
The point dipole approximation has been proven to successfully mimic the phase behaviour of polarised cubes at different densities and strengths of the external electric fields. The Clausius-Mossotti relationship is used to express the particle polarizability:
\begin{equation}
    \alpha = 3 V_p \epsilon_o \epsilon_s \left(\frac{\epsilon_p - \epsilon_s} {\epsilon_p + 2\epsilon_s}\right) = \frac{3 V_p \epsilon_s}{4 \pi k} \left(\frac{a - 1} {a + 2}\right)
\end{equation}
where $k \equiv 1 / (4 \pi \epsilon_0)$ is the Coulomb constant, set as unit in our simulations, $V_p = \sigma^3$ is the volume of one cubic particle, $\epsilon_0$ is the vacuum permittivity and $a = \epsilon_p / \epsilon_s$ the ratio between permittivities of the particles and the solvent \cite{Clausius1879,Mossotti1850,Parthasarathy1996}. The dipolar interaction between two dipoles is:
\begin{equation}
    u_{ij,dip} = -\frac{k}{\epsilon_s}\frac{3(\vect{p}_i \cdot \vers{r}_{ij})(\vect{p}_j \cdot \vers{r}_{ij}) - \vect{p}_i \cdot \vect{p}_j}{r^3_{ij}},
    \label{eq:udip_full}
\end{equation}
where $r_{ij}\vers{r}_{ij}$ is the distance vector between particles $i$ and $j$. One can simplify Eq.~\ref{eq:udip_full} considering that $\vect{p}_{i} \cdot \vect{p}_{j} = p^2$ and $\vect{p}_i \cdot \vers{r}_{ij} = \vect{p}_i \cdot \vers{r}_{ij} = p^2 \cos(\theta_{ij})$, where $\cos(\theta_{ij}) = \vers{E} \cdot \vers{r}_{ij}$. We can finally express Eq.~\ref{eq:udip_full} as a function of a dimensionless interaction parameter $\gamma$, as already defined in the work by Vutukuri \textit{et al.} \cite{Vutukuri2014}:
\begin{equation}
    \beta u_{ij,dip}(r_{ij},\theta_{ij}) = \frac{\gamma}{2} \bigg(\frac{\sigma}{r_{ij}}\bigg)^3
    \big[1 - 3\cos^2(\theta_{ij})\big],
    \label{eq:udip}
\end{equation}
where 
\begin{equation}
    \gamma = \frac{2 k \beta p^2}{\epsilon_s \sigma^3}.
\end{equation}
Figure~1 shows a schematic representation of the model dipolar interaction described by Eq.~\ref{eq:udip}, while Table~\ref{tab:units} lists the units of the physical quantities used in this work.

\begin{table}
    \caption{\label{tab:units} List of units used in this work.}
    \begin{ruledtabular}
        \begin{tabular}{cc}
            Quantity                &   Units  \\ \hline
            Length of cube edge     &   $\sigma$    \\
            Solvent viscosity       &   $\mu$    \\
            Thermodynamic beta              &   $\beta \equiv 1/ (k_B T)$   \\
            Time                &   $\tau \equiv \beta \mu \sigma^3$ \\
            Frequency           &   $1 / \tau$ \\
            Translational diffusivity & $\sigma^2 / \tau$ \\
            Rotational diffusivity & $\text{rad}^2 / \tau$ \\
            Pressure                &   $1/ (\beta \sigma^{3})$ \\
            Coulomb constant & $k \equiv 1 / (4 \pi \epsilon_0 )$ \\
            Magnitude of dipole moment & $\sqrt{\sigma^3  / (\beta k)}$ \\
            Polarizability & $\sigma^3 / k$ \\
            Intensity of external field & $\sqrt{k / (\beta \sigma^3 )}$ \\
        \end{tabular}
    \end{ruledtabular}
\end{table}

\begin{figure}[ht!]
    \centering
    \includegraphics{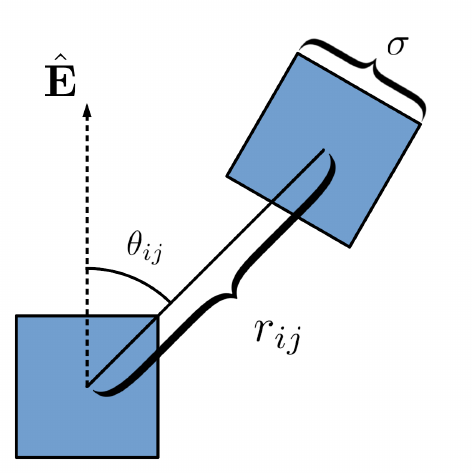}
    \caption{Model of dipolar interactions between nanocubes. The interaction potential $u_{ij,dip}(r_{ij},\theta_{ij})$ depends on the module of the vector $\vect{r}_{ij}$ between the centers of mass of particle $i$ and $j$, and $\theta_{ij}$, \textit{i.e.} the angle between $\vect{r}_{ij}$ and the external field with orientation $\vers{E}$.}
    \label{fig:model}
\end{figure}

Since simulations are performed in boxes with periodic boundaries, we employ the Ewald summation method for dipolar interactions to compute the long range contributions to the system's energy \cite{Tildesley2017}. In all simulations with field on, $\gamma$ is set equal to 13. At this field intensity, one can observe the formation of isolated strings that do not percolate through the simulation box and do not cluster. For this value of the dipole-dipole interaction strength, Vutukuri \textit{et al.} \cite{Vutukuri2014} determined that no phase transition has been observed within the packing fractions we investigated.

For simulations of passive microrheology, where a spherical probe is added to the suspension of nanocubes, the tracer is assumed to be unaffected by the presence of the external field, and only hard-core interactions between the tracer and the cubic particles are considered. For each trial move of the tracer, collisions between the sphere and the cubes are checked by using the OCSI algorithm \cite{Tonti2021}.

%%%%%% DMC simulations

\subsection{Dynamic Monte Carlo simulations}

We perform simulations in the $NVT$ ensemble using the dynamic Monte Carlo (DMC) technique to investigate the Brownian motion of nanocubes. Each trial consists of one translation of the center of mass and one body-centered rotation of a randomly
selected particle. Random moves are sampled uniformly over $[-\delta\xi_k : \delta\xi_k]$, where $\delta\xi_k$ are defined by the Einstein relation
\begin{equation}
    \delta\xi_k^2 = 2 D_{kk} {\delta}t_\text{MC},
    \label{eq:dmc_dmax}
\end{equation}
being $D_{kk}$ the diagonal element of the diffusion
tensor of the particle, and $\delta t_\text{MC}$ the MC
timescale of the particle. One time step in our simulations corresponds to 1 MC cycle ($N$ attempts to displace particles). In out-of-equilibrium DMC simulations, the physical time is recovered using Eq.~\ref{eq:dmc_out_scale}
\begin{equation}
    t_\text{BD} = \delta t_\text{MC} \sum_{c=0}^{\mathscr{C}_\text{MC}}
    \frac{\mathscr{A}_c}{3},
    \label{eq:dmc_out_scale}
\end{equation}
where $\mathscr{A}_c$ indicates the average acceptance of trial moves in one MC cycle, and $\mathscr{C}_\text{MC}$ the total number of MC cycles simulated \cite{Corbett2018}. We note that if the system is at equilibrium, the acceptance rate is constant over the the entire simulation and Eq.~\ref{eq:dmc_out_scale} can be further simplified to
\begin{equation}
    t_\text{BD} = \frac{\mathscr{A}}{3} \delta t_\text{MC} \mathscr{C}_\text{MC},
    \label{eq:dmc_eq_scale}
\end{equation}
Particles with different sizes and shapes exhibit different diffusivities and acceptance rates. Accordingly, since every particle holds its own MC
timescale, they all have to be balanced with their
respective acceptance rates to recover a unique timescale. In our microrheology simulations, $\delta_{\text{MC},sphere}$
and $\delta_{\text{MC},cube}$ are balanced using Eq.~\ref{eq:dmc_teqv}
\begin{equation}
     \mathscr{A}_{sphere} \delta t_{\text{MC},sphere} = 
     \mathscr{A}_{cube} \delta t_{\text{MC},cube}.
     \label{eq:dmc_teqv}
\end{equation}
As in this case the systems are in equilibrium, $\mathscr{A}_{sphere}$ and $\mathscr{A}_{cube}$ will be constant over time. The validity of Eq.~\ref{eq:dmc_teqv} is ensured before
running the DMC simulations by fixing one timescale and recalculating the other one using the estimated acceptance rate in a preliminary trial-and-error simulation \cite{Cuetos2015, tonti2021_2}.

The dynamics and kinetics of string formation are investigated by performing simulations of a system of $N_{cube}=1500$ nanocubes in cubic
simulation boxes with periodic boundary conditions, at packing a fraction $\eta = 0.02$. We perform 30 independent simulations, setting $\delta t_{\text{MC},cube} = 10^{-3} \, \tau$.
The diffusion tensor of the cube is estimated using
the software Hydro++ \cite{Garcia2007}, from which we
obtain translational and rotational diffusivities,
$D_{t,cube} = 8.35 \times 10^{-2}\:\sigma^2\tau^{-1}$, and $D_{r,cube} = 1.48 \times 10^{-1}\:\text{rad}^2 \tau^{-1}$.
We simulate 3 consecutive sequences where the external
electric field is turned on and off. In a single sequence, starting from a perfect isotropic phase (field off), the external field is switched on for 2$\times 10^6$ time steps and then turned off for $3\times 10^5$ time steps.

The study of microrheology is performed in suspensions of $N_{cube} = 1500$ cubes of side $\sigma$ and 1 spherical tracer with diameter $d_{sphere} = 3\,\sigma$ at a packing fraction $\eta = 0.2$, at equilibrium states when the external field is on and off. The diffusivity of the tracer at infinite dilution is $D_{t,sphere} = \left(1/9\pi\right)\sigma^2\tau^{-1}$, and attempted moves to displace it are sampled through trial translations in all the tree spatial directions. We set the nanocubes time step to $\delta t_{\text{MC},cube} = 5.0 \times 10^{-3} \, \tau$
and perform preliminary equilibration runs to recover $\delta t_{\text{MC},sphere}$ according to Eq.~\ref{eq:dmc_teqv}. While in isotropic phases (field off) $\delta t_{\text{MC},sphere,OFF} = 5.307 \times 10^{-3} \,\tau$, in string-like phases (field on) $\delta t_{\text{MC},sphere,ON} = 3.516 \times 10^{-3}\,\tau$. We compute the rheological properties of the host phase from 1000
independent trajectories of the system.

%%%%%% structural properties

\subsection{Structural properties}

The positional pair correlation functions are employed to investigate the structural properties of the suspension. Due to the system's anisotropy, we decompose the analysis in the direction parallel and perpendicular to the external field. We define $r_{ij}=\|\vect{r}_{ij}\|$,
$r_{ij,\parallel}=|\vect{r}_{ij}\cdot \vers{E}|$ and 
$r_{ij,\perp}= \|\vect{r}_{ij} - ( \vect{r}_{ij} \cdot \vers{E} )\vers{E}\|$
as the moduli of the relative distance between particles $i$
and $j$ for the total, parallel and perpendicular directions to the field, respectively. In isotropic phases, we compute the classical radial distribution function \cite{Hansen2013}
\begin{equation}
    g(r) = \frac{1}{N v_{r} \rho}\sum_{i \neq j} \bigg\langle
    \delta \Big( r - r_{ij}\Big) \bigg\rangle,
    \label{eq:grad}
\end{equation}
where $v_{r} = 4 \pi [(r + \Delta r)^3 - r^3] / 3$ is the volume of a hollow sphere of radii
$r$ and $r+\Delta r$. By contrast, for the field-induced string-like phase, we compute
the parallel and perpendicular pair correlation functions, defined as follows
\begin{equation}
    g_{\parallel}(r) = \frac{\sum\limits_{i \neq j} \bigg\langle H
    \left( R_{\parallel} - r_{ij,\perp} \right) \delta \left( r - r_{ij,\parallel} \right) \bigg\rangle}{N v_{\parallel} \rho}
    \label{eq:gpar}
\end{equation}
\begin{equation}
     g_{\perp}(r) = \frac{\sum\limits_{i \neq j} \bigg\langle H
    \left( h_{\perp} - r_{ij,\parallel} \right)
    \delta \left( r -  r_{ij,\perp} \right) \bigg\rangle}{N v_{\perp} \rho},
    \label{eq:gper}
\end{equation}
where $H$ is the Heaviside step function, $v_{\parallel} = \pi R^2_{\parallel} \Delta r$ is the volume of a cylinder of radius $R_{\parallel}$ and height $\Delta r$, and $v_{\perp} = \pi [ (r+\Delta r)^2 - r^2 ] h_{\perp}$ is the volume of a cylindrical annulus of height $h_{\perp}$ and thickness $\Delta r$ \cite{Busselez2014, Morillo2019}.

We assess the degree of orientational order with respect to the field by computing a specific uniaxial order parameter for each particle $i$, that takes into account the cubic symmetry of the particles
\begin{equation}
    S_i = \frac{\pi\Big( 3\max \big((\vers{E} \cdot \vers{e}_{k,i})^2\big)-1\Big) -2\sqrt{3}}{2\pi-2\sqrt{3}},
    \label{eq:op_cube}
\end{equation}
where $\vers{e}_{k,i}$ are the three axes of orientation of particle $i$, for $k=1,2,3$. A detailed proof for the normalisation of the order parameter is reported in Section S2 of the Supplementary Information. Since, by definition, each $S_i$ does not depend on the orientation of particles $j \neq i$, we could compute different order parameters depending on the set of particles considered for the average: the total order parameter is obtained by averaging over all the cubes in the system, \textit{i.e.} $S^{(tot)} = \langle S_i \rangle_{\forall i}$; the average order parameter of cubes that belong to chains is $S^{(c)} = \langle S_i \rangle_{\forall i \in c}$, for $c$ the set of cubes in clusters.

To investigate the formation of string-like structures we perform a cluster analysis on the simulated trajectories.
Given that $\beta u_{max}$ is our choice of the threshold energy for the cluster definition, two polarised cubes $i$ and $j$ are considered to form a cluster if $u_{ij,dip} \leq u_{max}$. The cluster analysis of the trajectories is performed
following the algorithm described by Sevick and co-workers \cite{Sevick1988}. To properly optimize the threshold parameter $\beta u_{max}$, we apply a density-based clustering algorithm to independent configurations of cubes in the string-like state. A detailed description of the method is reported in Section S1
of the Supplementary Information. From the cluster analysis, we estimate the number of clusters of size $l$ at time $t$, $N(l,t)$. The molar fraction of strings of size $l$ (free cubes
are labeled as $l=1$) reads
\begin{equation}
    X(l,t) = \frac{N(l,t)}{N_{cube}}.
\end{equation}
From the above expression we can recover the molar fraction of cubes that belong to strings as $X_c = \sum_{l=2}^{\infty} lX(l,t)$. Numerical and weighted average chain lengths read
\begin{equation}
    \langle l \rangle_n (t) = \frac{\sum\limits_{l=2}^{\infty} lN(l,t)}{\sum\limits_{l=2}^{\infty} N(l,t)}
    \label{eq:lcn}
\end{equation}
\begin{equation}
    \langle l \rangle_w (t)= \frac{\sum\limits_{l=2}^{\infty} l^2N(l,t)}{\sum\limits_{l=2}^{\infty} lN(l,t)}.
    \label{eq:lcw}
\end{equation}

%%%%%% Passive Microrheology

\subsection{Microrheological properties}

In passive microrheology, the viscoelastic properties of complex systems can be computed from the MSD of a tracer particle embedded in the host phase \cite{MASON1995,mason1997,mason2000,Evans2009,Nishi2018}. More specifically, according to Mason \cite{mason2000}, the complex shear modulus $G^{*}(\omega) = G'(\omega) + iG''(\omega)$, where $G'(\omega)$ and $G''(\omega)$ are the elastic and viscous moduli, respectively, can be written as
\begin{equation}
  |G^{*}(\omega)|=\frac{2 \sigma^3}{\pi d_{sphere} \langle \Delta r^2_t \left(1/\omega\right)\rangle\Gamma[1+\chi(\omega)]} \frac{1}{\beta\sigma^{3}},
  \label{eq:MR1}
\end{equation}
where $\chi(\omega)=d\ln(\Delta r^2_t(t))/d\ln (t)|_{t=\omega^{-1}}$ indicates the local exponent of tracer's MSD , $\Gamma$ is the gamma function and the modulus is in units of pressure, expressed as $1/(\beta\sigma^{3})$. Consequently, the elastic and viscous moduli are computed as follows
\begin{align}
  G'(\omega) &= |G^{*}(\omega)|\cos\left(\frac{\pi\chi(\omega)}{2}\right)\label{eq:MR2}\\
  G''(\omega) &= |G^{*}(\omega)|\sin\left(\frac{\pi\chi(\omega)}{2}\right).
  \label{eq:MR3}
\end{align}
While in a viscous system the motion of the particles is mainly diffusive ($\chi(\omega)\approx 1$) and $G''(\omega)$ is larger that $G'(\omega)$, in a elastic system the tracer motion is hindered by the local distribution of surrounding bath particles ($\chi(\omega) \ll 1$) and $G'(\omega)$ dominates over $G''(\omega)$. It should be pointed out that in addition to the Fourier approximation, other methods have been proposed to calculate the viscoelastic response of the bath from the tracer dynamics, including Laplace transform \cite{MASON1995} and compliance-based \cite{Evans2009,Nishi2018} methods among others. In Fig.~7 of the Supplementary Information, we compare the viscoelastic moduli calculated from the Fourier-based method (Eqs.~\ref{eq:MR1}-\ref{eq:MR3}) and the compliance approach proposed by Evans \textit{et al.} \cite{Evans2009} and show that they are in excellent quantitative agreement.

%%%%%%%%%%%%%%%%%%%%%%%%%%%%%%%%%%%%%%%%%%%%%%%%%
%%%%%%%%%%%%%%%%% RESULTS %%%%%%%%%%%%%%%%%%%%%%%
%%%%%%%%%%%%%%%%%%%%%%%%%%%%%%%%%%%%%%%%%%%%%%%%%

\section{Results and Discussion}

In the following, we first discuss the effect of applying an electric field to a suspension of polarisable nanocubes and what structural changes it determines. Then, we analyse
the nanocubes’ dynamics in the transitory unsteady state to
gain an insight into the mechanisms underpinning the process
of formation of chains. Finally, we employ passive MR to infer
the local viscoelastic behaviour of our model ER fluid and
compare it to the case where no external stimuli are applied.
As a reference system, we consider a suspension of hard nanocubes,
whose packing fraction \textit{vs} field intensity phase
diagram has been calculated by MC simulations by Dijkstra and co-workers \cite{Vutukuri2014}.

%%%%%% Effect of external field

\subsection{Effect of external field on system conformation}

\begin{figure}[ht!]
    \centering
    \includegraphics{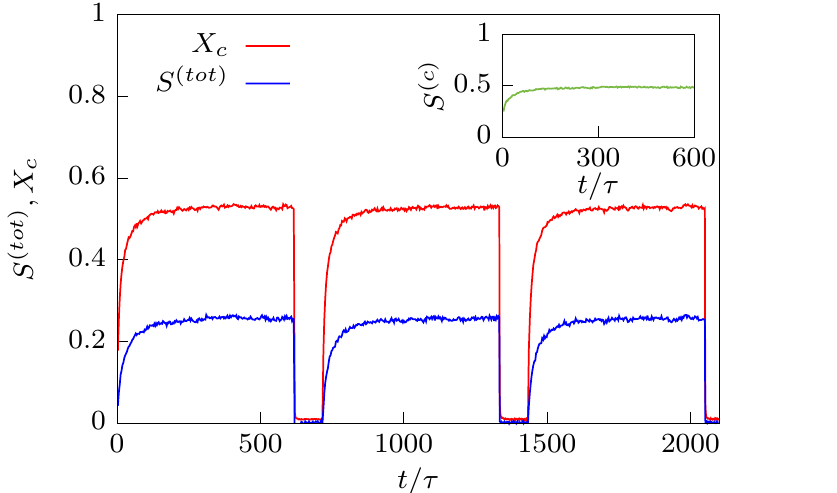}
    \caption{Average molar fraction of nanocubes in chains ($X_c$) and uniaxial order parameter ($S^{(tot)}$) over three on/off switching cycles. The inset  depicts the average uniaxial order parameter that only considers the nanocubes in chains ($S^{(c)}$), for the first on/off cycle (from $t=0$ to $600\,\tau$). The packing fraction and field strength are $\eta=0.02$ and $\gamma = 13$, respectively.}
    \label{fig:seq}
\end{figure}

Figure~\ref{fig:seq} shows the time evolution of the molar fraction $X_c$ of nanocubes in string-like clusters, and the uniaxial order parameter $S^{(tot)}$ averaged over all nanocubes, for three on/off field-switching cycles, at $\eta=0.02$ and $\gamma = 13$. This order parameter is specifically proposed for particles with cubic symmetry (see Eq.~\ref{eq:op_cube}). 
To identify the nanocubes belonging to the same cluster, we define an energy-based cluster criterion whose optimal parameterisation was achieved \textit{via} a density-based cluster analysis (see Methods and Section S1 of Supplementary Information). When the external field is switched on, $X_c$ increases up to $\sim 0.50$, indicating that nearly half of the nanocubes are assembled in strings. The clustering is completed at a response time $t_r \approx 150\,\tau$. We notice that, when the field is off, nanocubes displace an average distance approximately equal to $9\,\sigma$ over the same period of time. This suggests that, in terms of the nanocubes' ability to diffuse in an isotropic phase, the response time $t_r$ is relatively long.

\begin{figure}[ht!]
    \centering
    \includegraphics{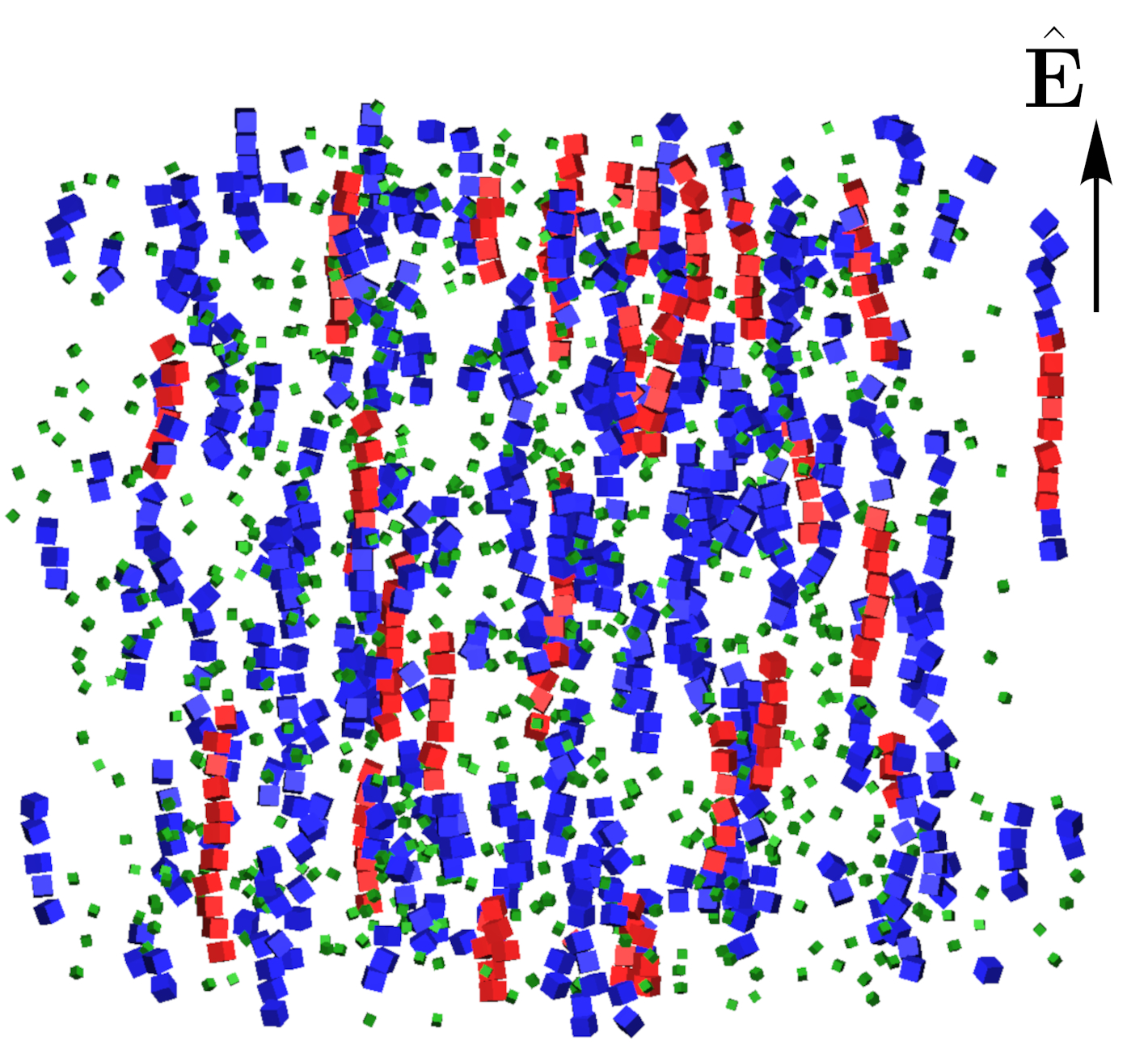}
    \caption{Suspensions of $N_{cube}=1500$ nanocubes upon the application of an external electric field. The orientation of the field $\vers{E}$ is shown with a black arrow on the top right of the Figure. Blue chains contain less than 6 nanocubes, while red chains contain at least 6 of them. Isolated nanocubes are shown in green and reduced in size for clarity. Packing fraction and field strength are $\eta=0.02$ and $\gamma = 13$, respectively.}
    \label{fig:state}
\end{figure}

Figure~\ref{fig:state} shows a typical snapshot of the system in the string-like state, with the strings of nanocubes oriented along $\vers{E}$. Most string-like clusters comprise between 2 and 4 nanocubes, while longer chains are less likely to be observed.
Our simulation results also reveal the presence of isolated nanocubes (reduced in size for clarity in Figure~\ref{fig:state}) that are free to assume random orientations with respect to the field direction, in agreement with former experiments and simulations \cite{Vutukuri2014}. In Figure~\ref{fig:seq}, we compare the total uniaxial order parameter $S^{(tot)}$ with the same parameter averaged exclusively over particles that belong to strings $S^{(c)}$ (see inset). In particular, $S^{(c)} =\langle S_i \rangle_{i\in c}$ is defined as the average contribution of particles in chains to $S^{(tot)}$, while $S^{(free)} = \langle S_i \rangle_{i\in free}$ is the contribution of free nanocubes. According to Figure~\ref{fig:seq}, during the switching on, the nanocubes in chains ($X_c \sim 50\%$) lead to $S^{(c)} \sim 0.50$ whereas $S^{(tot)}\sim 0.25$. Thereby, the contribution of free particles ($S^{(free)}$) to $S^{(tot)}$ is negligible since $S^{(tot)} \approx (S^{(c)} + S^{(free)})/2 \approx S^{(c)}/2$. One should notice that the dipolar interaction only depends on the distance between nanocubes and not on their space orientation. However, this interaction becomes very attractive when the nanocubes are face-to-face piled, minimizing their mutual distance and thus forming strings aligned with $\vect{E}$ as shown in Figure~\ref{fig:state}. 

When the field is switched off the system relaxes nearly instantaneously from the string-like state. At equilibrium, the isotropic configurations where both $X_c \approx 0$ and $S_c \approx 0$ are fully recovered. Their sudden decay is due to the approximations in our model, where the particle polarisation is assumed to be exclusively triggered by the external field. It is worth noting that a more general definition of particle polarisation should involve the contributions of both the external field and the particle-particle polarisation i.e. $\vect{p}_i = \alpha(\vect{E} + \sum_{i \neq j} \vect{E}_{j})$. Nevertheless, the approximation $\vect{p} = \alpha \vect{E}$ is often found in simulations of ER fluids \cite{Klingenberg1989,Whittle1990,Segovia2013,Belijar2016}. Incorporating both local and external fields into the particles' dynamics would require either the inversion of a $3N\times 3N$ matrix or the use of iterative procedures \cite{Xie2009,Bernardo1994,Vesely1977}, resulting in very demanding calculations. However, we notice that, for moderate electric fields, simulations of dilute ER fluids of spheres reveal that the particle-particle polarisation may contribute up to $5\%$ of the total dipole moment, and no tangible effects on dynamics are reported \cite{Fertig2021a}.

One may suggest that point-dipole approximations play a role in determining the dynamics of cubes when an electric field is applied. Vutukuri \textit{et al.} estimated the energy differences for pairs of aligned and misaligned cubes at face-to-face contact, being discretized in smaller cubes each containing one point dipole \cite{Vutukuri2014}. These authors found that energy differences level off to a limiting value of $0.01 \gamma k_{\rm{B}} T$ and, when comparing simulation results on single point-dipole cubes to those on multiple point-dipole cubes, they did not observe any notable difference in cubes alignment for field intensities below $\gamma=30$. Moreover, Kwaadgras \textit{et al.} applied the Couple Dipole Method to estimate the exact self-consistent dipolar interaction between two axis-aligned cubes polarised by an external field, as a function of their relative position in space, and compared it to the two aforementioned models \cite{Kwaadgras2014}. Both single and multiple point-dipole approximations showed similar discrepancies with respect to the exact model of interaction, underestimating attractions and overestimating repulsion. In conclusion, according to the findings reported in these works, the point-dipole approximation used here is not expected to be determinant in the formation of chains, at least for the field strength investigated in this work. Nonetheless, a more detailed description of the potential may highlight intriguing anisotropic phenomena occurring in the dynamics of cubes, especially at short distances, that cannot be observed when modelling the polarisation as a point-dipole.

%%%%%%%%%%%%%%%%%%%%%%%%%%%%%%%%%%%%
%%%%%% Structure of string-like state

\subsection{Structural properties at equilibrium}

\begin{figure}[ht!]
    \centering
    \includegraphics{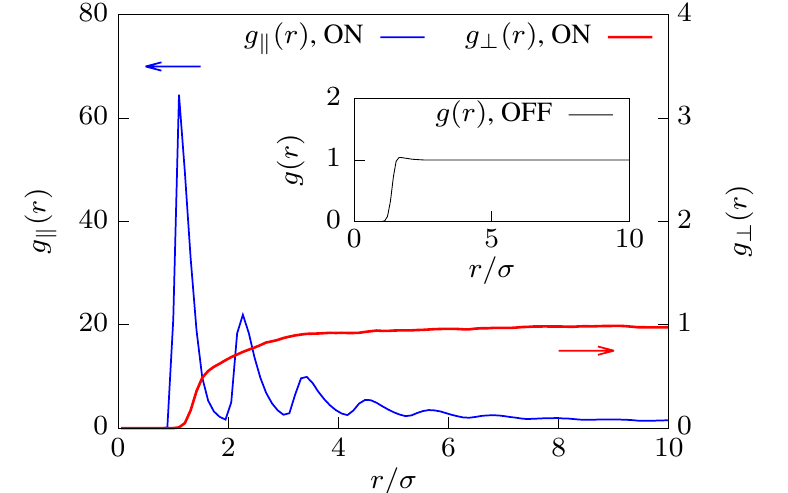}
    \caption{Pair correlation functions of nanocubes in the presence of an external field, in the direction parallel (blue line) and perpendicular (red line) to the field. For comparison, the radial distribution function when the field is off is also shown in the inset.
    }
    \label{fig:gofr}
\end{figure}

The structural organisation of nanocubes in suspension is inferred from the pair correlation functions parallel ($g_{\parallel}(r)$, see Eq.~\ref{eq:gpar}) and perpendicular ($g_{\perp}(r)$, see Eq.~\ref{eq:gper}) to the electric field. In Figure~\ref{fig:gofr}, we present $g_{\parallel}(r)$ and $g_{\perp}(r)$. For comparison, we also include the radial distribution function, $g(r)$, of the nanocubes in the isotropic phase, that is when the electric field is off.
The peaks observed in $g_{\parallel}(r)$ indicate a strong ordering of particles at short distances, arising from the piling of nanocubes on top of each other and the subsequent formation of string-like structures oriented along $\vers{E}$. This order vanishes for $r > 6\,\sigma$, suggesting that the occurrence of strings comprising more than 6 nanocubes is very unlikely. Moreover, we confirm that the string-like clusters do not percolate through the box boundaries, as $g_{\parallel}(r)$ decays at distances $r<l_{box}/2$, with $l_{box}\sim 42.2\, \sigma$ the box length. By contrast, no structural order is observed in the perpendicular direction. Since $g_{\perp}(r)<1$ at short distances, the chains are usually separated from each other and from isolated nanocubes by a long distance. Consequently, at $\eta=0.02$ and $\gamma=13$ the suspension is characterised by strong density heterogeneities that are not found in the isotropic phase, as illustrated by $g(r)$ in the inset of  Figure~\ref{fig:gofr}. 

%%%%%% Kinetics of off-on transition

\subsection{Dynamics of chain formation in transitory state}

In this section, we study the kinetics of string formation. To this end, we estimate the 2-particle connectivity with the aid of an energy threshold parameter, $\beta u_{max}=-3.2$. More specifically, two nanocubes $i$ and $j$ are considered to be connected if $\beta u_{ij,dip} < \beta u_{max}$. The cluster parameter $\beta u_{max}$ has been calculated by applying a density-based cluster analysis (DBSCAN) \cite{Hahsler2019,Schubert2017} on equilibrated configurations of strings of nanocubes. Details of these calculations are given in Section S1 of Supplementary Information.

\begin{figure*}[ht!]
    \centering
    \includegraphics{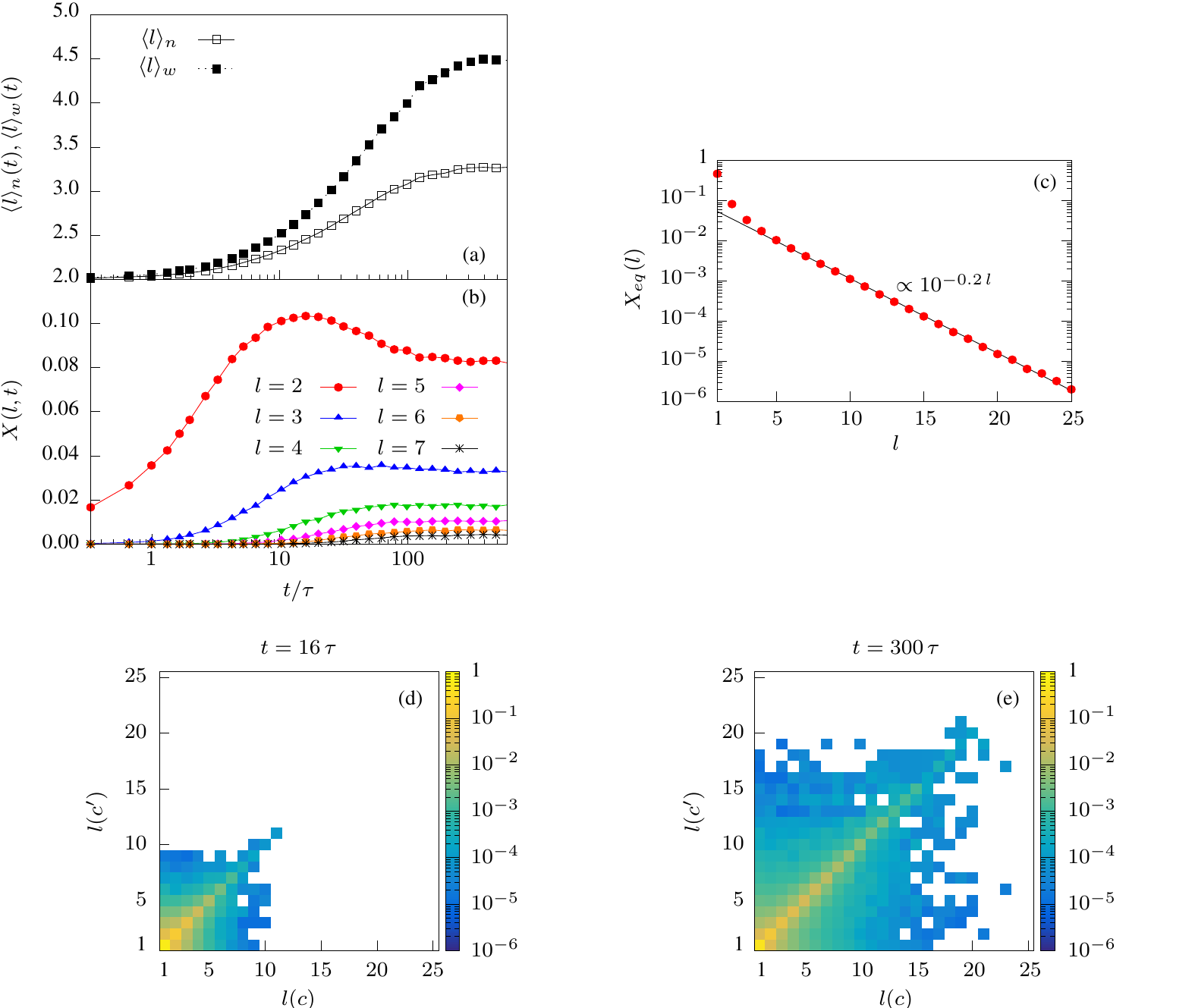}
    \caption{(a) Profiles of numerical ($\langle l \rangle_n(t)$) and weighted ($\langle l \rangle_w(t)$) average chain lengths over time, for the isotropic (field off) to string-like (field on) phase. (b) Time dependent profiles of molar fractions of chains of length $2\leq l \leq 7$, starting from the isotropic phase and switching the field on. (c) Equilibrium distribution of molar fraction of chain lengths
    in the string-like state from simulations (red points), together with the exponential decay obtained by nonlinear regression of data (black solid line). (d, e) Probability transition matrix of a nanocube moving from an initial cluster $c$ of size $l(c)$ to a cluster $c'$ of size $l(c')$ after $0.3\,\tau$, (d) at time $16\,\tau$, corresponding to the maximum of $X(2,t)$ and (e) at $300\,\tau$, when the system reaches the string-like state. Transition probabilities lower than $10^{-6}$ are in white.}
    \label{fig:kinetics}
\end{figure*}
Figure~\ref{fig:kinetics} reports the results obtained from the cluster analysis in suspensions of nanocubes at $\eta=0.02$ starting from an isotropic phase (field off) and ending into a phase of string-like clusters (field on). 
Numerical (Eq.~\ref{eq:lcn}) and weighted (Eq.~\ref{eq:lcw}) averaged chain lengths are reported in Figure~\ref{fig:kinetics}(a). Similarly, Figure~\ref{fig:kinetics}(b) presents the time dependence of the molar fraction of chains whose length ranges between $l=2$ and $l=7$. At short times ($t/\tau<1$), strings of sizes 2 and 3 are more likely to form. While the concentration of these short strings displays a local maximum at intermediate times ($10 < t/\tau < 50$), that of longer strings, with $l \geq 4$, is smoothly increasing until a plateau at times $t > t_r$ is achieved. Interestingly, although the molar fraction, $X_c$, of cubes in chains reaches a steady value at $t_r \sim 150\,\tau$, numerical ($\langle l \rangle_n$) and weighted ($\langle l \rangle_w$) lengths are still relaxing at this time. This suggests that strings of different sizes are continuously breaking and merging while reaching the equilibrium distribution regardless the fact that the percentage of nanocubes in chains remains basically constant for $t > t_r$ (see Figure~\ref{fig:seq}).

Figure~\ref{fig:kinetics}(c) contains the resulting equilibrium distribution of molar fractions $X_{eq}(t)$ for the different sizes of the string-like clusters. Our simulation results (red circles) are compared to a fitting model (black line) that assumes an exponential decay of $X_{eq}(t)$. According to the first-order Wertheim's perturbation theory of associating fluids, if the probability of two particles to be connected is constant and equal to $p_b$, a probability distribution of clusters at equilibrium can be defined as follows \cite{Wertheim1984,Wertheim1984_2,Wertheim1986,Smallenburg2012}
\begin{equation}
    X_{eq}(l) = (1- p_b)^2 p_b^{l-1}.
    \label{eq:wertheim}
\end{equation}
From Eq.\,\ref{eq:wertheim}, the linear dependence $\ln (X_{eq}) \propto l$ can be recovered. In our case, however, concentrations of free nanocubes and chains of 2 and 3 particles are significantly larger than those predicted by Wertheim's theory. We hypothesise that this discrepancy may arise from the crucial assumption underlying Wertheim's theory that considers particle connectivity to be independent from the cluster size. This approximation is not directly applicable to our system as dipolar interactions are long-ranged. Indeed, the mechanisms involved in the formation of string-like clusters of polarised particles are more complex than those in simple associating fluids. More specifically, the kinetics of formation of chains consisting of polarised nanoparticles is governed by the dipole moment of each particle, excluded-volume interactions and thermal fluctuations. 

To gain better insights into the kinetics of formation of string-like clusters of nanocubes, we compute the transition probability matrix of a cubic particle to belong to chain $c$ of size $l(c)$ at time $t_1$ and be in chain $c'$ of size $l(c')$ at time $t_2=t_1 + \delta t$, where $\delta t$ is set to $0.3\,\tau$ \cite{Chiricotto2019}. Figure~\ref{fig:kinetics}(d) and (e) present the transition matrices computed at times $t=16\,\tau$ and $t=300\,\tau$, corresponding, respectively, to the maximum and stationary values of $X(2,t)$ in Figure~\ref{fig:kinetics}(b). Since the largest probabilities have a tendency to lie on the diagonal ($l(c) = l(c')$), at long and short times most nanocubes prefer to remain in chain $c$ rather than diffusing to chain $c'$. At short times, strings contain no more than 10 nanocubes, while they grow up to $l\sim 20$ at longer times. Nonetheless, at the stationary state ($t=300\,\tau$), clusters with more than $25$ nanocubes have been occasionally seen, with $X_{eq}(l \geq 25) < 10^{-6}$ (Figure~\ref{fig:kinetics}(c)). It is interesting to note that, at both short and long time scales, a smooth decay of the transition probabilities is observed from short to long strings along the diagonal elements of the matrix, \textit{i.e.} where $l(c) = l(c')$, and also where $|l(c) - l(c')|$ gradually becomes larger. It has to be pointed out, however, that the transition matrices are not weighted by the size of the chains since their individual elements are computed tracking the single nanocubes rather than the cluster themselves. Nevertheless, a wide variety of transitions between chains of different lengths is still observed. We would expect that the kinetic rates characterising these transitions were string-size dependent and, what we actually observe is that aggregation and disaggregation mechanisms involving individual cubes or small clusters are more likely to occur rather than those involving larger strings. Equally interesting is the change of the mechanical properties of ER fluids upon switching the electric field on/off. The change in the viscoelastic behaviour of nanocubes in the presence of the external field is analysed in the following section by applying passive microrheology.

%%%%%% Passive Microrheology

\subsection{DMC Microrheology of ER fluids}

In passive MR, one can obtain the viscoelastic response of a soft material from the mean square displacement (MSD) of a tracer particle embedded in it. By estimating the complex shear modulus $G^{*}$, where $\omega$ is the frequency of interest, it is possible to identify the viscous ($G''$) and elastic ($G'$) moduli of the system from $G^{*}(\omega) = G'(\omega) + iG''(\omega)$. A number of methods have been proposed to determine $G^{*}$ from the dynamics of a tracer particle \cite{MASON1995,mason1997,mason2000,Evans2009,Nishi2018}. In this work, we apply the method developed by Mason \cite{mason2000} where the Fourier transform of tracer's MSD serves to estimate the viscoelastic properties of the bath. 

\begin{figure}[ht!]
    \centering
    \includegraphics{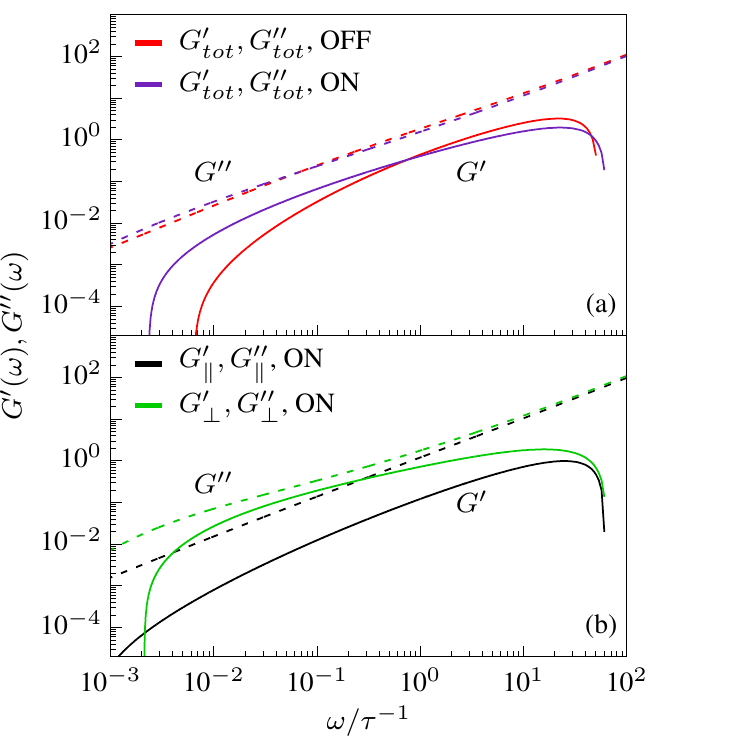}
    \caption{Viscous $G''$ (dashed lines) and elastic $G'$ (solid lines) moduli of a suspension of hard cubes at a packing fraction $\eta = 0.2$ containing a spherical tracer of diameter $3\,\sigma$. (a) Viscous and elastic moduli calculated in the three spatial coordinates with fields off (red curves) and on (blue curves). (b) Viscous and elastic moduli with the field on in the direction parallel (black curves) and perpendicular (green curves) to the field.}
    \label{fig:mr_parper}
\end{figure}

A preliminary MR analysis of a bath of nanocubes at a packing fraction $\eta = 0.02$ revealed no visible differences in the viscoelastic properties of the system when the external field is either on or off. Essentially, the very low concentration of bath particles does not affect the mobility of the tracer even when field-induced strings are formed, as the full equivalence of the MSDs reported in the Supplementary Information (Fig.~5, top frame) suggests. It follows that both viscous and elastic moduli in the presence of the field are the same as those calculated with no field applied. Consequently, to shed light on how the formation of chains affects the viscoelastic properties of the bath of nanocubes, we have increased the system packing by one order of magnitude. In Fig.~\ref{fig:mr_parper}, we report the viscous and elastic moduli for a system of nanocubes containing a spherical tracer of size $3\,\sigma$ at packing fraction $\eta = 0.20$. The top frame reports $G'$ and $G''$ calculated from the tracer's total MSD when the field is on and off. In both scenarios, the system exhibits a dominant viscous response, which becomes more viscoelastic at low and intermediate frequencies upon switching the field on (solid blue curve \textit{vs} solid red curve).   
Figure~\ref{fig:mr_parper}(a) shows the system's viscous and elastic moduli calculated from the total MSD of the tracer. While $G''$ is substantially unaffected by the presence of the electric field, $G'$ increases at short to intermediate frequencies ($\omega\tau<1$) when the field is switched on. The increase in $G'$, sparked by the presence of oriented strings that slow down the tracer's mobility at relatively long times (see Fig.~6(a) in the Supplementary Information), indicates that the material is significantly more viscoelastic compared to the field-off scenario. More intriguing is the viscoelastic response of the system in the directions parallel and perpendicular to the applied field, reported in Fig.~\ref{fig:mr_parper}(b). In particular, the perpendicular components of both $G'$ and $G''$ are larger than their counterparts along the direction of the field. This difference has its origin in the anisotropic diffusion of the tracer when the field is on (see the MSDs in Fig.~6(b) of the Supplementary Information). 
\begin{figure}[ht!]
    \centering
    \includegraphics{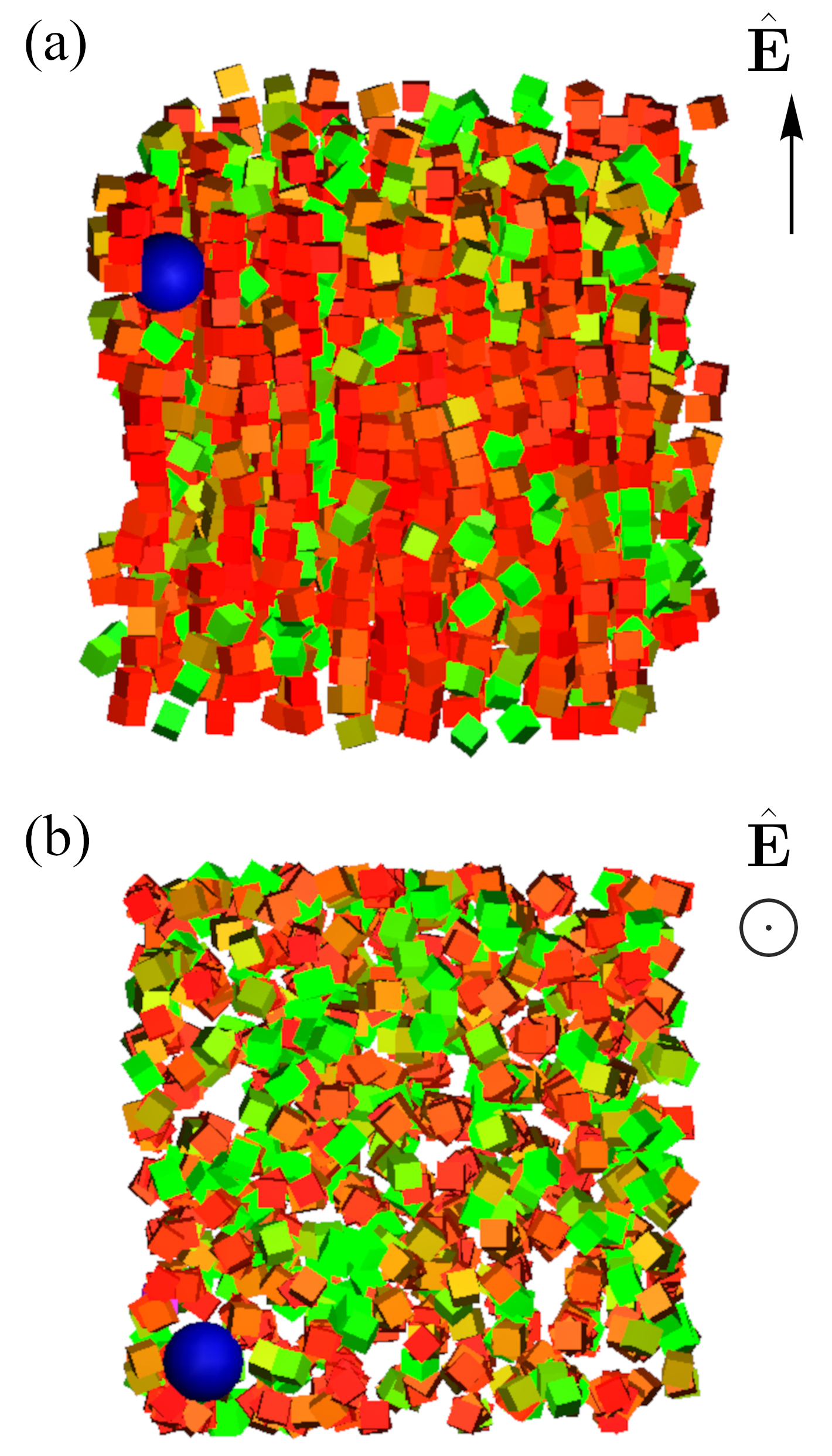}
    \caption{(Color on-line). (a) Side and (b) top view of a typical configuration of $N_{cube}=1500$ nanocubes of edge $\sigma$ and one spherical tracer of diameter $d_{sphere}=3\,\sigma$, under the application of the external field. The orientation of the field $\vers{E}$ is shown at the top right of each panel. Nanocubes are coloured with a gradient from green to red, depending on their orientation with respect to the external field. Packing fraction and field strength are $\eta = 0.20$ and $\gamma=13$, respectively.}
    \label{fig:snap_PMR}
\end{figure}
More specifically, the formation of chains redistributes the nanocubes in space by creating low-density vertical channels for the tracer diffusion in the direction of $\vers{E}$ and consequently an increase in density perpendicularly to $\vers{E}$. This particle redistribution, resulting into a density anisotropy, can be appreciated in the top and bottom frames of Fig.~\ref{fig:snap_PMR}. Accordingly, the viscoelastic properties assume an anisotropic behaviour with $G''_{\perp} > G''_{\parallel}$ and $G'_{\perp} > G'_{\parallel}$, as indicated in Figure~\ref{fig:mr_parper}(b). It follows that string-like phases of nanocubes exhibit a viscoelastic response in planes perpendicular to the field, but are substantially viscous along the field direction.

%%%%%%%%%%%%%%%%%%%%%%%%%%%%%%%%%%%%%%%%%%%%%%%%%
%%%%%%%%%%%%%%%% CONCLUSIONS %%%%%%%%%%%%%%%%%%%%
%%%%%%%%%%%%%%%%%%%%%%%%%%%%%%%%%%%%%%%%%%%%%%%%%

\section{Conclusions}

In summary, we employed DMC simulation to model the behaviour of colloidal suspensions of dielectric nanocubes upon application of an electric field. We also applied machine learning techniques to define a robust cluster criterion for the study of strings aggregation and breaking over time. Our results show that, at a volume fraction $\eta=0.02$ and field strength $\gamma = 13$, the suspension reaches a steady state where $\sim 50\%$ of particles are organized in aligned string-like clusters, with steady lengths and dispersity, or in individual cubes with no preferential orientation. Over the isotropic-to-string-like transition, one first observes the formation of relatively short strings, comprising 2 or 3 nanocubes. Subsequently, while clusters are still breaking and merging, the fraction of particles in chains converges to a steady-state value, which is fully achieved at $t_r \sim 150\,\tau$. By contrast, the concentration of chains with more than 3 nanocubes increases monotonically until equilibrium is reached. 

The kinetics of formation of clusters has been further investigated by analysing the transition probability of a nanocube to move from one cluster to another. Most nanocubes remain attached to a string over times shorter than the elementary time step and only a few of them move to clusters of different sizes. In fact, the larger the difference in size between the cluster a nanocube leaves and the cluster it joins, the lower the transition probability. This suggests that any aggregation and fragmentation mechanism can potentially occur, with larger probabilities for mechanisms that involve string-like clusters of similar sizes.

Finally, we have investigated how the viscoelastic response of our model suspensions is affected by the field-driven assembly of nanocubes in strings. To this end, we incorporated a spherical tracer into the system, calculate its total and directional MSDs and thus obtain the viscous and elastic moduli of the fluid. While very dilute ER fluids display the same viscoelastic response regardless of whether an electric field is applied or not, sufficiently dense ER fluids exhibit an increase in their viscoelasticity due to the field-induced string-like clusters. Such an enhanced viscoelastic response has an intriguing anisotropic signature with the viscous and elastic moduli significantly larger in the direction perpendicular to the electric field than parallel to it.    

All the simulation results reported in this work have been obtained neglecting the fluid-mediated hydrodynamic interactions (HI) between the nanocubes in suspension. Different works investigating the dynamics of colloids have shown quantitative discrepancies between simulations with and without HI, while retaining similar qualitative features \cite{Lettinga2010, Kwon2014, Pryamitsyn2008}. The same characteristics have been observed in the analysis of rheological properties of electrorheological \cite{Parthasarathy1999} and magnetorheological \cite{Segovia2013} fluids of spherical particles. HI have also been found to play a non-negligible role in active microrheology \cite{Weeber2012, Khair2006} and their inclusion in simulation generally results into a better agreement with experiments \cite{Torre2009}. Currently, a study on how HI can be incorporated into DMC simulation is in progress. Models of ER fluids can be further improved also by (\textit{i}) incorporating mutual polarisation of bath particles and (\textit{ii}) studying their behaviour under confinement to better reproduce an experimental setup.

% If in two-column mode, this environment will change to single-column format so that long equations can be displayed. 
% Use only when necessary.
%\begin{widetext}
%$$\mbox{put long equation here}$$
%\end{widetext}

% Figures should be put into the text as floats. 
% Use the graphics or graphicx packages (distributed with LaTeX2e).
% See the LaTeX Graphics Companion by Michel Goosens, Sebastian Rahtz, and Frank Mittelbach for examples. 
%
% Here is an example of the general form of a figure:
% Fill in the caption in the braces of the \caption{} command. 
% Put the label that you will use with \ref{} command in the braces of the \label{} command.
%
% \begin{figure}
% \includegraphics{}%
% \caption{\label{}}%
% \end{figure}

% Tables may be be put in the text as floats.
% Here is an example of the general form of a table:
% Fill in the caption in the braces of the \caption{} command. Put the label
% that you will use with \ref{} command in the braces of the \label{} command.
% Insert the column specifiers (l, r, c, d, etc.) in the empty braces of the
% \begin{tabular}{} command.
%
% \begin{table}
% \caption{\label{} }
% \begin{tabular}{}
% \end{tabular}
% \end{table}

%%% SUPPLEMENTARY MATERIAL

\section{Supplementary Material}

See the Supplementary Material for: \textit{(S1)} a description of the method developed to optimise the parameter used for the cluster definition; \textit{(S2)} a formal proof of the definition of the orientational order parameter $S$, normalised for particles with cubic symmetry; \textit{(S3)} the mean square displacement of a spherical tracer in a bath of nanocubes at $\eta = 0.02, 0.2$, in the field-on and field-off scenarios; \textit{(S4)} the benchmarking of the Fourier and compliance-based methods used to calculate the elastic and viscous moduli of a bath of nanocubes.

% If you have acknowledgments, this puts in the proper section head.
\begin{acknowledgments}
The authors acknowledge the Leverhulme Trust Research Project Grant RPG-2018-415 and the use of Computational Shared Facility at the University of Manchester. A.P. is supported by a ‘‘Maria Zambrano Senior” distinguished researcher fellowship, financed by the European Union within the NextGenerationEU program and the Spanish Ministry of Universities.
\end{acknowledgments}

% Create the reference section using BibTeX:
\bibliography{references}

%merlin.mbs aipnum4-1.bst 2010-07-25 4.21a (PWD, AO, DPC) hacked
%Control: key (0)
%Control: author (8) initials jnrlst
%Control: editor formatted (1) identically to author
%Control: production of article title (0) allowed
%Control: page (1) range
%Control: year (1) truncated
%Control: production of eprint (0) enabled
\providecommand{\noopsort}[1]{}\providecommand{\singleletter}[1]{#1}%
\begin{thebibliography}{92}%
\makeatletter
\providecommand \@ifxundefined [1]{%
 \@ifx{#1\undefined}
}%
\providecommand \@ifnum [1]{%
 \ifnum #1\expandafter \@firstoftwo
 \else \expandafter \@secondoftwo
 \fi
}%
\providecommand \@ifx [1]{%
 \ifx #1\expandafter \@firstoftwo
 \else \expandafter \@secondoftwo
 \fi
}%
\providecommand \natexlab [1]{#1}%
\providecommand \enquote  [1]{``#1''}%
\providecommand \bibnamefont  [1]{#1}%
\providecommand \bibfnamefont [1]{#1}%
\providecommand \citenamefont [1]{#1}%
\providecommand \href@noop [0]{\@secondoftwo}%
\providecommand \href [0]{\begingroup \@sanitize@url \@href}%
\providecommand \@href[1]{\@@startlink{#1}\@@href}%
\providecommand \@@href[1]{\endgroup#1\@@endlink}%
\providecommand \@sanitize@url [0]{\catcode `\\12\catcode `\$12\catcode
  `\&12\catcode `\#12\catcode `\^12\catcode `\_12\catcode `\%12\relax}%
\providecommand \@@startlink[1]{}%
\providecommand \@@endlink[0]{}%
\providecommand \url  [0]{\begingroup\@sanitize@url \@url }%
\providecommand \@url [1]{\endgroup\@href {#1}{\urlprefix }}%
\providecommand \urlprefix  [0]{URL }%
\providecommand \Eprint [0]{\href }%
\providecommand \doibase [0]{http://dx.doi.org/}%
\providecommand \selectlanguage [0]{\@gobble}%
\providecommand \bibinfo  [0]{\@secondoftwo}%
\providecommand \bibfield  [0]{\@secondoftwo}%
\providecommand \translation [1]{[#1]}%
\providecommand \BibitemOpen [0]{}%
\providecommand \bibitemStop [0]{}%
\providecommand \bibitemNoStop [0]{.\EOS\space}%
\providecommand \EOS [0]{\spacefactor3000\relax}%
\providecommand \BibitemShut  [1]{\csname bibitem#1\endcsname}%
\let\auto@bib@innerbib\@empty
%</preamble>
\bibitem [{\citenamefont {Winslow}(1949)}]{Winslow1949}%
  \BibitemOpen
  \bibfield  {author} {\bibinfo {author} {\bibfnamefont {W.~M.}\ \bibnamefont
  {Winslow}},\ }\bibfield  {title} {\enquote {\bibinfo {title} {{Induced
  Fibration of Suspensions}},}\ }\href {\doibase 10.1063/1.1698285} {\bibfield
  {journal} {\bibinfo  {journal} {J. Appl. Phys.}\ }\textbf {\bibinfo {volume}
  {20}},\ \bibinfo {pages} {1137--1140} (\bibinfo {year} {1949})}\BibitemShut
  {NoStop}%
\bibitem [{\citenamefont {Hao}(2001)}]{hao2001}%
  \BibitemOpen
  \bibfield  {author} {\bibinfo {author} {\bibfnamefont {T.}~\bibnamefont
  {Hao}},\ }\bibfield  {title} {\enquote {\bibinfo {title} {Electrorheological
  fluids},}\ }\href {\doibase
  10.1002/1521-4095(200112)13:24<1847::AID-ADMA1847>3.0.CO;2-A} {\bibfield
  {journal} {\bibinfo  {journal} {Advanced Materials}\ }\textbf {\bibinfo
  {volume} {13}},\ \bibinfo {pages} {1847--1857} (\bibinfo {year}
  {2001})}\BibitemShut {NoStop}%
\bibitem [{\citenamefont {Zhang}, \citenamefont {Liu},\ and\ \citenamefont
  {Choi}(2012)}]{zhang2012}%
  \BibitemOpen
  \bibfield  {author} {\bibinfo {author} {\bibfnamefont {K.}~\bibnamefont
  {Zhang}}, \bibinfo {author} {\bibfnamefont {Y.~D.}\ \bibnamefont {Liu}}, \
  and\ \bibinfo {author} {\bibfnamefont {H.~J.}\ \bibnamefont {Choi}},\
  }\bibfield  {title} {\enquote {\bibinfo {title} {{Carbon nanotube coated
  snowman-like particles and their electro-responsive characteristics}},}\
  }\href {\doibase 10.1039/C1CC16140G} {\bibfield  {journal} {\bibinfo
  {journal} {Chem. Commun.}\ }\textbf {\bibinfo {volume} {48}},\ \bibinfo
  {pages} {136--138} (\bibinfo {year} {2012})}\BibitemShut {NoStop}%
\bibitem [{\citenamefont {Dong}, \citenamefont {Seo},\ and\ \citenamefont
  {Choi}(2019)}]{dong2019}%
  \BibitemOpen
  \bibfield  {author} {\bibinfo {author} {\bibfnamefont {Y.~Z.}\ \bibnamefont
  {Dong}}, \bibinfo {author} {\bibfnamefont {Y.}~\bibnamefont {Seo}}, \ and\
  \bibinfo {author} {\bibfnamefont {H.~J.}\ \bibnamefont {Choi}},\ }\bibfield
  {title} {\enquote {\bibinfo {title} {{Recent development of
  electro-responsive smart electrorheological fluids}},}\ }\href {\doibase
  10.1039/C9SM00210C} {\bibfield  {journal} {\bibinfo  {journal} {Soft Matter}\
  }\textbf {\bibinfo {volume} {15}},\ \bibinfo {pages} {3473--3486} (\bibinfo
  {year} {2019})}\BibitemShut {NoStop}%
\bibitem [{\citenamefont {Clausius}(1879)}]{Clausius1879}%
  \BibitemOpen
  \bibfield  {author} {\bibinfo {author} {\bibfnamefont {R.}~\bibnamefont
  {Clausius}},\ }\enquote {\bibinfo {title} {Behandlung dielectrischer
  medien},}\ in\ \href {\doibase 10.1007/978-3-663-20232-5_3} {\emph {\bibinfo
  {booktitle} {Die Mechanische Behandlung der Electricität}}}\ (\bibinfo
  {publisher} {Vieweg+Teubner Verlag},\ \bibinfo {address} {Wiesbaden},\
  \bibinfo {year} {1879})\ pp.\ \bibinfo {pages} {62--97}\BibitemShut {NoStop}%
\bibitem [{\citenamefont {Mossotti}(1850)}]{Mossotti1850}%
  \BibitemOpen
  \bibfield  {author} {\bibinfo {author} {\bibfnamefont {O.~F.}\ \bibnamefont
  {Mossotti}},\ }\enquote {\bibinfo {title} {{Discussione analitica
  sull’influenza che l’azione di un mezzo dielettrico ha sulla
  distribuzione dell’elettricità alla superficie di più corpi elettrici
  disseminati in esso}},}\ in\ \href@noop {} {\emph {\bibinfo {booktitle}
  {Memorie Mat. Fis. Soc. ital. Sci}}}\ (\bibinfo  {publisher} {Società
  Italiana delle Scienze},\ \bibinfo {address} {Modena},\ \bibinfo {year}
  {1850})\ pp.\ \bibinfo {pages} {49--74}\BibitemShut {NoStop}%
\bibitem [{\citenamefont {Toor}(1993)}]{Toor1993}%
  \BibitemOpen
  \bibfield  {author} {\bibinfo {author} {\bibfnamefont {W.~R.}\ \bibnamefont
  {Toor}},\ }\bibfield  {title} {\enquote {\bibinfo {title} {{Structure
  Formation in Electrorheological Fluids}},}\ }\href {\doibase
  10.1006/jcis.1993.1121} {\bibfield  {journal} {\bibinfo  {journal} {J. Coll.
  Inter. Sci.}\ }\textbf {\bibinfo {volume} {156}},\ \bibinfo {pages}
  {335--349} (\bibinfo {year} {1993})}\BibitemShut {NoStop}%
\bibitem [{\citenamefont {Martin}, \citenamefont {Odinek},\ and\ \citenamefont
  {Halsey}(1992)}]{Martin1992}%
  \BibitemOpen
  \bibfield  {author} {\bibinfo {author} {\bibfnamefont {J.~E.}\ \bibnamefont
  {Martin}}, \bibinfo {author} {\bibfnamefont {J.}~\bibnamefont {Odinek}}, \
  and\ \bibinfo {author} {\bibfnamefont {T.~C.}\ \bibnamefont {Halsey}},\
  }\bibfield  {title} {\enquote {\bibinfo {title} {{Evolution of structure in a
  quiescent electrorheological fluid}},}\ }\href {\doibase
  10.1103/PhysRevLett.69.1524} {\bibfield  {journal} {\bibinfo  {journal}
  {Phys. Rev. Lett.}\ }\textbf {\bibinfo {volume} {69}},\ \bibinfo {pages}
  {1524--1527} (\bibinfo {year} {1992})}\BibitemShut {NoStop}%
\bibitem [{\citenamefont {Martin}\ \emph {et~al.}(1998)\citenamefont {Martin},
  \citenamefont {Odinek}, \citenamefont {Halsey},\ and\ \citenamefont
  {Kamien}}]{Martin1998}%
  \BibitemOpen
  \bibfield  {author} {\bibinfo {author} {\bibfnamefont {J.~E.}\ \bibnamefont
  {Martin}}, \bibinfo {author} {\bibfnamefont {J.}~\bibnamefont {Odinek}},
  \bibinfo {author} {\bibfnamefont {T.~C.}\ \bibnamefont {Halsey}}, \ and\
  \bibinfo {author} {\bibfnamefont {R.}~\bibnamefont {Kamien}},\ }\bibfield
  {title} {\enquote {\bibinfo {title} {{Structure and dynamics of
  electrorheological fluids}},}\ }\href {\doibase 10.1103/PhysRevE.57.756}
  {\bibfield  {journal} {\bibinfo  {journal} {Phys. Rev. E}\ }\textbf {\bibinfo
  {volume} {57}},\ \bibinfo {pages} {756--775} (\bibinfo {year}
  {1998})}\BibitemShut {NoStop}%
\bibitem [{\citenamefont {Cao}, \citenamefont {Huang},\ and\ \citenamefont
  {Zhou}(2006)}]{Cao2006}%
  \BibitemOpen
  \bibfield  {author} {\bibinfo {author} {\bibfnamefont {J.~G.}\ \bibnamefont
  {Cao}}, \bibinfo {author} {\bibfnamefont {J.~P.}\ \bibnamefont {Huang}}, \
  and\ \bibinfo {author} {\bibfnamefont {L.~W.}\ \bibnamefont {Zhou}},\
  }\bibfield  {title} {\enquote {\bibinfo {title} {{Structure of
  Electrorheological Fluids under an Electric Field and a Shear Flow:
  Experiment and Computer Simulation}},}\ }\href {\doibase 10.1021/jp0611774}
  {\bibfield  {journal} {\bibinfo  {journal} {J. Phys. Chem. B}\ }\textbf
  {\bibinfo {volume} {110}},\ \bibinfo {pages} {11635--11639} (\bibinfo {year}
  {2006})}\BibitemShut {NoStop}%
\bibitem [{\citenamefont {Horv{\'a}th}\ and\ \citenamefont
  {Szalai}(2012)}]{Horvat2012}%
  \BibitemOpen
  \bibfield  {author} {\bibinfo {author} {\bibfnamefont {B.}~\bibnamefont
  {Horv{\'a}th}}\ and\ \bibinfo {author} {\bibfnamefont {I.}~\bibnamefont
  {Szalai}},\ }\bibfield  {title} {\enquote {\bibinfo {title} {{Structure of
  electrorheological fluids: A dielectric study of chain formation}},}\ }\href
  {\doibase 10.1103/PhysRevE.86.061403} {\bibfield  {journal} {\bibinfo
  {journal} {Phys. Rev. E}\ }\textbf {\bibinfo {volume} {86}},\ \bibinfo
  {pages} {061403} (\bibinfo {year} {2012})}\BibitemShut {NoStop}%
\bibitem [{\citenamefont {Belijar}\ \emph {et~al.}(2016)\citenamefont
  {Belijar}, \citenamefont {Valdez-Nava}, \citenamefont {Diaham}, \citenamefont
  {Laudebat}, \citenamefont {Jones},\ and\ \citenamefont
  {Lebey}}]{Belijar2016}%
  \BibitemOpen
  \bibfield  {author} {\bibinfo {author} {\bibfnamefont {G.}~\bibnamefont
  {Belijar}}, \bibinfo {author} {\bibfnamefont {Z.}~\bibnamefont
  {Valdez-Nava}}, \bibinfo {author} {\bibfnamefont {S.}~\bibnamefont {Diaham}},
  \bibinfo {author} {\bibfnamefont {L.}~\bibnamefont {Laudebat}}, \bibinfo
  {author} {\bibfnamefont {T.~B.}\ \bibnamefont {Jones}}, \ and\ \bibinfo
  {author} {\bibfnamefont {T.}~\bibnamefont {Lebey}},\ }\bibfield  {title}
  {\enquote {\bibinfo {title} {{Dynamics of particle chain formation in a
  liquid polymer under ac electric field: modeling and experiments}},}\ }\href
  {\doibase 10.1088/1361-6463/50/2/025303} {\bibfield  {journal} {\bibinfo
  {journal} {J. Phys. D: Appl. Phys.}\ }\textbf {\bibinfo {volume} {50}},\
  \bibinfo {pages} {025303} (\bibinfo {year} {2016})}\BibitemShut {NoStop}%
\bibitem [{\citenamefont {Baxter‐Drayton}\ and\ \citenamefont
  {Brady}(1996)}]{Baxter1996}%
  \BibitemOpen
  \bibfield  {author} {\bibinfo {author} {\bibfnamefont {Y.}~\bibnamefont
  {Baxter‐Drayton}}\ and\ \bibinfo {author} {\bibfnamefont {J.~F.}\
  \bibnamefont {Brady}},\ }\bibfield  {title} {\enquote {\bibinfo {title}
  {{Brownian electrorheological fluids as a model for flocculated
  dispersions}},}\ }\href {\doibase 10.1122/1.550772} {\bibfield  {journal}
  {\bibinfo  {journal} {J. Rheo.}\ }\textbf {\bibinfo {volume} {40}},\ \bibinfo
  {pages} {1027--1056} (\bibinfo {year} {1996})}\BibitemShut {NoStop}%
\bibitem [{\citenamefont {Klingenberg}, \citenamefont {{van Swol}},\ and\
  \citenamefont {Zukoski}(1989)}]{Klingenberg1989}%
  \BibitemOpen
  \bibfield  {author} {\bibinfo {author} {\bibfnamefont {D.~J.}\ \bibnamefont
  {Klingenberg}}, \bibinfo {author} {\bibfnamefont {F.}~\bibnamefont {{van
  Swol}}}, \ and\ \bibinfo {author} {\bibfnamefont {C.~F.}\ \bibnamefont
  {Zukoski}},\ }\bibfield  {title} {\enquote {\bibinfo {title} {{Dynamic
  simulation of electrorheological suspensions}},}\ }\href {\doibase
  10.1063/1.457256} {\bibfield  {journal} {\bibinfo  {journal} {J. Chem.
  Phys.}\ }\textbf {\bibinfo {volume} {91}},\ \bibinfo {pages} {7888--7895}
  (\bibinfo {year} {1989})}\BibitemShut {NoStop}%
\bibitem [{\citenamefont {Whittle}(1990)}]{Whittle1990}%
  \BibitemOpen
  \bibfield  {author} {\bibinfo {author} {\bibfnamefont {M.}~\bibnamefont
  {Whittle}},\ }\bibfield  {title} {\enquote {\bibinfo {title} {{Computer
  simulation of an electrorheological fluid}},}\ }\href {\doibase
  10.1016/0377-0257(90)90007-X} {\bibfield  {journal} {\bibinfo  {journal} {J.
  Non-Newtonian Fluid Mech.}\ }\textbf {\bibinfo {volume} {37}},\ \bibinfo
  {pages} {233--263} (\bibinfo {year} {1990})}\BibitemShut {NoStop}%
\bibitem [{\citenamefont {Hass}(1993)}]{Hass1993}%
  \BibitemOpen
  \bibfield  {author} {\bibinfo {author} {\bibfnamefont {K.~C.}\ \bibnamefont
  {Hass}},\ }\bibfield  {title} {\enquote {\bibinfo {title} {{Computer
  simulations of nonequilibrium structure formation in electrorheological
  fluids}},}\ }\href {\doibase 10.1103/PhysRevE.47.3362} {\bibfield  {journal}
  {\bibinfo  {journal} {Phys. Rev. E}\ }\textbf {\bibinfo {volume} {47}},\
  \bibinfo {pages} {3362--3373} (\bibinfo {year} {1993})}\BibitemShut {NoStop}%
\bibitem [{\citenamefont {Fertig}, \citenamefont {Boda},\ and\ \citenamefont
  {Szalai}(2021{\natexlab{a}})}]{Fertig2021}%
  \BibitemOpen
  \bibfield  {author} {\bibinfo {author} {\bibfnamefont {D.}~\bibnamefont
  {Fertig}}, \bibinfo {author} {\bibfnamefont {D.}~\bibnamefont {Boda}}, \ and\
  \bibinfo {author} {\bibfnamefont {I.}~\bibnamefont {Szalai}},\ }\bibfield
  {title} {\enquote {\bibinfo {title} {{Induced permittivity increment of
  electrorheological fluids in an applied electric field in association with
  chain formation: A Brownian dynamics simulation study}},}\ }\href {\doibase
  10.1103/PhysRevE.103.062608} {\bibfield  {journal} {\bibinfo  {journal}
  {Phys. Rev. E}\ }\textbf {\bibinfo {volume} {103}},\ \bibinfo {pages}
  {062608} (\bibinfo {year} {2021}{\natexlab{a}})}\BibitemShut {NoStop}%
\bibitem [{\citenamefont {Fertig}, \citenamefont {Boda},\ and\ \citenamefont
  {Szalai}(2021{\natexlab{b}})}]{Fertig2021a}%
  \BibitemOpen
  \bibfield  {author} {\bibinfo {author} {\bibfnamefont {D.}~\bibnamefont
  {Fertig}}, \bibinfo {author} {\bibfnamefont {D.}~\bibnamefont {Boda}}, \ and\
  \bibinfo {author} {\bibfnamefont {I.}~\bibnamefont {Szalai}},\ }\bibfield
  {title} {\enquote {\bibinfo {title} {{A systematic study of the dynamics of
  chain formation in electrorheological fluids}},}\ }\href {\doibase
  10.1063/5.0037985} {\bibfield  {journal} {\bibinfo  {journal} {AIP Advances}\
  }\textbf {\bibinfo {volume} {11}},\ \bibinfo {pages} {25243} (\bibinfo {year}
  {2021}{\natexlab{b}})}\BibitemShut {NoStop}%
\bibitem [{\citenamefont {Sun}\ and\ \citenamefont {Xia}(2002)}]{sun2002}%
  \BibitemOpen
  \bibfield  {author} {\bibinfo {author} {\bibfnamefont {Y.}~\bibnamefont
  {Sun}}\ and\ \bibinfo {author} {\bibfnamefont {Y.}~\bibnamefont {Xia}},\
  }\bibfield  {title} {\enquote {\bibinfo {title} {{Shape-Controlled Synthesis
  of Gold and Silver Nanoparticles}},}\ }\href {\doibase
  10.1126/science.1077229} {\bibfield  {journal} {\bibinfo  {journal}
  {Science}\ }\textbf {\bibinfo {volume} {298}},\ \bibinfo {pages} {2176--2179}
  (\bibinfo {year} {2002})}\BibitemShut {NoStop}%
\bibitem [{\citenamefont {Manoharan}, \citenamefont {Elsesser},\ and\
  \citenamefont {Pine}(2003)}]{Manoharan2003}%
  \BibitemOpen
  \bibfield  {author} {\bibinfo {author} {\bibfnamefont {V.~N.}\ \bibnamefont
  {Manoharan}}, \bibinfo {author} {\bibfnamefont {M.~T.}\ \bibnamefont
  {Elsesser}}, \ and\ \bibinfo {author} {\bibfnamefont {D.~J.}\ \bibnamefont
  {Pine}},\ }\bibfield  {title} {\enquote {\bibinfo {title} {{Dense Packing and
  Symmetry in Small Clusters of Microspheres}},}\ }\href {\doibase
  10.1126/science.1086189} {\bibfield  {journal} {\bibinfo  {journal}
  {Science}\ }\textbf {\bibinfo {volume} {301}},\ \bibinfo {pages} {483--487}
  (\bibinfo {year} {2003})}\BibitemShut {NoStop}%
\bibitem [{\citenamefont {Shankar}\ \emph {et~al.}(2004)\citenamefont
  {Shankar}, \citenamefont {Rai}, \citenamefont {Ankamwar}, \citenamefont
  {Singh}, \citenamefont {Ahmad},\ and\ \citenamefont {Sastry}}]{Shankar2004}%
  \BibitemOpen
  \bibfield  {author} {\bibinfo {author} {\bibfnamefont {S.~S.}\ \bibnamefont
  {Shankar}}, \bibinfo {author} {\bibfnamefont {A.}~\bibnamefont {Rai}},
  \bibinfo {author} {\bibfnamefont {B.}~\bibnamefont {Ankamwar}}, \bibinfo
  {author} {\bibfnamefont {A.}~\bibnamefont {Singh}}, \bibinfo {author}
  {\bibfnamefont {A.}~\bibnamefont {Ahmad}}, \ and\ \bibinfo {author}
  {\bibfnamefont {M.}~\bibnamefont {Sastry}},\ }\bibfield  {title} {\enquote
  {\bibinfo {title} {{Biological synthesis of triangular gold nanoprisms}},}\
  }\href {\doibase 10.1038/nmat1152} {\bibfield  {journal} {\bibinfo  {journal}
  {Nature Materials}\ }\textbf {\bibinfo {volume} {3}},\ \bibinfo {pages}
  {482--488} (\bibinfo {year} {2004})}\BibitemShut {NoStop}%
\bibitem [{\citenamefont {Xiang}\ \emph {et~al.}(2006)\citenamefont {Xiang},
  \citenamefont {Wu}, \citenamefont {Liu}, \citenamefont {Jiang}, \citenamefont
  {Chu}, \citenamefont {Li}, \citenamefont {Ma}, \citenamefont {Zhou},\ and\
  \citenamefont {Xie}}]{xiang2006}%
  \BibitemOpen
  \bibfield  {author} {\bibinfo {author} {\bibfnamefont {Y.}~\bibnamefont
  {Xiang}}, \bibinfo {author} {\bibfnamefont {X.}~\bibnamefont {Wu}}, \bibinfo
  {author} {\bibfnamefont {D.}~\bibnamefont {Liu}}, \bibinfo {author}
  {\bibfnamefont {X.}~\bibnamefont {Jiang}}, \bibinfo {author} {\bibfnamefont
  {W.}~\bibnamefont {Chu}}, \bibinfo {author} {\bibfnamefont {Z.}~\bibnamefont
  {Li}}, \bibinfo {author} {\bibfnamefont {Y.}~\bibnamefont {Ma}}, \bibinfo
  {author} {\bibfnamefont {W.}~\bibnamefont {Zhou}}, \ and\ \bibinfo {author}
  {\bibfnamefont {S.}~\bibnamefont {Xie}},\ }\bibfield  {title} {\enquote
  {\bibinfo {title} {{Formation of Rectangularly Shaped Pd/Au Bimetallic
  Nanorods: Evidence for Competing Growth of the Pd Shell between the {110} and
  {100} Side Facets of Au Nanorods}},}\ }\href {\doibase 10.1021/nl061722c}
  {\bibfield  {journal} {\bibinfo  {journal} {Nano Letters}\ }\textbf {\bibinfo
  {volume} {6}},\ \bibinfo {pages} {2290--2294} (\bibinfo {year}
  {2006})}\BibitemShut {NoStop}%
\bibitem [{\citenamefont {Sacanna}\ \emph {et~al.}(2010)\citenamefont
  {Sacanna}, \citenamefont {Irvine}, \citenamefont {Chaikin},\ and\
  \citenamefont {Pine}}]{sacanna2010}%
  \BibitemOpen
  \bibfield  {author} {\bibinfo {author} {\bibfnamefont {S.}~\bibnamefont
  {Sacanna}}, \bibinfo {author} {\bibfnamefont {W.~T.~M.}\ \bibnamefont
  {Irvine}}, \bibinfo {author} {\bibfnamefont {P.~M.}\ \bibnamefont {Chaikin}},
  \ and\ \bibinfo {author} {\bibfnamefont {D.~J.}\ \bibnamefont {Pine}},\
  }\bibfield  {title} {\enquote {\bibinfo {title} {{Lock and key colloids}},}\
  }\href {\doibase 10.1038/nature08906} {\bibfield  {journal} {\bibinfo
  {journal} {Nature}\ }\textbf {\bibinfo {volume} {464}},\ \bibinfo {pages}
  {575--578} (\bibinfo {year} {2010})}\BibitemShut {NoStop}%
\bibitem [{\citenamefont {Okuno}\ \emph {et~al.}(2010)\citenamefont {Okuno},
  \citenamefont {Nishioka}, \citenamefont {Kiya}, \citenamefont {Nakashima},
  \citenamefont {Ishibashi},\ and\ \citenamefont {Niidome}}]{okuno2010}%
  \BibitemOpen
  \bibfield  {author} {\bibinfo {author} {\bibfnamefont {Y.}~\bibnamefont
  {Okuno}}, \bibinfo {author} {\bibfnamefont {K.}~\bibnamefont {Nishioka}},
  \bibinfo {author} {\bibfnamefont {A.}~\bibnamefont {Kiya}}, \bibinfo {author}
  {\bibfnamefont {N.}~\bibnamefont {Nakashima}}, \bibinfo {author}
  {\bibfnamefont {A.}~\bibnamefont {Ishibashi}}, \ and\ \bibinfo {author}
  {\bibfnamefont {Y.}~\bibnamefont {Niidome}},\ }\bibfield  {title} {\enquote
  {\bibinfo {title} {{Uniform and controllable preparation of Au–Ag
  core–shell nanorods using anisotropic silver shell formation on gold
  nanorods}},}\ }\href {\doibase 10.1039/C0NR00130A} {\bibfield  {journal}
  {\bibinfo  {journal} {Nanoscale}\ }\textbf {\bibinfo {volume} {2}},\ \bibinfo
  {pages} {1489--1493} (\bibinfo {year} {2010})}\BibitemShut {NoStop}%
\bibitem [{\citenamefont {Cortie}\ \emph {et~al.}(2012)\citenamefont {Cortie},
  \citenamefont {Liu}, \citenamefont {Arnold},\ and\ \citenamefont
  {Niidome}}]{cortie2012}%
  \BibitemOpen
  \bibfield  {author} {\bibinfo {author} {\bibfnamefont {M.~B.}\ \bibnamefont
  {Cortie}}, \bibinfo {author} {\bibfnamefont {F.}~\bibnamefont {Liu}},
  \bibinfo {author} {\bibfnamefont {M.~D.}\ \bibnamefont {Arnold}}, \ and\
  \bibinfo {author} {\bibfnamefont {Y.}~\bibnamefont {Niidome}},\ }\bibfield
  {title} {\enquote {\bibinfo {title} {{Multimode Resonances in Silver
  Nanocuboids}},}\ }\href {\doibase 10.1021/la300407u} {\bibfield  {journal}
  {\bibinfo  {journal} {Langmuir}\ }\textbf {\bibinfo {volume} {28}},\ \bibinfo
  {pages} {9103--9112} (\bibinfo {year} {2012})}\BibitemShut {NoStop}%
\bibitem [{\citenamefont {Sacanna}\ \emph {et~al.}(2013)\citenamefont
  {Sacanna}, \citenamefont {Korpics}, \citenamefont {Rodriguez}, \citenamefont
  {Col{\'o}n-Mel{\'e}ndez}, \citenamefont {Kim}, \citenamefont {Pine},\ and\
  \citenamefont {Yi}}]{sacanna2013}%
  \BibitemOpen
  \bibfield  {author} {\bibinfo {author} {\bibfnamefont {S.}~\bibnamefont
  {Sacanna}}, \bibinfo {author} {\bibfnamefont {M.}~\bibnamefont {Korpics}},
  \bibinfo {author} {\bibfnamefont {K.}~\bibnamefont {Rodriguez}}, \bibinfo
  {author} {\bibfnamefont {L.}~\bibnamefont {Col{\'o}n-Mel{\'e}ndez}}, \bibinfo
  {author} {\bibfnamefont {S.-H.}\ \bibnamefont {Kim}}, \bibinfo {author}
  {\bibfnamefont {D.~J.}\ \bibnamefont {Pine}}, \ and\ \bibinfo {author}
  {\bibfnamefont {G.-R.}\ \bibnamefont {Yi}},\ }\bibfield  {title} {\enquote
  {\bibinfo {title} {Shaping colloids for self-assembly},}\ }\href {\doibase
  10.1038/ncomms2694} {\bibfield  {journal} {\bibinfo  {journal} {Nature
  Communications}\ }\textbf {\bibinfo {volume} {4}},\ \bibinfo {pages} {1688}
  (\bibinfo {year} {2013})}\BibitemShut {NoStop}%
\bibitem [{\citenamefont {Khlebtsov}\ \emph {et~al.}(2015)\citenamefont
  {Khlebtsov}, \citenamefont {Liu}, \citenamefont {Ye},\ and\ \citenamefont
  {Khlebtsov}}]{Khlebtsov2015}%
  \BibitemOpen
  \bibfield  {author} {\bibinfo {author} {\bibfnamefont {B.~N.}\ \bibnamefont
  {Khlebtsov}}, \bibinfo {author} {\bibfnamefont {Z.}~\bibnamefont {Liu}},
  \bibinfo {author} {\bibfnamefont {J.}~\bibnamefont {Ye}}, \ and\ \bibinfo
  {author} {\bibfnamefont {N.~G.}\ \bibnamefont {Khlebtsov}},\ }\bibfield
  {title} {\enquote {\bibinfo {title} {{Au@Ag core/shell cuboids and dumbbells:
  Optical properties and SERS response}},}\ }\href {\doibase
  10.1016/j.jqsrt.2015.07.024} {\bibfield  {journal} {\bibinfo  {journal}
  {Journal of Quantitative Spectroscopy and Radiative Transfer}\ }\textbf
  {\bibinfo {volume} {167}},\ \bibinfo {pages} {64--75} (\bibinfo {year}
  {2015})}\BibitemShut {NoStop}%
\bibitem [{\citenamefont {Rossi}\ \emph {et~al.}(2015)\citenamefont {Rossi},
  \citenamefont {Soni}, \citenamefont {Ashton}, \citenamefont {Pine},
  \citenamefont {Philipse}, \citenamefont {Chaikin}, \citenamefont {Dijkstra},
  \citenamefont {Sacanna},\ and\ \citenamefont {Irvine}}]{rossi2015}%
  \BibitemOpen
  \bibfield  {author} {\bibinfo {author} {\bibfnamefont {L.}~\bibnamefont
  {Rossi}}, \bibinfo {author} {\bibfnamefont {V.}~\bibnamefont {Soni}},
  \bibinfo {author} {\bibfnamefont {D.~J.}\ \bibnamefont {Ashton}}, \bibinfo
  {author} {\bibfnamefont {D.~J.}\ \bibnamefont {Pine}}, \bibinfo {author}
  {\bibfnamefont {A.~P.}\ \bibnamefont {Philipse}}, \bibinfo {author}
  {\bibfnamefont {P.~M.}\ \bibnamefont {Chaikin}}, \bibinfo {author}
  {\bibfnamefont {M.}~\bibnamefont {Dijkstra}}, \bibinfo {author}
  {\bibfnamefont {S.}~\bibnamefont {Sacanna}}, \ and\ \bibinfo {author}
  {\bibfnamefont {W.~T.~M.}\ \bibnamefont {Irvine}},\ }\bibfield  {title}
  {\enquote {\bibinfo {title} {{Shape-sensitive crystallization in colloidal
  superball fluids}},}\ }\href {\doibase 10.1073/pnas.1415467112} {\bibfield
  {journal} {\bibinfo  {journal} {Proceedings of the National Academy of
  Sciences}\ }\textbf {\bibinfo {volume} {112}},\ \bibinfo {pages} {5286--5290}
  (\bibinfo {year} {2015})}\BibitemShut {NoStop}%
\bibitem [{\citenamefont {Glotzer}\ and\ \citenamefont
  {Solomon}(2007)}]{Glotzer2007}%
  \BibitemOpen
  \bibfield  {author} {\bibinfo {author} {\bibfnamefont {S.~C.}\ \bibnamefont
  {Glotzer}}\ and\ \bibinfo {author} {\bibfnamefont {M.}~\bibnamefont
  {Solomon}},\ }\bibfield  {title} {\enquote {\bibinfo {title} {{Anisotropy of
  building blocks and their assembly into complex structures}},}\ }\href
  {\doibase 10.1038/nmat1949} {\bibfield  {journal} {\bibinfo  {journal} {Nat.
  Mater.}\ }\textbf {\bibinfo {volume} {6}},\ \bibinfo {pages} {557–562}
  (\bibinfo {year} {2007})}\BibitemShut {NoStop}%
\bibitem [{\citenamefont {Zhang}, \citenamefont {Zhang},\ and\ \citenamefont
  {Glotzer}(2007)}]{Xi2007}%
  \BibitemOpen
  \bibfield  {author} {\bibinfo {author} {\bibfnamefont {X.}~\bibnamefont
  {Zhang}}, \bibinfo {author} {\bibfnamefont {Z.}~\bibnamefont {Zhang}}, \ and\
  \bibinfo {author} {\bibfnamefont {S.~C.}\ \bibnamefont {Glotzer}},\
  }\bibfield  {title} {\enquote {\bibinfo {title} {{Simulation Study of
  Dipole-Induced Self-Assembly of Nanocubes}},}\ }\href {\doibase
  10.1021/jp065953j} {\bibfield  {journal} {\bibinfo  {journal} {J. Phys. Chem.
  C}\ }\textbf {\bibinfo {volume} {111}},\ \bibinfo {pages} {4132--4137}
  (\bibinfo {year} {2007})}\BibitemShut {NoStop}%
\bibitem [{\citenamefont {Batten}, \citenamefont {Stillinger},\ and\
  \citenamefont {Torquato}(2010)}]{Torquato2010}%
  \BibitemOpen
  \bibfield  {author} {\bibinfo {author} {\bibfnamefont {R.~D.}\ \bibnamefont
  {Batten}}, \bibinfo {author} {\bibfnamefont {F.~H.}\ \bibnamefont
  {Stillinger}}, \ and\ \bibinfo {author} {\bibfnamefont {S.}~\bibnamefont
  {Torquato}},\ }\bibfield  {title} {\enquote {\bibinfo {title} {{Phase
  behavior of colloidal superballs: Shape interpolation from spheres to
  cubes}},}\ }\href {\doibase 10.1103/PhysRevE.81.061105} {\bibfield  {journal}
  {\bibinfo  {journal} {Phys. Rev. E}\ }\textbf {\bibinfo {volume} {81}},\
  \bibinfo {pages} {061105} (\bibinfo {year} {2010})}\BibitemShut {NoStop}%
\bibitem [{\citenamefont {Damasceno}, \citenamefont {Engel},\ and\
  \citenamefont {Glotzer}(2012)}]{Damasceno2012}%
  \BibitemOpen
  \bibfield  {author} {\bibinfo {author} {\bibfnamefont {P.~F.}\ \bibnamefont
  {Damasceno}}, \bibinfo {author} {\bibfnamefont {M.}~\bibnamefont {Engel}}, \
  and\ \bibinfo {author} {\bibfnamefont {S.~C.}\ \bibnamefont {Glotzer}},\
  }\bibfield  {title} {\enquote {\bibinfo {title} {{Predictive Self-Assembly of
  Polyhedra into Complex Structures}},}\ }\href {\doibase
  10.1126/science.1220869} {\bibfield  {journal} {\bibinfo  {journal}
  {Science}\ }\textbf {\bibinfo {volume} {337}},\ \bibinfo {pages} {453--457}
  (\bibinfo {year} {2012})}\BibitemShut {NoStop}%
\bibitem [{\citenamefont {Donaldson}, \citenamefont {Linse},\ and\
  \citenamefont {Kantorovich}(2017)}]{Donaldson2017}%
  \BibitemOpen
  \bibfield  {author} {\bibinfo {author} {\bibfnamefont {J.~G.}\ \bibnamefont
  {Donaldson}}, \bibinfo {author} {\bibfnamefont {P.}~\bibnamefont {Linse}}, \
  and\ \bibinfo {author} {\bibfnamefont {S.~S.}\ \bibnamefont {Kantorovich}},\
  }\bibfield  {title} {\enquote {\bibinfo {title} {{How cube-like must magnetic
  nanoparticles be to modify their self-assembly?}}}\ }\href {\doibase
  10.1039/C7NR01245D} {\bibfield  {journal} {\bibinfo  {journal} {Nanoscale}\
  }\textbf {\bibinfo {volume} {9}},\ \bibinfo {pages} {6448--6462} (\bibinfo
  {year} {2017})}\BibitemShut {NoStop}%
\bibitem [{\citenamefont {Cuetos}\ \emph {et~al.}(2017)\citenamefont {Cuetos},
  \citenamefont {Dennison}, \citenamefont {Masters},\ and\ \citenamefont
  {Patti}}]{Dennison2017}%
  \BibitemOpen
  \bibfield  {author} {\bibinfo {author} {\bibfnamefont {A.}~\bibnamefont
  {Cuetos}}, \bibinfo {author} {\bibfnamefont {M.}~\bibnamefont {Dennison}},
  \bibinfo {author} {\bibfnamefont {A.}~\bibnamefont {Masters}}, \ and\
  \bibinfo {author} {\bibfnamefont {A.}~\bibnamefont {Patti}},\ }\bibfield
  {title} {\enquote {\bibinfo {title} {{Phase behaviour of hard board-like
  particles}},}\ }\href {\doibase 10.1039/C7SM00726D} {\bibfield  {journal}
  {\bibinfo  {journal} {Soft Matter}\ }\textbf {\bibinfo {volume} {13}},\
  \bibinfo {pages} {4720--4732} (\bibinfo {year} {2017})}\BibitemShut {NoStop}%
\bibitem [{\citenamefont {Donaldson}, \citenamefont {Pyanzina},\ and\
  \citenamefont {Kantorovich}(2017)}]{Joe2017}%
  \BibitemOpen
  \bibfield  {author} {\bibinfo {author} {\bibfnamefont {J.~G.}\ \bibnamefont
  {Donaldson}}, \bibinfo {author} {\bibfnamefont {E.~S.}\ \bibnamefont
  {Pyanzina}}, \ and\ \bibinfo {author} {\bibfnamefont {S.~S.}\ \bibnamefont
  {Kantorovich}},\ }\bibfield  {title} {\enquote {\bibinfo {title}
  {{Nanoparticle Shape Influences the Magnetic Response of Ferro-Colloids}},}\
  }\href {\doibase 10.1021/acsnano.7b03064} {\bibfield  {journal} {\bibinfo
  {journal} {ACS Nano}\ }\textbf {\bibinfo {volume} {11}},\ \bibinfo {pages}
  {8153--8166} (\bibinfo {year} {2017})}\BibitemShut {NoStop}%
\bibitem [{\citenamefont {Patti}\ and\ \citenamefont
  {Cuetos}(2018)}]{Patti2018}%
  \BibitemOpen
  \bibfield  {author} {\bibinfo {author} {\bibfnamefont {A.}~\bibnamefont
  {Patti}}\ and\ \bibinfo {author} {\bibfnamefont {A.}~\bibnamefont {Cuetos}},\
  }\bibfield  {title} {\enquote {\bibinfo {title} {{Monte Carlo simulation of
  binary mixtures of hard colloidal cuboids}},}\ }\href {\doibase
  10.1080/08927022.2017.1402307} {\bibfield  {journal} {\bibinfo  {journal}
  {Molecular Simulation}\ }\textbf {\bibinfo {volume} {44}},\ \bibinfo {pages}
  {516--522} (\bibinfo {year} {2018})}\BibitemShut {NoStop}%
\bibitem [{\citenamefont {Chiappini}\ \emph {et~al.}(2019)\citenamefont
  {Chiappini}, \citenamefont {Drwenski}, \citenamefont {{van Roij}},\ and\
  \citenamefont {Dijkstra}}]{Chiappini2019}%
  \BibitemOpen
  \bibfield  {author} {\bibinfo {author} {\bibfnamefont {M.}~\bibnamefont
  {Chiappini}}, \bibinfo {author} {\bibfnamefont {T.}~\bibnamefont {Drwenski}},
  \bibinfo {author} {\bibfnamefont {R.}~\bibnamefont {{van Roij}}}, \ and\
  \bibinfo {author} {\bibfnamefont {M.}~\bibnamefont {Dijkstra}},\ }\bibfield
  {title} {\enquote {\bibinfo {title} {{Biaxial, Twist-bend, and Splay-bend
  Nematic Phases of Banana-shaped Particles Revealed by Lifting the ``Smectic
  Blanket''}},}\ }\href {\doibase 10.1103/PhysRevLett.123.068001} {\bibfield
  {journal} {\bibinfo  {journal} {Phys. Rev. Lett.}\ }\textbf {\bibinfo
  {volume} {123}},\ \bibinfo {pages} {068001} (\bibinfo {year}
  {2019})}\BibitemShut {NoStop}%
\bibitem [{\citenamefont {Cuetos}\ \emph {et~al.}(2019)\citenamefont {Cuetos},
  \citenamefont {{Mirzad Rafael}}, \citenamefont {Corbett},\ and\ \citenamefont
  {Patti}}]{Cuetos2019}%
  \BibitemOpen
  \bibfield  {author} {\bibinfo {author} {\bibfnamefont {A.}~\bibnamefont
  {Cuetos}}, \bibinfo {author} {\bibfnamefont {E.}~\bibnamefont {{Mirzad
  Rafael}}}, \bibinfo {author} {\bibfnamefont {D.}~\bibnamefont {Corbett}}, \
  and\ \bibinfo {author} {\bibfnamefont {A.}~\bibnamefont {Patti}},\ }\bibfield
   {title} {\enquote {\bibinfo {title} {{Biaxial nematics of hard cuboids in an
  external field}},}\ }\href {\doibase 10.1039/C8SM02283F} {\bibfield
  {journal} {\bibinfo  {journal} {Soft Matter}\ }\textbf {\bibinfo {volume}
  {15}},\ \bibinfo {pages} {1922--1926} (\bibinfo {year} {2019})}\BibitemShut
  {NoStop}%
\bibitem [{\citenamefont {Casquilho}\ and\ \citenamefont
  {Figueirinhas}(2021)}]{Casquilho2021}%
  \BibitemOpen
  \bibfield  {author} {\bibinfo {author} {\bibfnamefont {J.~P.}\ \bibnamefont
  {Casquilho}}\ and\ \bibinfo {author} {\bibfnamefont {J.~L.}\ \bibnamefont
  {Figueirinhas}},\ }\bibfield  {title} {\enquote {\bibinfo {title} {{Lattice
  Monte Carlo study of orientational order in a confined system of biaxial
  particles: Effect of an external electric field}},}\ }\href {\doibase
  10.1103/PhysRevE.103.032701} {\bibfield  {journal} {\bibinfo  {journal}
  {Phys. Rev. E}\ }\textbf {\bibinfo {volume} {103}},\ \bibinfo {pages}
  {032701} (\bibinfo {year} {2021})}\BibitemShut {NoStop}%
\bibitem [{\citenamefont {Yan}\ \emph {et~al.}(2013)\citenamefont {Yan},
  \citenamefont {Chaudhary}, \citenamefont {{Chul Bae}}, \citenamefont
  {Lewis},\ and\ \citenamefont {Granick}}]{Yan2013}%
  \BibitemOpen
  \bibfield  {author} {\bibinfo {author} {\bibfnamefont {J.}~\bibnamefont
  {Yan}}, \bibinfo {author} {\bibfnamefont {K.}~\bibnamefont {Chaudhary}},
  \bibinfo {author} {\bibfnamefont {S.}~\bibnamefont {{Chul Bae}}}, \bibinfo
  {author} {\bibfnamefont {J.~A.}\ \bibnamefont {Lewis}}, \ and\ \bibinfo
  {author} {\bibfnamefont {S.}~\bibnamefont {Granick}},\ }\bibfield  {title}
  {\enquote {\bibinfo {title} {{Colloidal ribbons and rings from Janus magnetic
  rods}},}\ }\href {\doibase 10.1038/ncomms2520} {\bibfield  {journal}
  {\bibinfo  {journal} {Nat. Commun.}\ }\textbf {\bibinfo {volume} {4}},\
  \bibinfo {pages} {1516} (\bibinfo {year} {2013})}\BibitemShut {NoStop}%
\bibitem [{\citenamefont {Okada}\ and\ \citenamefont
  {Satoh}(2018)}]{Okada2018}%
  \BibitemOpen
  \bibfield  {author} {\bibinfo {author} {\bibfnamefont {K.}~\bibnamefont
  {Okada}}\ and\ \bibinfo {author} {\bibfnamefont {A.}~\bibnamefont {Satoh}},\
  }\bibfield  {title} {\enquote {\bibinfo {title} {{3D Monte Carlo simulations
  on the aggregate structures of a suspension composed of cubic hematite
  particles}},}\ }\href {\doibase 10.1080/00268976.2018.1478138} {\bibfield
  {journal} {\bibinfo  {journal} {Molecular Physics}\ }\textbf {\bibinfo
  {volume} {116}},\ \bibinfo {pages} {2300--2309} (\bibinfo {year}
  {2018})}\BibitemShut {NoStop}%
\bibitem [{\citenamefont {Kuijk}\ \emph {et~al.}(2014)\citenamefont {Kuijk},
  \citenamefont {Troppenz}, \citenamefont {Filion}, \citenamefont {Imhof},
  \citenamefont {{van Roij}}, \citenamefont {Dijkstra},\ and\ \citenamefont
  {{van Blaaderen}}}]{Anke2014}%
  \BibitemOpen
  \bibfield  {author} {\bibinfo {author} {\bibfnamefont {A.}~\bibnamefont
  {Kuijk}}, \bibinfo {author} {\bibfnamefont {T.}~\bibnamefont {Troppenz}},
  \bibinfo {author} {\bibfnamefont {L.}~\bibnamefont {Filion}}, \bibinfo
  {author} {\bibfnamefont {A.}~\bibnamefont {Imhof}}, \bibinfo {author}
  {\bibfnamefont {R.}~\bibnamefont {{van Roij}}}, \bibinfo {author}
  {\bibfnamefont {M.}~\bibnamefont {Dijkstra}}, \ and\ \bibinfo {author}
  {\bibfnamefont {A.}~\bibnamefont {{van Blaaderen}}},\ }\bibfield  {title}
  {\enquote {\bibinfo {title} {{Effect of external electric fields on the phase
  behavior of colloidal silica rods}},}\ }\href {\doibase 10.1039/C4SM00957F}
  {\bibfield  {journal} {\bibinfo  {journal} {Soft Matter}\ }\textbf {\bibinfo
  {volume} {10}},\ \bibinfo {pages} {6249--6255} (\bibinfo {year}
  {2014})}\BibitemShut {NoStop}%
\bibitem [{\citenamefont {Smallenburg}\ \emph {et~al.}(2012)\citenamefont
  {Smallenburg}, \citenamefont {Vutukuri}, \citenamefont {Imhof}, \citenamefont
  {{van Blaaderen}},\ and\ \citenamefont {Dijkstra}}]{Smallenburg2012}%
  \BibitemOpen
  \bibfield  {author} {\bibinfo {author} {\bibfnamefont {F.}~\bibnamefont
  {Smallenburg}}, \bibinfo {author} {\bibfnamefont {H.~R.}\ \bibnamefont
  {Vutukuri}}, \bibinfo {author} {\bibfnamefont {A.}~\bibnamefont {Imhof}},
  \bibinfo {author} {\bibfnamefont {A.}~\bibnamefont {{van Blaaderen}}}, \ and\
  \bibinfo {author} {\bibfnamefont {M.}~\bibnamefont {Dijkstra}},\ }\bibfield
  {title} {\enquote {\bibinfo {title} {{Self-assembly of colloidal particles
  into strings in a homogeneous external electric or magnetic field}},}\ }\href
  {\doibase 10.1088/0953-8984/24/46/464113} {\bibfield  {journal} {\bibinfo
  {journal} {Journal of Physics: Condensed Matter}\ }\textbf {\bibinfo {volume}
  {24}},\ \bibinfo {pages} {464113} (\bibinfo {year} {2012})}\BibitemShut
  {NoStop}%
\bibitem [{\citenamefont {Hynninen}\ and\ \citenamefont
  {Dijkstra}(2005)}]{Hynniyen2005}%
  \BibitemOpen
  \bibfield  {author} {\bibinfo {author} {\bibfnamefont {A.-P.}\ \bibnamefont
  {Hynninen}}\ and\ \bibinfo {author} {\bibfnamefont {M.}~\bibnamefont
  {Dijkstra}},\ }\bibfield  {title} {\enquote {\bibinfo {title} {{Phase Diagram
  of Dipolar Hard and Soft Spheres: Manipulation of Colloidal Crystal
  Structures by an External Field}},}\ }\href {\doibase
  10.1103/PhysRevLett.94.138303} {\bibfield  {journal} {\bibinfo  {journal}
  {Phys. Rev. Lett.}\ }\textbf {\bibinfo {volume} {94}},\ \bibinfo {pages}
  {138303} (\bibinfo {year} {2005})}\BibitemShut {NoStop}%
\bibitem [{\citenamefont {Kanu}\ and\ \citenamefont {Shaw}(1998)}]{Rex1998}%
  \BibitemOpen
  \bibfield  {author} {\bibinfo {author} {\bibfnamefont {R.~C.}\ \bibnamefont
  {Kanu}}\ and\ \bibinfo {author} {\bibfnamefont {M.~T.}\ \bibnamefont
  {Shaw}},\ }\bibfield  {title} {\enquote {\bibinfo {title} {{Enhanced
  electrorheological fluids using anisotropic particles}},}\ }\href {\doibase
  10.1122/1.550944} {\bibfield  {journal} {\bibinfo  {journal} {J. Rheo.}\
  }\textbf {\bibinfo {volume} {42}},\ \bibinfo {pages} {657--670} (\bibinfo
  {year} {1998})}\BibitemShut {NoStop}%
\bibitem [{\citenamefont {Qi}\ and\ \citenamefont {Wen}(2002)}]{Qi2002}%
  \BibitemOpen
  \bibfield  {author} {\bibinfo {author} {\bibfnamefont {Y.}~\bibnamefont
  {Qi}}\ and\ \bibinfo {author} {\bibfnamefont {W.}~\bibnamefont {Wen}},\
  }\bibfield  {title} {\enquote {\bibinfo {title} {{Influences of geometry of
  particles on electrorheological fluids}},}\ }\href {\doibase
  10.1088/0022-3727/35/17/322} {\bibfield  {journal} {\bibinfo  {journal} {J.
  Phys. D: Appl. Phys.}\ }\textbf {\bibinfo {volume} {35}},\ \bibinfo {pages}
  {2231--2235} (\bibinfo {year} {2002})}\BibitemShut {NoStop}%
\bibitem [{\citenamefont {He}\ \emph {et~al.}(2017)\citenamefont {He},
  \citenamefont {Wen}, \citenamefont {Wang}, \citenamefont {Wang},
  \citenamefont {Yu}, \citenamefont {Hao},\ and\ \citenamefont
  {Chen}}]{Kai2017}%
  \BibitemOpen
  \bibfield  {author} {\bibinfo {author} {\bibfnamefont {K.}~\bibnamefont
  {He}}, \bibinfo {author} {\bibfnamefont {Q.}~\bibnamefont {Wen}}, \bibinfo
  {author} {\bibfnamefont {C.}~\bibnamefont {Wang}}, \bibinfo {author}
  {\bibfnamefont {B.}~\bibnamefont {Wang}}, \bibinfo {author} {\bibfnamefont
  {S.}~\bibnamefont {Yu}}, \bibinfo {author} {\bibfnamefont {C.}~\bibnamefont
  {Hao}}, \ and\ \bibinfo {author} {\bibfnamefont {K.}~\bibnamefont {Chen}},\
  }\bibfield  {title} {\enquote {\bibinfo {title} {{Synthesis of anatase TiO2
  with exposed (100) facets and enhanced electrorheological activity}},}\
  }\href {\doibase 10.1039/C7SM01422H} {\bibfield  {journal} {\bibinfo
  {journal} {Soft Matter}\ }\textbf {\bibinfo {volume} {13}},\ \bibinfo {pages}
  {7879--7889} (\bibinfo {year} {2017})}\BibitemShut {NoStop}%
\bibitem [{\citenamefont {Vutukuri}\ \emph {et~al.}(2012)\citenamefont
  {Vutukuri}, \citenamefont {Demir{\"o}rs}, \citenamefont {Peng}, \citenamefont
  {{van Oostrum}}, \citenamefont {Imhof},\ and\ \citenamefont {{van
  Blaaderen}}}]{Vutukuri2012}%
  \BibitemOpen
  \bibfield  {author} {\bibinfo {author} {\bibfnamefont {H.~R.}\ \bibnamefont
  {Vutukuri}}, \bibinfo {author} {\bibfnamefont {A.~F.}\ \bibnamefont
  {Demir{\"o}rs}}, \bibinfo {author} {\bibfnamefont {B.}~\bibnamefont {Peng}},
  \bibinfo {author} {\bibfnamefont {P.~D.~J.}\ \bibnamefont {{van Oostrum}}},
  \bibinfo {author} {\bibfnamefont {A.}~\bibnamefont {Imhof}}, \ and\ \bibinfo
  {author} {\bibfnamefont {A.}~\bibnamefont {{van Blaaderen}}},\ }\bibfield
  {title} {\enquote {\bibinfo {title} {{Colloidal Analogues of Charged and
  Uncharged Polymer Chains with Tunable Stiffness}},}\ }\href {\doibase
  10.1002/anie.201202592} {\bibfield  {journal} {\bibinfo  {journal}
  {Angewandte Chemie International Edition}\ }\textbf {\bibinfo {volume}
  {51}},\ \bibinfo {pages} {11249--11253} (\bibinfo {year} {2012})}\BibitemShut
  {NoStop}%
\bibitem [{\citenamefont {Vutukuri}\ \emph {et~al.}(2017)\citenamefont
  {Vutukuri}, \citenamefont {Bet}, \citenamefont {{van Roij}}, \citenamefont
  {Dijkstra},\ and\ \citenamefont {Huck}}]{Vutukuri2017}%
  \BibitemOpen
  \bibfield  {author} {\bibinfo {author} {\bibfnamefont {H.~R.}\ \bibnamefont
  {Vutukuri}}, \bibinfo {author} {\bibfnamefont {B.}~\bibnamefont {Bet}},
  \bibinfo {author} {\bibfnamefont {R.}~\bibnamefont {{van Roij}}}, \bibinfo
  {author} {\bibfnamefont {M.}~\bibnamefont {Dijkstra}}, \ and\ \bibinfo
  {author} {\bibfnamefont {W.~T.~S.}\ \bibnamefont {Huck}},\ }\bibfield
  {title} {\enquote {\bibinfo {title} {{Rational design and dynamics of
  self-propelled colloidal bead chains: from rotators to flagella}},}\ }\href
  {\doibase 10.1038/s41598-017-16731-5} {\bibfield  {journal} {\bibinfo
  {journal} {Sci. Rep.}\ }\textbf {\bibinfo {volume} {7}},\ \bibinfo {pages}
  {16758} (\bibinfo {year} {2017})}\BibitemShut {NoStop}%
\bibitem [{\citenamefont {Huang}\ \emph {et~al.}(2022)\citenamefont {Huang},
  \citenamefont {Li}, \citenamefont {Gao}, \citenamefont {Gao}, \citenamefont
  {Xu}, \citenamefont {Lin},\ and\ \citenamefont {Cai}}]{Huang2022}%
  \BibitemOpen
  \bibfield  {author} {\bibinfo {author} {\bibfnamefont {S.}~\bibnamefont
  {Huang}}, \bibinfo {author} {\bibfnamefont {Z.}~\bibnamefont {Li}}, \bibinfo
  {author} {\bibfnamefont {L.}~\bibnamefont {Gao}}, \bibinfo {author}
  {\bibfnamefont {H.}~\bibnamefont {Gao}}, \bibinfo {author} {\bibfnamefont
  {Z.}~\bibnamefont {Xu}}, \bibinfo {author} {\bibfnamefont {J.}~\bibnamefont
  {Lin}}, \ and\ \bibinfo {author} {\bibfnamefont {C.}~\bibnamefont {Cai}},\
  }\bibfield  {title} {\enquote {\bibinfo {title} {{Colloidal Polymers of Iron
  Oxide Cubes Prepared by Dipolar-Driven Assembly and In Situ Welding with
  Silica}},}\ }\href {\doibase 10.1016/j.giant.2021.100083} {\bibfield
  {journal} {\bibinfo  {journal} {Giant}\ }\textbf {\bibinfo {volume} {9}},\
  \bibinfo {pages} {100083} (\bibinfo {year} {2022})}\BibitemShut {NoStop}%
\bibitem [{\citenamefont {Patti}\ and\ \citenamefont
  {Cuetos}(2012)}]{Patti2012}%
  \BibitemOpen
  \bibfield  {author} {\bibinfo {author} {\bibfnamefont {A.}~\bibnamefont
  {Patti}}\ and\ \bibinfo {author} {\bibfnamefont {A.}~\bibnamefont {Cuetos}},\
  }\bibfield  {title} {\enquote {\bibinfo {title} {{Brownian dynamics and
  dynamic Monte Carlo simulations of isotropic and liquid crystal phases of
  anisotropic colloidal particles: A comparative study}},}\ }\href {\doibase
  10.1103/PhysRevE.86.011403} {\bibfield  {journal} {\bibinfo  {journal} {Phys.
  Rev. E}\ }\textbf {\bibinfo {volume} {86}},\ \bibinfo {pages} {011403}
  (\bibinfo {year} {2012})}\BibitemShut {NoStop}%
\bibitem [{\citenamefont {Cuetos}\ and\ \citenamefont
  {Patti}(2015)}]{Cuetos2015}%
  \BibitemOpen
  \bibfield  {author} {\bibinfo {author} {\bibfnamefont {A.}~\bibnamefont
  {Cuetos}}\ and\ \bibinfo {author} {\bibfnamefont {A.}~\bibnamefont {Patti}},\
  }\bibfield  {title} {\enquote {\bibinfo {title} {{Equivalence of Brownian
  dynamics and dynamic Monte Carlo simulations in multicomponent colloidal
  suspensions}},}\ }\href {\doibase 10.1103/PhysRevE.92.022302} {\bibfield
  {journal} {\bibinfo  {journal} {Phys. Rev. E}\ }\textbf {\bibinfo {volume}
  {92}},\ \bibinfo {pages} {022302} (\bibinfo {year} {2015})}\BibitemShut
  {NoStop}%
\bibitem [{\citenamefont {Corbett}\ \emph {et~al.}(2018)\citenamefont
  {Corbett}, \citenamefont {Cuetos}, \citenamefont {Dennison},\ and\
  \citenamefont {Patti}}]{Corbett2018}%
  \BibitemOpen
  \bibfield  {author} {\bibinfo {author} {\bibfnamefont {D.}~\bibnamefont
  {Corbett}}, \bibinfo {author} {\bibfnamefont {A.}~\bibnamefont {Cuetos}},
  \bibinfo {author} {\bibfnamefont {M.}~\bibnamefont {Dennison}}, \ and\
  \bibinfo {author} {\bibfnamefont {A.}~\bibnamefont {Patti}},\ }\bibfield
  {title} {\enquote {\bibinfo {title} {{Dynamic Monte Carlo algorithm for
  out-of-equilibrium processes in colloidal dispersions}},}\ }\href {\doibase
  10.1039/C8CP02415D} {\bibfield  {journal} {\bibinfo  {journal} {Phys. Chem.
  Chem. Phys.}\ }\textbf {\bibinfo {volume} {20}},\ \bibinfo {pages}
  {15118--15127} (\bibinfo {year} {2018})}\BibitemShut {NoStop}%
\bibitem [{\citenamefont {{Garc{\'i}a Daza}}, \citenamefont {Cuetos},\ and\
  \citenamefont {Patti}(2020)}]{Daza2020}%
  \BibitemOpen
  \bibfield  {author} {\bibinfo {author} {\bibfnamefont {F.~A.}\ \bibnamefont
  {{Garc{\'i}a Daza}}}, \bibinfo {author} {\bibfnamefont {A.}~\bibnamefont
  {Cuetos}}, \ and\ \bibinfo {author} {\bibfnamefont {A.}~\bibnamefont
  {Patti}},\ }\bibfield  {title} {\enquote {\bibinfo {title} {{Dynamic Monte
  Carlo simulations of inhomogeneous colloidal suspensions}},}\ }\href
  {\doibase 10.1103/PhysRevE.102.013302} {\bibfield  {journal} {\bibinfo
  {journal} {Phys. Rev. E}\ }\textbf {\bibinfo {volume} {102}},\ \bibinfo
  {pages} {013302} (\bibinfo {year} {2020})}\BibitemShut {NoStop}%
\bibitem [{\citenamefont {Chiappini}, \citenamefont {Patti},\ and\
  \citenamefont {Dijkstra}(2020)}]{chiappini2020}%
  \BibitemOpen
  \bibfield  {author} {\bibinfo {author} {\bibfnamefont {M.}~\bibnamefont
  {Chiappini}}, \bibinfo {author} {\bibfnamefont {A.}~\bibnamefont {Patti}}, \
  and\ \bibinfo {author} {\bibfnamefont {M.}~\bibnamefont {Dijkstra}},\
  }\bibfield  {title} {\enquote {\bibinfo {title} {{Helicoidal dynamics of
  biaxial curved rods in twist-bend nematic phases unveiled by unsupervised
  machine learning techniques}},}\ }\href {\doibase
  10.1103/PhysRevE.102.040601} {\bibfield  {journal} {\bibinfo  {journal}
  {Phys. Rev. E}\ }\textbf {\bibinfo {volume} {102}},\ \bibinfo {pages}
  {040601} (\bibinfo {year} {2020})}\BibitemShut {NoStop}%
\bibitem [{\citenamefont {Patti}\ and\ \citenamefont
  {Cuetos}(2021)}]{patti2021}%
  \BibitemOpen
  \bibfield  {author} {\bibinfo {author} {\bibfnamefont {A.}~\bibnamefont
  {Patti}}\ and\ \bibinfo {author} {\bibfnamefont {A.}~\bibnamefont {Cuetos}},\
  }\bibfield  {title} {\enquote {\bibinfo {title} {{Dynamics of colloidal cubes
  and cuboids in cylindrical nanopores}},}\ }\href {\doibase 10.1063/5.0063152}
  {\bibfield  {journal} {\bibinfo  {journal} {Physics of Fluids}\ }\textbf
  {\bibinfo {volume} {33}},\ \bibinfo {pages} {097103} (\bibinfo {year}
  {2021})}\BibitemShut {NoStop}%
\bibitem [{\citenamefont {{Garc{\'i}a Daza}}\ \emph {et~al.}(2022)\citenamefont
  {{Garc{\'i}a Daza}}, \citenamefont {Puertas}, \citenamefont {Cuetos},\ and\
  \citenamefont {Patti}}]{Daza2022}%
  \BibitemOpen
  \bibfield  {author} {\bibinfo {author} {\bibfnamefont {F.~A.}\ \bibnamefont
  {{Garc{\'i}a Daza}}}, \bibinfo {author} {\bibfnamefont {A.~M.}\ \bibnamefont
  {Puertas}}, \bibinfo {author} {\bibfnamefont {A.}~\bibnamefont {Cuetos}}, \
  and\ \bibinfo {author} {\bibfnamefont {A.}~\bibnamefont {Patti}},\ }\bibfield
   {title} {\enquote {\bibinfo {title} {{Microrheology of colloidal suspensions
  via dynamic Monte Carlo simulations}},}\ }\href {\doibase
  10.1016/j.jcis.2021.07.088} {\bibfield  {journal} {\bibinfo  {journal} {J.
  Col. Inter. Sci.}\ }\textbf {\bibinfo {volume} {605}},\ \bibinfo {pages}
  {182--192} (\bibinfo {year} {2022})}\BibitemShut {NoStop}%
\bibitem [{\citenamefont {Vutukuri}\ \emph {et~al.}(2014)\citenamefont
  {Vutukuri}, \citenamefont {Smallenburg}, \citenamefont {Badaire},
  \citenamefont {Imhof}, \citenamefont {Dijkstra},\ and\ \citenamefont {{Van
  Blaaderen}}}]{Vutukuri2014}%
  \BibitemOpen
  \bibfield  {author} {\bibinfo {author} {\bibfnamefont {H.~R.}\ \bibnamefont
  {Vutukuri}}, \bibinfo {author} {\bibfnamefont {F.}~\bibnamefont
  {Smallenburg}}, \bibinfo {author} {\bibfnamefont {S.}~\bibnamefont
  {Badaire}}, \bibinfo {author} {\bibfnamefont {A.}~\bibnamefont {Imhof}},
  \bibinfo {author} {\bibfnamefont {M.}~\bibnamefont {Dijkstra}}, \ and\
  \bibinfo {author} {\bibfnamefont {A.}~\bibnamefont {{Van Blaaderen}}},\
  }\bibfield  {title} {\enquote {\bibinfo {title} {{An experimental and
  simulation study on the self-assembly of colloidal cubes in external electric
  fields}},}\ }\href {\doibase 10.1039/c4sm01778a} {\bibfield  {journal}
  {\bibinfo  {journal} {Soft Matter}\ }\textbf {\bibinfo {volume} {10}},\
  \bibinfo {pages} {9110--9119} (\bibinfo {year} {2014})}\BibitemShut {NoStop}%
\bibitem [{\citenamefont {Gottschalk}, \citenamefont {Lin},\ and\ \citenamefont
  {Manocha}(1996)}]{Gottschalk1996}%
  \BibitemOpen
  \bibfield  {author} {\bibinfo {author} {\bibfnamefont {S.}~\bibnamefont
  {Gottschalk}}, \bibinfo {author} {\bibfnamefont {M.}~\bibnamefont {Lin}}, \
  and\ \bibinfo {author} {\bibfnamefont {D.}~\bibnamefont {Manocha}},\
  }\bibfield  {title} {\enquote {\bibinfo {title} {{OBBTree: A Hierarchical
  Structure for Rapid Interference Detection}},}\ }in\ \href@noop {} {\emph
  {\bibinfo {booktitle} {23rd Annual Conference on Computer Graphics and
  Interactive Techniques}}}\ (\bibinfo {address} {New Orleans, LA, USA},\
  \bibinfo {year} {1996})\ pp.\ \bibinfo {pages} {171--180}\BibitemShut
  {NoStop}%
\bibitem [{\citenamefont {Rosenberg}\ \emph {et~al.}(2020)\citenamefont
  {Rosenberg}, \citenamefont {Dekker}, \citenamefont {Donaldson}, \citenamefont
  {Philipse},\ and\ \citenamefont {Kantorovich}}]{Rosenberg2020}%
  \BibitemOpen
  \bibfield  {author} {\bibinfo {author} {\bibfnamefont {M.}~\bibnamefont
  {Rosenberg}}, \bibinfo {author} {\bibfnamefont {F.}~\bibnamefont {Dekker}},
  \bibinfo {author} {\bibfnamefont {J.~G.}\ \bibnamefont {Donaldson}}, \bibinfo
  {author} {\bibfnamefont {A.~P.}\ \bibnamefont {Philipse}}, \ and\ \bibinfo
  {author} {\bibfnamefont {S.~S.}\ \bibnamefont {Kantorovich}},\ }\bibfield
  {title} {\enquote {\bibinfo {title} {{Self-assembly of charged colloidal
  cubes}},}\ }\href {\doibase 10.1039/C9SM02189B} {\bibfield  {journal}
  {\bibinfo  {journal} {Soft Matter}\ }\textbf {\bibinfo {volume} {16}},\
  \bibinfo {pages} {4451--4461} (\bibinfo {year} {2020})}\BibitemShut {NoStop}%
\bibitem [{\citenamefont {Parthasarathy}\ and\ \citenamefont
  {Klingenberg}(1996)}]{Parthasarathy1996}%
  \BibitemOpen
  \bibfield  {author} {\bibinfo {author} {\bibfnamefont {M.}~\bibnamefont
  {Parthasarathy}}\ and\ \bibinfo {author} {\bibfnamefont {D.~J.}\ \bibnamefont
  {Klingenberg}},\ }\bibfield  {title} {\enquote {\bibinfo {title}
  {{Electrorheology: Mechanisms and models}},}\ }\href {\doibase
  10.1016/0927-796X(96)00191-X} {\bibfield  {journal} {\bibinfo  {journal}
  {Materials Science and Engineering: R: Reports}\ }\textbf {\bibinfo {volume}
  {17}},\ \bibinfo {pages} {57--103} (\bibinfo {year} {1996})}\BibitemShut
  {NoStop}%
\bibitem [{\citenamefont {Allen}\ and\ \citenamefont
  {Tildesley}(2017)}]{Tildesley2017}%
  \BibitemOpen
  \bibfield  {author} {\bibinfo {author} {\bibfnamefont {M.~P.}\ \bibnamefont
  {Allen}}\ and\ \bibinfo {author} {\bibfnamefont {D.~J.}\ \bibnamefont
  {Tildesley}},\ }\bibfield  {title} {\enquote {\bibinfo {title} {{Chapter 6 -
  Long-range forces}},}\ }in\ \href {\doibase
  10.1093/oso/9780198803195.003.0006} {\emph {\bibinfo {booktitle} {Computer
  simulation of liquids}}},\ \bibinfo {editor} {edited by\ \bibinfo {editor}
  {\bibfnamefont {M.~P.}\ \bibnamefont {Allen}}\ and\ \bibinfo {editor}
  {\bibfnamefont {D.~J.}\ \bibnamefont {Tildesley}}}\ (\bibinfo  {publisher}
  {Oxford University Press},\ \bibinfo {address} {Oxford},\ \bibinfo {year}
  {2017})\ \bibinfo {edition} {second edition}\ ed.,\ pp.\ \bibinfo {pages}
  {216--257}\BibitemShut {NoStop}%
\bibitem [{\citenamefont {Tonti}\ and\ \citenamefont
  {Patti}(2021)}]{Tonti2021}%
  \BibitemOpen
  \bibfield  {author} {\bibinfo {author} {\bibfnamefont {L.}~\bibnamefont
  {Tonti}}\ and\ \bibinfo {author} {\bibfnamefont {A.}~\bibnamefont {Patti}},\
  }\bibfield  {title} {\enquote {\bibinfo {title} {{Fast Overlap Detection
  between Hard-Core Colloidal Cuboids and Spheres. The OCSI Algorithm}},}\
  }\href {\doibase 10.3390/a14030072} {\bibfield  {journal} {\bibinfo
  {journal} {Algorithms}\ }\textbf {\bibinfo {volume} {14}},\ \bibinfo {pages}
  {72} (\bibinfo {year} {2021})}\BibitemShut {NoStop}%
\bibitem [{\citenamefont {Tonti}, \citenamefont {{Garc{\'i}a Daza}},\ and\
  \citenamefont {Patti}(2021)}]{tonti2021_2}%
  \BibitemOpen
  \bibfield  {author} {\bibinfo {author} {\bibfnamefont {L.}~\bibnamefont
  {Tonti}}, \bibinfo {author} {\bibfnamefont {F.~A.}\ \bibnamefont {{Garc{\'i}a
  Daza}}}, \ and\ \bibinfo {author} {\bibfnamefont {A.}~\bibnamefont {Patti}},\
  }\bibfield  {title} {\enquote {\bibinfo {title} {{Diffusion of globular
  macromolecules in liquid crystals of colloidal cuboids}},}\ }\href {\doibase
  10.1016/j.molliq.2021.116640} {\bibfield  {journal} {\bibinfo  {journal}
  {Journal of Molecular Liquids}\ }\textbf {\bibinfo {volume} {338}},\ \bibinfo
  {pages} {116640} (\bibinfo {year} {2021})}\BibitemShut {NoStop}%
\bibitem [{\citenamefont {{Garc{\'i}a de la Torre}}, \citenamefont {{del Rio
  Echenique}},\ and\ \citenamefont {Ortega}(2007)}]{Garcia2007}%
  \BibitemOpen
  \bibfield  {author} {\bibinfo {author} {\bibfnamefont {J.~G.}\ \bibnamefont
  {{Garc{\'i}a de la Torre}}}, \bibinfo {author} {\bibfnamefont
  {G.}~\bibnamefont {{del Rio Echenique}}}, \ and\ \bibinfo {author}
  {\bibfnamefont {A.}~\bibnamefont {Ortega}},\ }\bibfield  {title} {\enquote
  {\bibinfo {title} {{Improved Calculation of Rotational Diffusion and
  Intrinsic Viscosity of Bead Models for Macromolecules and Nanoparticles}},}\
  }\href {\doibase 10.1021/jp0647941} {\bibfield  {journal} {\bibinfo
  {journal} {J. Phys. Chem. B}\ }\textbf {\bibinfo {volume} {111}},\ \bibinfo
  {pages} {955--961} (\bibinfo {year} {2007})}\BibitemShut {NoStop}%
\bibitem [{\citenamefont {Hansen}\ and\ \citenamefont
  {McDonald}(2013)}]{Hansen2013}%
  \BibitemOpen
  \bibfield  {author} {\bibinfo {author} {\bibfnamefont {J.-P.}\ \bibnamefont
  {Hansen}}\ and\ \bibinfo {author} {\bibfnamefont {I.~R.}\ \bibnamefont
  {McDonald}},\ }\bibfield  {title} {\enquote {\bibinfo {title} {{Chapter 2 -
  Statistical Mechanics}},}\ }in\ \href {\doibase
  https://doi.org/10.1016/B978-0-12-387032-2.00002-7} {\emph {\bibinfo
  {booktitle} {Theory of Simple Liquids (Fourth Edition)}}},\ \bibinfo {editor}
  {edited by\ \bibinfo {editor} {\bibfnamefont {J.-P.}\ \bibnamefont {Hansen}}\
  and\ \bibinfo {editor} {\bibfnamefont {I.~R.}\ \bibnamefont {McDonald}}}\
  (\bibinfo  {publisher} {Academic Press},\ \bibinfo {address} {Oxford},\
  \bibinfo {year} {2013})\ \bibinfo {edition} {fourth edition}\ ed.,\ pp.\
  \bibinfo {pages} {13--59}\BibitemShut {NoStop}%
\bibitem [{\citenamefont {Busselez}\ \emph {et~al.}(2014)\citenamefont
  {Busselez}, \citenamefont {Cerclier}, \citenamefont {Ndao}, \citenamefont
  {Ghoufi}, \citenamefont {Lefort},\ and\ \citenamefont
  {Morineau}}]{Busselez2014}%
  \BibitemOpen
  \bibfield  {author} {\bibinfo {author} {\bibfnamefont {R.}~\bibnamefont
  {Busselez}}, \bibinfo {author} {\bibfnamefont {C.~V.}\ \bibnamefont
  {Cerclier}}, \bibinfo {author} {\bibfnamefont {M.}~\bibnamefont {Ndao}},
  \bibinfo {author} {\bibfnamefont {A.}~\bibnamefont {Ghoufi}}, \bibinfo
  {author} {\bibfnamefont {R.}~\bibnamefont {Lefort}}, \ and\ \bibinfo {author}
  {\bibfnamefont {D.}~\bibnamefont {Morineau}},\ }\bibfield  {title} {\enquote
  {\bibinfo {title} {{Discotic columnar liquid crystal studied in the bulk and
  nanoconfined states by molecular dynamics simulation}},}\ }\href {\doibase
  10.1063/1.4896052} {\bibfield  {journal} {\bibinfo  {journal} {J. Chem.
  Phys.}\ }\textbf {\bibinfo {volume} {141}},\ \bibinfo {pages} {134902}
  (\bibinfo {year} {2014})}\BibitemShut {NoStop}%
\bibitem [{\citenamefont {Morillo}(2019)}]{Morillo2019}%
  \BibitemOpen
  \bibfield  {author} {\bibinfo {author} {\bibfnamefont {N.}~\bibnamefont
  {Morillo}},\ }\emph {\bibinfo {title} {{Anisotropy \& self-assembly. A walk
  through intricate free-energy landscapes}}},\ \href@noop {} {Ph.D. thesis},\
  \bibinfo  {school} {Department of Physical, Chemical and Natural System of
  the Pablo de Olavide University}, \bibinfo {address} {Seville, Spain}
  (\bibinfo {year} {2019})\BibitemShut {NoStop}%
\bibitem [{\citenamefont {Sevick}, \citenamefont {Monson},\ and\ \citenamefont
  {Ottino}(1988)}]{Sevick1988}%
  \BibitemOpen
  \bibfield  {author} {\bibinfo {author} {\bibfnamefont {E.~M.}\ \bibnamefont
  {Sevick}}, \bibinfo {author} {\bibfnamefont {P.~A.}\ \bibnamefont {Monson}},
  \ and\ \bibinfo {author} {\bibfnamefont {J.~M.}\ \bibnamefont {Ottino}},\
  }\bibfield  {title} {\enquote {\bibinfo {title} {{Monte Carlo calculations of
  cluster statistics in continuum models of composite morphology}},}\ }\href
  {\doibase 10.1063/1.454720} {\bibfield  {journal} {\bibinfo  {journal} {J.
  Chem. Phys.}\ }\textbf {\bibinfo {volume} {88}},\ \bibinfo {pages}
  {1198--1206} (\bibinfo {year} {1988})}\BibitemShut {NoStop}%
\bibitem [{\citenamefont {Mason}\ and\ \citenamefont
  {Weitz}(1995)}]{MASON1995}%
  \BibitemOpen
  \bibfield  {author} {\bibinfo {author} {\bibfnamefont {T.~G.}\ \bibnamefont
  {Mason}}\ and\ \bibinfo {author} {\bibfnamefont {D.~A.}\ \bibnamefont
  {Weitz}},\ }\bibfield  {title} {\enquote {\bibinfo {title} {{Optical
  Measurements of Frequency-Dependent Linear Viscoelastic Moduli of Complex
  Fluids}},}\ }\href {\doibase 10.1103/PhysRevLett.74.1250} {\bibfield
  {journal} {\bibinfo  {journal} {Phys. Rev. Lett.}\ }\textbf {\bibinfo
  {volume} {74}},\ \bibinfo {pages} {1250--1253} (\bibinfo {year}
  {1995})}\BibitemShut {NoStop}%
\bibitem [{\citenamefont {Mason}\ \emph {et~al.}(1997)\citenamefont {Mason},
  \citenamefont {Ganesan}, \citenamefont {van Zanten}, \citenamefont {Wirtz},\
  and\ \citenamefont {Kuo}}]{mason1997}%
  \BibitemOpen
  \bibfield  {author} {\bibinfo {author} {\bibfnamefont {T.~G.}\ \bibnamefont
  {Mason}}, \bibinfo {author} {\bibfnamefont {K.}~\bibnamefont {Ganesan}},
  \bibinfo {author} {\bibfnamefont {J.~H.}\ \bibnamefont {van Zanten}},
  \bibinfo {author} {\bibfnamefont {D.}~\bibnamefont {Wirtz}}, \ and\ \bibinfo
  {author} {\bibfnamefont {S.~C.}\ \bibnamefont {Kuo}},\ }\bibfield  {title}
  {\enquote {\bibinfo {title} {{Particle Tracking Microrheology of Complex
  Fluids}},}\ }\href {\doibase 10.1103/PhysRevLett.79.3282} {\bibfield
  {journal} {\bibinfo  {journal} {Phys. Rev. Lett.}\ }\textbf {\bibinfo
  {volume} {79}},\ \bibinfo {pages} {3282--3285} (\bibinfo {year}
  {1997})}\BibitemShut {NoStop}%
\bibitem [{\citenamefont {Mason}(2000)}]{mason2000}%
  \BibitemOpen
  \bibfield  {author} {\bibinfo {author} {\bibfnamefont {T.~G.}\ \bibnamefont
  {Mason}},\ }\bibfield  {title} {\enquote {\bibinfo {title} {{Estimating the
  viscoelastic moduli of complex fluids using the generalized Stokes--Einstein
  equation}},}\ }\href {\doibase 10.1007/s003970000094} {\bibfield  {journal}
  {\bibinfo  {journal} {Rheol. Acta}\ }\textbf {\bibinfo {volume} {39}},\
  \bibinfo {pages} {371--378} (\bibinfo {year} {2000})}\BibitemShut {NoStop}%
\bibitem [{\citenamefont {Evans}\ \emph {et~al.}(2009)\citenamefont {Evans},
  \citenamefont {Tassieri}, \citenamefont {Auhl},\ and\ \citenamefont
  {Waigh}}]{Evans2009}%
  \BibitemOpen
  \bibfield  {author} {\bibinfo {author} {\bibfnamefont {R.~M.~L.}\
  \bibnamefont {Evans}}, \bibinfo {author} {\bibfnamefont {M.}~\bibnamefont
  {Tassieri}}, \bibinfo {author} {\bibfnamefont {D.}~\bibnamefont {Auhl}}, \
  and\ \bibinfo {author} {\bibfnamefont {T.~A.}\ \bibnamefont {Waigh}},\
  }\bibfield  {title} {\enquote {\bibinfo {title} {{Direct conversion of
  rheological compliance measurements into storage and loss moduli}},}\ }\href
  {\doibase 10.1103/PhysRevE.80.012501} {\bibfield  {journal} {\bibinfo
  {journal} {Phys. Rev. E}\ }\textbf {\bibinfo {volume} {80}},\ \bibinfo
  {pages} {012501} (\bibinfo {year} {2009})}\BibitemShut {NoStop}%
\bibitem [{\citenamefont {Nishi}\ \emph {et~al.}(2018)\citenamefont {Nishi},
  \citenamefont {Kilfoil}, \citenamefont {Schmidt},\ and\ \citenamefont
  {MacKintosh}}]{Nishi2018}%
  \BibitemOpen
  \bibfield  {author} {\bibinfo {author} {\bibfnamefont {K.}~\bibnamefont
  {Nishi}}, \bibinfo {author} {\bibfnamefont {M.~L.}\ \bibnamefont {Kilfoil}},
  \bibinfo {author} {\bibfnamefont {C.~F.}\ \bibnamefont {Schmidt}}, \ and\
  \bibinfo {author} {\bibfnamefont {F.~C.}\ \bibnamefont {MacKintosh}},\
  }\bibfield  {title} {\enquote {\bibinfo {title} {{A symmetrical method to
  obtain shear moduli from microrheology}},}\ }\href {\doibase
  10.1039/C7SM02499A} {\bibfield  {journal} {\bibinfo  {journal} {Soft Matter}\
  }\textbf {\bibinfo {volume} {14}},\ \bibinfo {pages} {3716--3723} (\bibinfo
  {year} {2018})}\BibitemShut {NoStop}%
\bibitem [{\citenamefont {Segovia-Guti{\'e}rrez}\ \emph
  {et~al.}(2013)\citenamefont {Segovia-Guti{\'e}rrez}, \citenamefont {{de
  Vicente}}, \citenamefont {Hidalgo-{\'A}lvarez},\ and\ \citenamefont
  {Puertas}}]{Segovia2013}%
  \BibitemOpen
  \bibfield  {author} {\bibinfo {author} {\bibfnamefont {J.~P.}\ \bibnamefont
  {Segovia-Guti{\'e}rrez}}, \bibinfo {author} {\bibfnamefont {J.}~\bibnamefont
  {{de Vicente}}}, \bibinfo {author} {\bibfnamefont {R.}~\bibnamefont
  {Hidalgo-{\'A}lvarez}}, \ and\ \bibinfo {author} {\bibfnamefont {A.~M.}\
  \bibnamefont {Puertas}},\ }\bibfield  {title} {\enquote {\bibinfo {title}
  {{Brownian dynamics simulations in magnetorheology and comparison with
  experiments}},}\ }\href {\doibase 10.1039/C3SM00137G} {\bibfield  {journal}
  {\bibinfo  {journal} {Soft Matter}\ }\textbf {\bibinfo {volume} {9}},\
  \bibinfo {pages} {6970--6977} (\bibinfo {year} {2013})}\BibitemShut {NoStop}%
\bibitem [{\citenamefont {Xie}, \citenamefont {Pu},\ and\ \citenamefont
  {Gao}(2009)}]{Xie2009}%
  \BibitemOpen
  \bibfield  {author} {\bibinfo {author} {\bibfnamefont {W.}~\bibnamefont
  {Xie}}, \bibinfo {author} {\bibfnamefont {J.}~\bibnamefont {Pu}}, \ and\
  \bibinfo {author} {\bibfnamefont {J.}~\bibnamefont {Gao}},\ }\bibfield
  {title} {\enquote {\bibinfo {title} {{A Coupled Polarization-Matrix Inversion
  and Iteration Approach for Accelerating the Dipole Convergence in a
  Polarizable Potential Function}},}\ }\href {\doibase 10.1021/JP808952M}
  {\bibfield  {journal} {\bibinfo  {journal} {J. Phys. Chem. A}\ }\textbf
  {\bibinfo {volume} {113}},\ \bibinfo {pages} {2109} (\bibinfo {year}
  {2009})}\BibitemShut {NoStop}%
\bibitem [{\citenamefont {Bernardo}\ \emph {et~al.}(1994)\citenamefont
  {Bernardo}, \citenamefont {Ding}, \citenamefont {Krogh-Jespersen},\ and\
  \citenamefont {Levy}}]{Bernardo1994}%
  \BibitemOpen
  \bibfield  {author} {\bibinfo {author} {\bibfnamefont {D.~N.}\ \bibnamefont
  {Bernardo}}, \bibinfo {author} {\bibfnamefont {Y.}~\bibnamefont {Ding}},
  \bibinfo {author} {\bibfnamefont {K.}~\bibnamefont {Krogh-Jespersen}}, \ and\
  \bibinfo {author} {\bibfnamefont {R.~M.}\ \bibnamefont {Levy}},\ }\bibfield
  {title} {\enquote {\bibinfo {title} {{An Anisotropic Polarizable Water Model:
  Incorporation of All-Atom Polarizabilities into Molecular Mechanics Force
  Fields}},}\ }\href {\doibase 10.1021/J100066A043} {\bibfield  {journal}
  {\bibinfo  {journal} {J. Phys. Chem.}\ }\textbf {\bibinfo {volume} {98}},\
  \bibinfo {pages} {4180} (\bibinfo {year} {1994})}\BibitemShut {NoStop}%
\bibitem [{\citenamefont {Vesely}(1977)}]{Vesely1977}%
  \BibitemOpen
  \bibfield  {author} {\bibinfo {author} {\bibfnamefont {F.~J.}\ \bibnamefont
  {Vesely}},\ }\bibfield  {title} {\enquote {\bibinfo {title} {{N-particle
  dynamics of polarizable Stockmayer-type molecules}},}\ }\href {\doibase
  10.1016/0021-9991(77)90028-6} {\bibfield  {journal} {\bibinfo  {journal} {J.
  Comp. Phys.}\ }\textbf {\bibinfo {volume} {24}},\ \bibinfo {pages} {361}
  (\bibinfo {year} {1977})}\BibitemShut {NoStop}%
\bibitem [{\citenamefont {Kwaadgras}, \citenamefont {{Van Roij}},\ and\
  \citenamefont {Dijkstra}(2014)}]{Kwaadgras2014}%
  \BibitemOpen
  \bibfield  {author} {\bibinfo {author} {\bibfnamefont {B.~W.}\ \bibnamefont
  {Kwaadgras}}, \bibinfo {author} {\bibfnamefont {R.}~\bibnamefont {{Van
  Roij}}}, \ and\ \bibinfo {author} {\bibfnamefont {M.}~\bibnamefont
  {Dijkstra}},\ }\bibfield  {title} {\enquote {\bibinfo {title}
  {{Self-consistent electric field-induced dipole interaction of colloidal
  spheres, cubes, rods, and dumbbells}},}\ }\href {\doibase 10.1063/1.4870251}
  {\bibfield  {journal} {\bibinfo  {journal} {J. Chem. Phys.}\ }\textbf
  {\bibinfo {volume} {140}},\ \bibinfo {pages} {154901} (\bibinfo {year}
  {2014})}\BibitemShut {NoStop}%
\bibitem [{\citenamefont {Hahsler}, \citenamefont {Piekenbrock},\ and\
  \citenamefont {Doran}(2019)}]{Hahsler2019}%
  \BibitemOpen
  \bibfield  {author} {\bibinfo {author} {\bibfnamefont {M.}~\bibnamefont
  {Hahsler}}, \bibinfo {author} {\bibfnamefont {M.}~\bibnamefont
  {Piekenbrock}}, \ and\ \bibinfo {author} {\bibfnamefont {D.}~\bibnamefont
  {Doran}},\ }\bibfield  {title} {\enquote {\bibinfo {title} {{dbscan: Fast
  Density-Based Clustering with R}},}\ }\href {\doibase 10.18637/jss.v091.i01}
  {\bibfield  {journal} {\bibinfo  {journal} {Journal of Statistical Software}\
  }\textbf {\bibinfo {volume} {91}},\ \bibinfo {pages} {1–30} (\bibinfo
  {year} {2019})}\BibitemShut {NoStop}%
\bibitem [{\citenamefont {Schubert}\ \emph {et~al.}(2017)\citenamefont
  {Schubert}, \citenamefont {Sander}, \citenamefont {Ester}, \citenamefont
  {Kriegel},\ and\ \citenamefont {Xu}}]{Schubert2017}%
  \BibitemOpen
  \bibfield  {author} {\bibinfo {author} {\bibfnamefont {E.}~\bibnamefont
  {Schubert}}, \bibinfo {author} {\bibfnamefont {J.}~\bibnamefont {Sander}},
  \bibinfo {author} {\bibfnamefont {M.}~\bibnamefont {Ester}}, \bibinfo
  {author} {\bibfnamefont {H.~P.}\ \bibnamefont {Kriegel}}, \ and\ \bibinfo
  {author} {\bibfnamefont {X.}~\bibnamefont {Xu}},\ }\bibfield  {title}
  {\enquote {\bibinfo {title} {{DBSCAN Revisited, Revisited: Why and How You
  Should (Still) Use DBSCAN}},}\ }\href {\doibase 10.1145/3068335} {\bibfield
  {journal} {\bibinfo  {journal} {ACM Trans. Database Syst.}\ }\textbf
  {\bibinfo {volume} {42}},\ \bibinfo {pages} {19} (\bibinfo {year}
  {2017})}\BibitemShut {NoStop}%
\bibitem [{\citenamefont {Wertheim}(1984{\natexlab{a}})}]{Wertheim1984}%
  \BibitemOpen
  \bibfield  {author} {\bibinfo {author} {\bibfnamefont {M.}~\bibnamefont
  {Wertheim}},\ }\bibfield  {title} {\enquote {\bibinfo {title} {{Fluids with
  highly directional attractive forces. I. Statistical thermodynamics}},}\
  }\href {\doibase 10.1007/BF01017362} {\bibfield  {journal} {\bibinfo
  {journal} {J. Stat. Phys.}\ }\textbf {\bibinfo {volume} {35}},\ \bibinfo
  {pages} {19–34} (\bibinfo {year} {1984)}{\natexlab{a}})}\BibitemShut
  {NoStop}%
\bibitem [{\citenamefont {Wertheim}(1984{\natexlab{b}})}]{Wertheim1984_2}%
  \BibitemOpen
  \bibfield  {author} {\bibinfo {author} {\bibfnamefont {M.}~\bibnamefont
  {Wertheim}},\ }\bibfield  {title} {\enquote {\bibinfo {title} {{Fluids with
  highly directional attractive forces. II. Thermodynamic perturbation theory
  and integral equations}},}\ }\href {\doibase 10.1007/BF01017363} {\bibfield
  {journal} {\bibinfo  {journal} {J. Stat. Phys.}\ }\textbf {\bibinfo {volume}
  {35}},\ \bibinfo {pages} {35--47} (\bibinfo {year}
  {1984}{\natexlab{b}})}\BibitemShut {NoStop}%
\bibitem [{\citenamefont {Wertheim}(1986)}]{Wertheim1986}%
  \BibitemOpen
  \bibfield  {author} {\bibinfo {author} {\bibfnamefont {M.}~\bibnamefont
  {Wertheim}},\ }\bibfield  {title} {\enquote {\bibinfo {title} {{Fluids with
  highly directional attractive forces. III. Multiple attraction sites}},}\
  }\href {\doibase 10.1007/BF01127721} {\bibfield  {journal} {\bibinfo
  {journal} {J. Stat. Phys.}\ }\textbf {\bibinfo {volume} {42}},\ \bibinfo
  {pages} {459--476} (\bibinfo {year} {1986})}\BibitemShut {NoStop}%
\bibitem [{\citenamefont {Chiricotto}\ \emph {et~al.}(2019)\citenamefont
  {Chiricotto}, \citenamefont {Melchionna}, \citenamefont {Derreumaux},\ and\
  \citenamefont {Sterpone}}]{Chiricotto2019}%
  \BibitemOpen
  \bibfield  {author} {\bibinfo {author} {\bibfnamefont {M.}~\bibnamefont
  {Chiricotto}}, \bibinfo {author} {\bibfnamefont {S.}~\bibnamefont
  {Melchionna}}, \bibinfo {author} {\bibfnamefont {P.}~\bibnamefont
  {Derreumaux}}, \ and\ \bibinfo {author} {\bibfnamefont {F.}~\bibnamefont
  {Sterpone}},\ }\bibfield  {title} {\enquote {\bibinfo {title} {{Multiscale
  Aggregation of the Amyloid A$\beta$16–22 Peptide: From Disordered
  Coagulation and Lateral Branching to Amorphous Prefibrils}},}\ }\href
  {\doibase 10.1021/acs.jpclett.9b00423} {\bibfield  {journal} {\bibinfo
  {journal} {J. Phys. Chem. Lett.}\ }\textbf {\bibinfo {volume} {10}},\
  \bibinfo {pages} {1594--1599} (\bibinfo {year} {2019})}\BibitemShut {NoStop}%
\bibitem [{\citenamefont {Lettinga}\ \emph {et~al.}(2010)\citenamefont
  {Lettinga}, \citenamefont {Dhont}, \citenamefont {Zhang}, \citenamefont
  {Messlinger},\ and\ \citenamefont {Gompper}}]{Lettinga2010}%
  \BibitemOpen
  \bibfield  {author} {\bibinfo {author} {\bibfnamefont {M.~P.}\ \bibnamefont
  {Lettinga}}, \bibinfo {author} {\bibfnamefont {J.~K.~G.}\ \bibnamefont
  {Dhont}}, \bibinfo {author} {\bibfnamefont {Z.}~\bibnamefont {Zhang}},
  \bibinfo {author} {\bibfnamefont {S.}~\bibnamefont {Messlinger}}, \ and\
  \bibinfo {author} {\bibfnamefont {G.}~\bibnamefont {Gompper}},\ }\bibfield
  {title} {\enquote {\bibinfo {title} {{Hydrodynamic interactions in rod
  suspensions with orientational ordering}},}\ }\href {\doibase
  10.1039/C0SM00081G} {\bibfield  {journal} {\bibinfo  {journal} {Soft Matter}\
  }\textbf {\bibinfo {volume} {6}},\ \bibinfo {pages} {4556--4562} (\bibinfo
  {year} {2010})}\BibitemShut {NoStop}%
\bibitem [{\citenamefont {Kwon}, \citenamefont {Sung},\ and\ \citenamefont
  {Yethiraj}(2014)}]{Kwon2014}%
  \BibitemOpen
  \bibfield  {author} {\bibinfo {author} {\bibfnamefont {G.}~\bibnamefont
  {Kwon}}, \bibinfo {author} {\bibfnamefont {B.~J.}\ \bibnamefont {Sung}}, \
  and\ \bibinfo {author} {\bibfnamefont {A.}~\bibnamefont {Yethiraj}},\
  }\bibfield  {title} {\enquote {\bibinfo {title} {{Dynamics in Crowded
  Environments: Is Non-Gaussian Brownian Diffusion Normal?}}}\ }\href {\doibase
  10.1021/jp5011617} {\bibfield  {journal} {\bibinfo  {journal} {The Journal of
  Physical Chemistry B}\ }\textbf {\bibinfo {volume} {118}},\ \bibinfo {pages}
  {8128--8134} (\bibinfo {year} {2014})}\BibitemShut {NoStop}%
\bibitem [{\citenamefont {Pryamitsyn}\ and\ \citenamefont
  {Ganesan}(2008)}]{Pryamitsyn2008}%
  \BibitemOpen
  \bibfield  {author} {\bibinfo {author} {\bibfnamefont {V.}~\bibnamefont
  {Pryamitsyn}}\ and\ \bibinfo {author} {\bibfnamefont {V.}~\bibnamefont
  {Ganesan}},\ }\bibfield  {title} {\enquote {\bibinfo {title} {{Screening of
  hydrodynamic interactions in Brownian rod suspensions}},}\ }\href {\doibase
  10.1063/1.2842075} {\bibfield  {journal} {\bibinfo  {journal} {The Journal of
  Chemical Physics}\ }\textbf {\bibinfo {volume} {128}},\ \bibinfo {pages}
  {134901} (\bibinfo {year} {2008})}\BibitemShut {NoStop}%
\bibitem [{\citenamefont {Parthasarathy}\ and\ \citenamefont
  {Klingenberg}(1999)}]{Parthasarathy1999}%
  \BibitemOpen
  \bibfield  {author} {\bibinfo {author} {\bibfnamefont {M.}~\bibnamefont
  {Parthasarathy}}\ and\ \bibinfo {author} {\bibfnamefont {D.~J.}\ \bibnamefont
  {Klingenberg}},\ }\bibfield  {title} {\enquote {\bibinfo {title} {{Large
  amplitude oscillatory shear of ER suspensions}},}\ }\href {\doibase
  10.1016/S0377-0257(98)00096-2} {\bibfield  {journal} {\bibinfo  {journal}
  {Journal of Non-Newtonian Fluid Mechanics}\ }\textbf {\bibinfo {volume}
  {81}},\ \bibinfo {pages} {83--104} (\bibinfo {year} {1999})}\BibitemShut
  {NoStop}%
\bibitem [{\citenamefont {Weeber}\ and\ \citenamefont
  {Harting}(2012)}]{Weeber2012}%
  \BibitemOpen
  \bibfield  {author} {\bibinfo {author} {\bibfnamefont {R.}~\bibnamefont
  {Weeber}}\ and\ \bibinfo {author} {\bibfnamefont {J.}~\bibnamefont
  {Harting}},\ }\bibfield  {title} {\enquote {\bibinfo {title} {{Hydrodynamic
  interactions in active colloidal crystal microrheology}},}\ }\href {\doibase
  10.1103/PhysRevE.86.057302} {\bibfield  {journal} {\bibinfo  {journal} {Phys.
  Rev. E}\ }\textbf {\bibinfo {volume} {86}},\ \bibinfo {pages} {057302}
  (\bibinfo {year} {2012})}\BibitemShut {NoStop}%
\bibitem [{\citenamefont {Khair}\ and\ \citenamefont
  {Brady}(2006)}]{Khair2006}%
  \BibitemOpen
  \bibfield  {author} {\bibinfo {author} {\bibfnamefont {A.~S.}\ \bibnamefont
  {Khair}}\ and\ \bibinfo {author} {\bibfnamefont {J.~F.}\ \bibnamefont
  {Brady}},\ }\bibfield  {title} {\enquote {\bibinfo {title} {{Single particle
  motion in colloidal dispersions: a simple model for active and nonlinear
  microrheology}},}\ }\href {\doibase 10.1017/S0022112006009608} {\bibfield
  {journal} {\bibinfo  {journal} {Journal of Fluid Mechanics}\ }\textbf
  {\bibinfo {volume} {557}},\ \bibinfo {pages} {73–117} (\bibinfo {year}
  {2006})}\BibitemShut {NoStop}%
\bibitem [{\citenamefont {{Garc{\'i}a de la Torre}}\ \emph
  {et~al.}(2009)\citenamefont {{Garc{\'i}a de la Torre}}, \citenamefont
  {{Hern{\'a}ndez Cifre}}, \citenamefont {Ortega}, \citenamefont {Schmidt},
  \citenamefont {Fernandes}, \citenamefont {{P{\'e}rez S{\'a}nchez}},\ and\
  \citenamefont {Pamies}}]{Torre2009}%
  \BibitemOpen
  \bibfield  {author} {\bibinfo {author} {\bibfnamefont {J.}~\bibnamefont
  {{Garc{\'i}a de la Torre}}}, \bibinfo {author} {\bibfnamefont {J.~G.}\
  \bibnamefont {{Hern{\'a}ndez Cifre}}}, \bibinfo {author} {\bibfnamefont
  {{\'A}.}~\bibnamefont {Ortega}}, \bibinfo {author} {\bibfnamefont {R.~R.}\
  \bibnamefont {Schmidt}}, \bibinfo {author} {\bibfnamefont {M.~X.}\
  \bibnamefont {Fernandes}}, \bibinfo {author} {\bibfnamefont {H.~E.}\
  \bibnamefont {{P{\'e}rez S{\'a}nchez}}}, \ and\ \bibinfo {author}
  {\bibfnamefont {R.}~\bibnamefont {Pamies}},\ }\bibfield  {title} {\enquote
  {\bibinfo {title} {{SIMUFLEX: Algorithms and Tools for Simulation of the
  Conformation and Dynamics of Flexible Molecules and Nanoparticles in Dilute
  Solution}},}\ }\href {\doibase 10.1021/ct900269n} {\bibfield  {journal}
  {\bibinfo  {journal} {Journal of Chemical Theory and Computation}\ }\textbf
  {\bibinfo {volume} {5}},\ \bibinfo {pages} {2606--2618} (\bibinfo {year}
  {2009})}\BibitemShut {NoStop}%
\end{thebibliography}%


\begin{thebibliography}{1}
\expandafter\ifx\csname url\endcsname\relax
  \def\url#1{\texttt{#1}}\fi
\expandafter\ifx\csname urlprefix\endcsname\relax\def\urlprefix{URL }\fi
\providecommand{\bibinfo}[2]{#2}
\providecommand{\eprint}[2][]{\url{#2}}

\bibitem{Ester1996}
\bibinfo{author}{Ester, M.}, \bibinfo{author}{Kriegel, H.~P.},
  \bibinfo{author}{Sander, J.} \& \bibinfo{author}{Xu, X.}
\newblock \bibinfo{title}{{A Density-Based Algorithm for Discovering Clusters
  in Large Spatial Databases with Noise}}.
\newblock In \emph{\bibinfo{booktitle}{Proceedings of the Second International
  Conference on Knowledge Discovery and Data Mining}}, \bibinfo{pages}{226}
  (\bibinfo{year}{1996}).

\bibitem{Hahsler2019}
\bibinfo{author}{Hahsler, M.}, \bibinfo{author}{Piekenbrock, M.} \&
  \bibinfo{author}{Doran, D.}
\newblock \bibinfo{title}{{dbscan: Fast Density-Based Clustering with R}}.
\newblock \emph{\bibinfo{journal}{Journal of Statistical Software}}
  \textbf{\bibinfo{volume}{91}}, \bibinfo{pages}{1} (\bibinfo{year}{2019}).

\bibitem{Torquato2010}
\bibinfo{author}{Batten, R.}, \bibinfo{author}{Stillinger, F.~H.} \&
  \bibinfo{author}{Torquato, S.}
\newblock \bibinfo{title}{{Phase behavior of colloidal superballs: Shape
  interpolation from spheres to cubes}}.
\newblock \emph{\bibinfo{journal}{Phys. Rev. E}} \textbf{\bibinfo{volume}{81}},
  \bibinfo{pages}{061105} (\bibinfo{year}{2010}).

\bibitem{Evans2009}
\bibinfo{author}{Evans, R. M.~L.}, \bibinfo{author}{Tassieri, M.},
  \bibinfo{author}{Auhl, D.} \& \bibinfo{author}{Waigh, T.~A.}
\newblock \bibinfo{title}{{Direct conversion of rheological compliance
  measurements into storage and loss moduli}}.
\newblock \emph{\bibinfo{journal}{Phys. Rev. E}} \textbf{\bibinfo{volume}{80}},
  \bibinfo{pages}{012501} (\bibinfo{year}{2009}).

\bibitem{Nishi2018}
\bibinfo{author}{Nishi, K.}, \bibinfo{author}{Kilfoil, M.~L.},
  \bibinfo{author}{Schmidt, C.~F.} \& \bibinfo{author}{MacKintosh, F.~C.}
\newblock \bibinfo{title}{{A symmetrical method to obtain shear moduli from
  microrheology}}.
\newblock \emph{\bibinfo{journal}{Soft Matter}} \textbf{\bibinfo{volume}{14}},
  \bibinfo{pages}{3716} (\bibinfo{year}{2018}).

\bibitem{mason2000}
\bibinfo{author}{Mason, T.~G.}
\newblock \bibinfo{title}{{Estimating the viscoelastic moduli of complex fluids
  using the generalized Stokes--Einstein equation}}.
\newblock \emph{\bibinfo{journal}{Rheol. Acta}} \textbf{\bibinfo{volume}{39}},
  \bibinfo{pages}{371} (\bibinfo{year}{2000}).

\end{thebibliography}

\end{document}

\begin{center}
\textbf{{\Large Supplementary Information}}
\vskip 5mm

\textbf{{\large Kinetics of isotropic to string-like phase switching in \\electrorheological fluids of nanocubes}}
\normalsize
\vskip 5mm

Luca Tonti\textsuperscript{1}, Fabián A. García Daza\textsuperscript{1}, Alessandro Patti\textsuperscript{1,2,*}

\vskip 5mm

\textit{\textsuperscript{1}Department of Chemical Engineering, The University of Manchester, Manchester, M13 9PL, UK \\
\textsuperscript{2}Department of Applied Physics, University of Granada, Fuente Nueva s/n, 18071 Granada, Spain}

\vskip 4mm

\textsuperscript{*}apatti@ugr.es
\end{center}
\vskip 10mm

%%%%%%%%%%%%%%%%%%%%%%%%%%%%%%%%%%%%%%%%%%%%
%%%%%%%%%%%%%%%%%%%%%%%%%%%%%%%%%%%%%%%%%%%%
%%%%% SECTION 1: DBSCAN
%%%%%%%%%%%%%%%%%%%%%%%%%%%%%%%%%%%%%%%%%%%%
%%%%%%%%%%%%%%%%%%%%%%%%%%%%%%%%%%%%%%%%%%%%

\secti{S1. Cluster criterion optimisation via density-based cluster analysis}

\begin{figure}[ht!]
    \centering
    \includegraphics[width=\linewidth]{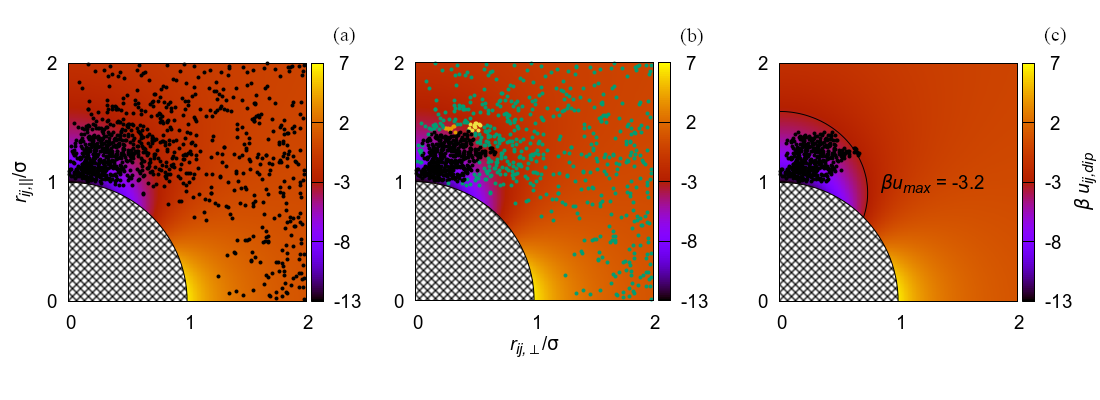}
    \caption{(a) Plot of the first and second closest particles $j$ with respect to reference particle $i$, decomposed in the direction parallel and perpendicular to the external field, taken from a configuration of an equilibrated string-like phase of nanocubes. (b) Resulting clusters obtained from the application of DBSCAN algorithm to the points plotted in panel (a); points of the same color belong to a cluster according to DBSCAN. (c) Plot of the points of the cluster closest to the minimum of energy (black points), together with the contour line in which $\beta u_{ij,dip} = \beta u_{max}$. The background of each plot shows the value of the dipolar interaction energy $\beta u_{ij,dip}$ in color gradient, as a function of $r_{ij,\perp}$ and $r_{ij,\parallel}$. The patterned areas in black and white contain all the points for which $\|\vect{r}_{ij}\|<\sigma$, which are inaccessible due to hard core interactions between nanocubes.}
    \label{fig:clus_example}
\end{figure}

Defining a criterion to recognize when cubes belong to chains is fundamental for the analysis of all the properties of chain-like structures. Due to the cylindrical symmetry of the chains and considering that their formation is driven only by dipolar interactions between the centers of mass of the particles, we characterized the connectivity between two particles $i$ and $j$ using their relative dipolar interaction energy: $i$ and $j$ are connected if $\beta u_{ij,dip} < \beta u_{max}$, where $\beta u_{max}$ is the cluster parameter. Since $\beta u_{ij,dip}$ depends only on the relative distance between particles and the alignment of $\vect{r}_{ij}$ with the external field, we optimized $\beta u_{max}$ from the analysis of the distribution of the first two smallest values of $\|\vect{r}_{ij}\|$ for each particle, named $\vect{r}_{ij,1}$ and $\vect{r}_{ij,2}$, extrapolated from an equilibrated string-like configuration of cubes, where $\vect{r}_{ij,1}$ and $\vect{r}_{ij,2}$ are decomposed in the direction parallel ($r_{ij,\parallel}$) and perpendicular ($r_{ij,\perp}$) to the orientation of the field. Most of the points of the data set are found in the proximity of the minimum of the interaction potential between two cubes, as expected for particles disposed one on top of the other and aligned with the external field (see Figure~\ref{fig:clus_example}(a) for a plot of one sample data set).

\begin{figure}[ht!]
    \centering
    \includegraphics[scale=0.80]{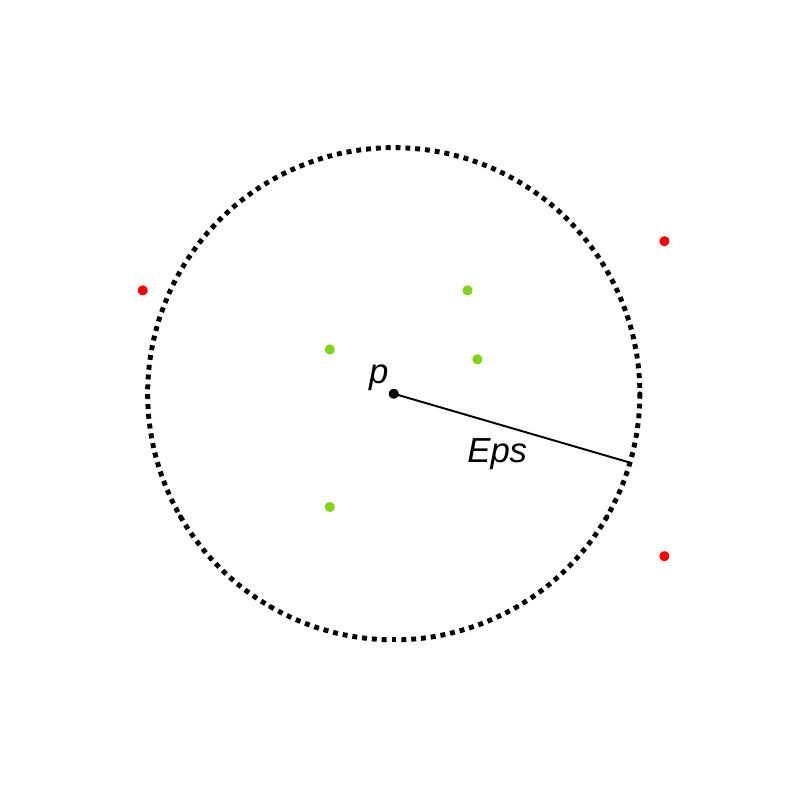}
    \caption{Representation of "$Eps$-neighbourhood" of $p$. All the points are elements of the data set. $p$ is the initial element, $Eps$ is the radius of the area in which we analyse the neighbourhood of $p$. Green points are elements that belong to the "$Eps$-neighbourhood" of $p$, while red points are outsiders. In one iteration, if the number of green points is $\ge MinPts$, then the green points plus $p$ form a cluster, and the green points are chosen as new starting points $p$ from which the cluster can be extended.}
    \label{fig:kn_example}
\end{figure}

In order to obtain an optimal value of $\beta u_{max}$ for the analysis of the connectivity, we applied the DBSCAN clustering algorithm to the data set of the distances $\vect{r}_{ij}$.
The DBSCAN algorithm analyses the local density of the data set, and assigns elements of the data set to a cluster if their density is higher than a threshold value~\cite{Ester1996}. It requires the definition of two parameters: $Eps$, which is the radius of the $n$-ball centered at element $p$; $MinPts$, which is the minimum number of elements $q$ contained in the $n$-ball with radius $Eps$ and center $p$. By definition, the set of points $q$ found in the $n$-ball centered at $p$ with radius $Eps$ is called "$Eps$-neighbourhood" of $p$, and the number of elements of this set is $N_{Eps}$:

\begin{equation}
    N_{Eps}(p) = \#\{q \in D\,|\, dist(p,q) \leq Eps \}
\end{equation}

where $D$ is the data set. The algorithm starts from a random element $p_0$ and computes all the neighbouring elements $q$. If $N_{Eps} \ge MinPts$, a new cluster is found and is made of $p_0$ plus all the neighbouring elements $q$. At this point, DBSCAN reiterates the computation of the "$Eps$-neighbourhood" for each element $q$, as long as new neighbor elements are found. The set of all the "$Eps$-neighbourhoods" found starting from $p_0$ is a cluster. When the iteration stops, the algorithm restarts the analysis from a random element left of the data set, until the entire data set is covered. A schematic representation of one step of the algorithm is shown in Figure~\ref{fig:kn_example}.

We performed density-based cluster analysis using the library DBSCAN available in R~\cite{Hahsler2019}. The algorithm was applied to 44 independent data sets, where each data set is extracted from uncorrelated configurations of an equilibrated chain-like phase of nanocubes. From each configuration, we computed the first ($\vect{r}_{ij,1}$) and second ($\vect{r}_{ij,2}$) closest distances between the centres of mass of two particles, with periodic boundary conditions, for all the particles of the configuration. Doubles of the distances $\vect{r}_{ij}$ were discarded. All the distances computed were saved as a 2D vector with components $(r_{ij,\parallel},r_{ij,\perp}$). Since we were interested only in a portion of space were the energy reaches a minimum, we limited the data sets only to the points within the range $0\leq r_{ij,\parallel,},r_{ij,\bot}\leq 2$. An example of resulting data set extrapolated from a configuration is shown in Figure~\ref{fig:clus_example}(a). We then applied DBSCAN to each data set independently, from which recovered only the points belonging to the cluster closest to the minimum of the energy. An example of application of the clustering algorithm to a data set is shown in Figure~\ref{fig:clus_example}(b), where the cluster of interest is highlighted in black. We finally computed the maximum value of the energy $\beta u_{ij,dip}$ for the points of each recovered cluster: all the points of each cluster belong to a portion of the ($r_{ij,\perp}$,$r_{ij,\parallel}$) plot delimited by a contour line in which the energy is constant. We finally set the average of all contour energies of each data set as our threshold energy parameter $\beta u_{max} = \langle \beta u_{max} \rangle_{set} = -3.20$. Figure~\ref{fig:clus_example}(c) shows the the elements of the desired cluster extrapolated with DBSCAN, contained in the area delimited by the countour line in which $\beta u_{ij,dip} = \beta u_{max}$.

\clearpage

%%%%%%%%%%%%%%%%%%%%%%%%%%%%%%%%%%%%%%%%%%%%
%%%%%%%%%%%%%%%%%%%%%%%%%%%%%%%%%%%%%%%%%%%%
%%%%% SECTION 2: UNIAXIAL OP
%%%%%%%%%%%%%%%%%%%%%%%%%%%%%%%%%%%%%%%%%%%%
%%%%%%%%%%%%%%%%%%%%%%%%%%%%%%%%%%%%%%%%%%%%

\secti{S2. Renormalisation of the uniaxial order parameter for particles with cubic symmetry}

The most favoured configuration of two polarised nanocubes is for their two centers of mass to be relatively aligned with the external field and to be as close as possible, thus having one of their faces in contact with each other, and with these faces orthogonal to the orientation of the field $\vers{E}$. For this reason, the orientational order of the cubic particles with respect to the external field, which can be evaluated with the uniaxial order parameter $U=\langle P_2 (\vers{E} \cdot \vers{e}_i ) \rangle$ ($P_2$ is the second Legendre polynomial), can indirectly give information on the formation of chain-like structures. However, due to the geometry of cubes, each of the three axes of orientation $\vers{e}_{1}$, $\vers{e}_{2}$, $\vers{e}_{3}$ of every particle can be equivalently aligned with $\vers{E}$ when they organise in chains. To overcome this issue, Batten \textit{et al.} \cite{Torquato2010}, in their work on the investigation of the phase behaviour of colloidal superballs, suggested to "relabel" particles axes of orientation according to their best alignment with the reference axes, hence for a reference axis $\vers{n}$ they computed $\langle P_{2}(\max((|\vers{n}\cdot{e_{k}}|),k=1,2,3)\rangle$. Since in our case the director along which particles should be aligned is known and it is the orientation of the external field $\vers{E}$, we computed the average of the alignment between $\vers{n}$ and the most aligned axis $\vers{e}_{k}$ of each cube, for $\vers{n} = \vers{E}$
\begin{equation}
    \Big\langle \max\big((\vers{n} \cdot \vers{e}_k)^2\big),i=\{1,2,3\} \Big\rangle
    \label{eq:op_cubic}
\end{equation}
Here we report the formal proof developed to normalize the order parameter in Eq.~\ref{eq:op_cubic}.

\begin{figure}[ht!]
    \centering
    \includegraphics[width=0.4\textwidth]{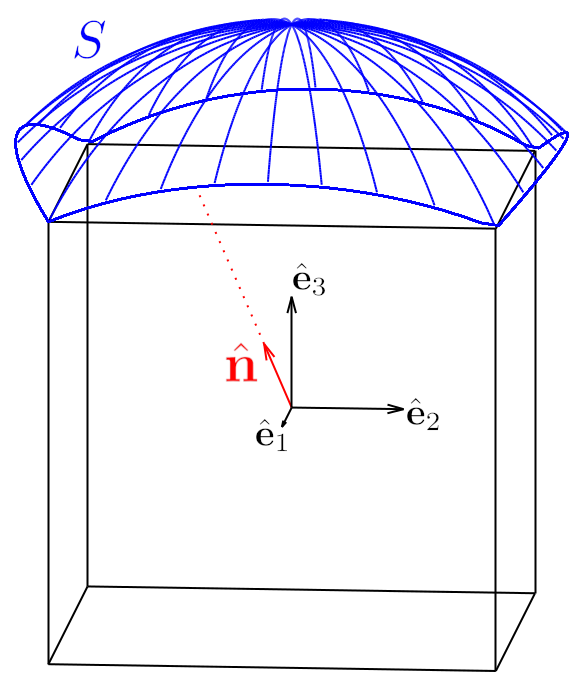}
    \caption{Cubic particle (black lines) topped with a portion of a sphere with surface $S$ (in blue). $\vers{e}_{1}$, $\vers{e}_{2}$, $\vers{e}_{3}$ are the three axes of orientation of the cube, originated from the center of mass. Red arrow shows one possible direction of the director $\vers{n}$ and the red dot line is its extension up to surface $S$. $S$ is the area in which integrate $(\vers{n} \cdot \vers{e}_{3})^2$ for normalisation of the order parameter, where $\vers{e}_{3}$ is here the most aligned axis with $\vers{n}$.}
    \label{fig:op_surf}
\end{figure}
\begin{figure}[ht!]
    \centering
    \includegraphics[width=0.80\linewidth]{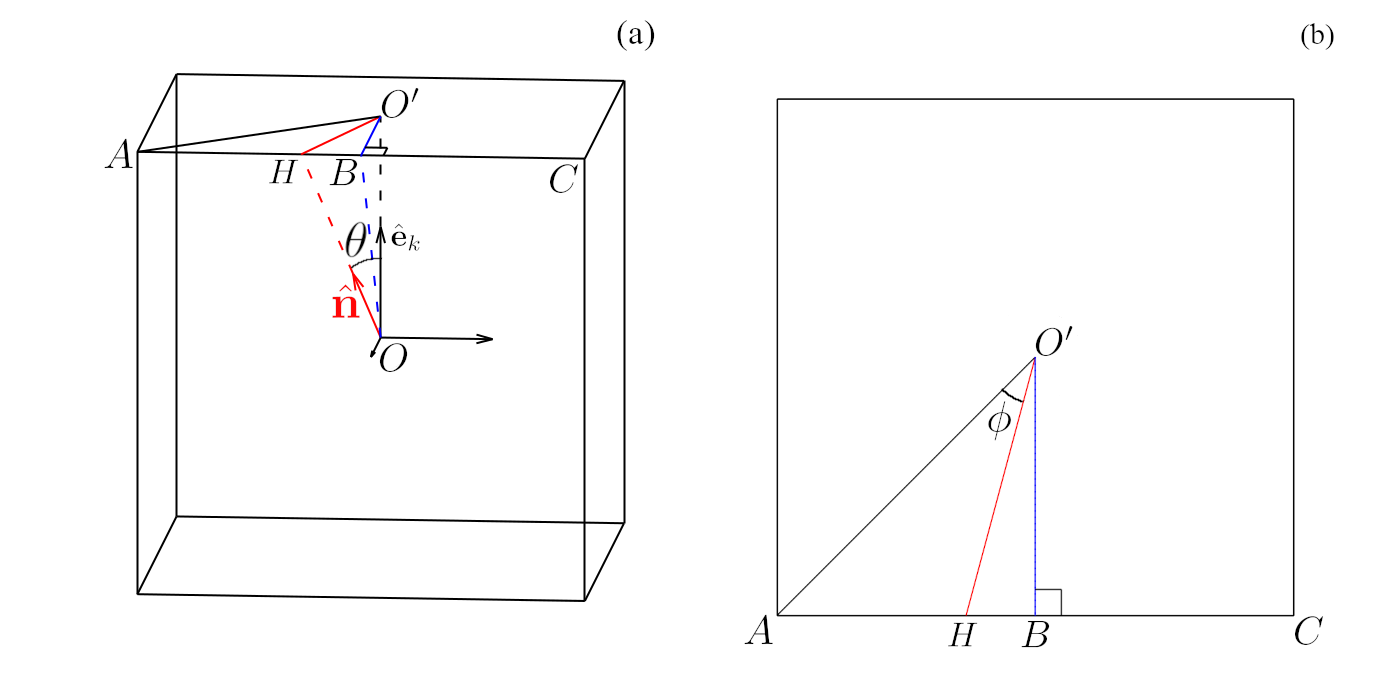}
    \caption{Geometric representation of the cube with the director $\vers{n}$. Panel (a, left) shows the cube and all the projections in 3D; panel (b, right) shows the view from the top face of the cube. $O$ is the center of mass of the cube. $O'$ is the projection of $O$ on the top face of the cube. $A$ and $C$ are two of the vertices of the cube. $\overline{AC}$ is one edge of the top face of the cube. $B$ is the midpoint of $\overline{AC}$. Dashed blue line connects the origin $O$ with the midpoint $B$, and straight blue line is the projection of $\overline{OB}$ on the top face of the cube. Both $\overline{OB}$ and $\overline{AO'}$ are half of the diagonal of one face of the cube. Both $\overline{O'B}$ and $\overline{OO'}$ are half of the length of one edge of the cube. Red arrow and dashed red line are, respectively, the unit vector $\vers{n}$ and its extension $\overline{OH}$ to the edge $\overline{AC}$; the extension of $\vers{n}$ intersect $\overline{AC}$ at $H$. Straight red line is the projection of $\overline{OH}$ on the top face of the cube. $\theta$ is the angle between $\vers{n}$ and $\vers{e}_{k}$. $\phi$ is the angle between the projection $\overline{AO'}$ and the projection $\overline{O'H}$.}
    \label{fig:op_projs}
\end{figure}

Figure~\ref{fig:op_surf} shows one cube, with its axes of orientation, here tkaed as reference axes, and the director $\vers{n}$ with one possible orientation with respect to $\vers{e}_{1}$, $\vers{e}_2$, and $\vers{e}_3$. The prolongation of $\vers{n}$ intersects always the surface of the cube. Due to the symmetry of the cube, for each possible direction of $\vers{n}$, the most aligned axis of orientation $\vers{e}_{k}$ with $\vers{n}$ will always be the one orthogonal to the face whom the prolongation of $\vers{n}$ intersects. This means that Eq.~\ref{eq:op_cubic} can be normalised integrating only over the possible orientations of $\vers{n}$ with respect to one of the faces of the cube: this normalisation corresponds to integrate $(\vers{n} \cdot \vers{e}_{k})^2$ over one sixth of the surface of a unit sphere, which is schematically represented in Figure~\ref{fig:op_surf} with blue lines, and all the angles between $\vers{n}$ and one $\vers{e}_{k}$, for which the extension of $\vers{n}$ intersects the edges of one face of the cube, are boundaries of the integral. We can define $\vers{n}$ in spherical coordinates as a function of two angles $\theta$ and $\phi$ with respect to the axes of orientation of the cube and recover the following definitions, according to the geometric representation in Figure~\ref{fig:op_projs} and assuming that the cube has unit length
\begin{gather*}
    \widehat{AO'H}=\phi \\
    \widehat{HOO'}=\theta \\
    \widehat{AO'B}=\frac{\pi}{4} \\
    \widehat{HBO}=\widehat{BO'O}=\widehat{ABO'}=\frac{\pi}{2} \\
    \overline{OH}=x \\
    \overline{HB}=y \\
    \overline{AO'}=\overline{OB}=\frac{1}{\sqrt{2}} \\
    \overline{O'B}=\overline{OO'}=\frac{1}{2}
\end{gather*}

Thanks to geometry, we can recover the link between $\theta$ and $\phi$ for the case when $\vers{n}$ intersect the edges of one of the faces of the cube. If $0 \leq \phi < \pi / 4$, the following equations hold
\begin{gather*}
    \cos(\theta) = \frac{\overline{OO'}}{\overline{OH}} = \frac{1}{2x} \\
    x = \overline{OH} = \sqrt{\overline{OB}^2 + \overline{HB}^2} = \sqrt{\Bigg(\frac{1}{\sqrt{2}}\Bigg)^2 + y^2} \\
    y = \overline{HB} = \overline{O'B}\tan\Big(\widehat{HO'B}\Big) = \frac{\tan\bigg(\dfrac{\pi}{4} - \phi\bigg)}{2}
\end{gather*}
\begin{equation}
    \cos(\theta) = \frac{1}{\sqrt{2+\tan^2 \bigg(\dfrac{\pi}{4} - \phi\bigg)}} = f^{(+)}(\phi)
    \label{eq:op+f+}
\end{equation}
For $\pi / 4 \leq \phi < \pi /2$, Eq.~\ref{eq:op+f+} holds since $\tan^2(x)$ is an even function. For $\pi /2 \leq \phi < 3\pi /4$, we obtain
\begin{equation*}
     y = \frac{\tan\bigg(\phi - \dfrac{\pi}{4} - \dfrac{\pi}{2}\bigg)}{2} = \frac{-\cot\bigg(\phi - \dfrac{\pi}{4}\bigg)}{2}
\end{equation*}
\begin{equation}
    \cos(\theta) = \frac{1}{\sqrt{2+\cot^2 \bigg(\dfrac{\pi}{4} - \phi\bigg)}}=f^{(-)}(\phi)
    \label{eq:op_f-}
\end{equation}
The validity of Eq.~\ref{eq:op_f-} is extended to $3\pi /4 \leq \phi < \pi$ for the same reason as Eq.~\ref{eq:op+f+}. Since both Eq.~\ref{eq:op+f+} and \ref{eq:op_f-} are periodic to $\pi$, we can define the link between $\theta$ and $\phi$ along the border of one face of the cube as follows
\begin{equation}
    \cos(\theta) = 
    \begin{cases}
    f^{(+)}(\phi) \quad if \quad 0 \leq \phi < \dfrac{\pi}{2} \cup \pi \leq \phi < \dfrac{3\pi}{2} \\
    f^{(-)}(\phi) \quad if \quad \dfrac{\pi}{2} \leq \phi < \pi  \cup \dfrac{3\pi}{2} \leq \phi < 2\pi
    \end{cases}
    \label{eq:op_cases}
\end{equation}
The average value of the order parameter for a uniform distribution of $\vers{n}$ can be estimated according to the following integral
\begin{equation}
    \Big\langle \max \Big((\vers{n} \cdot \vers{e}_k)^2\Big) \Big\rangle = \dfrac{\displaystyle\int\int_{S} \cos^2(\theta) \sin(\theta) d\theta d\phi}{\displaystyle\int\int_{S} \sin(\theta) d\theta d\phi}
    \label{eq:op_defi}
\end{equation}
where the limits of $S$ depend on Eq.~\ref{eq:op_cases}. We first compute the integral in the denominator of Eq.~\ref{eq:op_defi}
\begin{gather*}
    \int\int_{S} \sin(\theta) d\theta d\phi = \int_0^{\arccos(f^{(+)}(\phi))}\int_{0}^{\pi/2} \sin(\theta) d\theta d\phi +
    \int_0^{\arccos(f^{(-)}(\phi))}\int_{\pi/2}^{\pi} \sin(\theta) d\theta d\phi + \\ + \int_0^{\arccos(f^{(+)}(\phi))}\int_{\pi}^{3\pi/2} \sin(\theta) d\theta d\phi + \int_0^{\arccos(f^{(-)}(\phi))}\int_{3\pi/2}^{2\pi} \sin(\theta) d\theta d\phi
\end{gather*}
We substitute $\int\sin(\theta)d\theta=-\cos(\theta)$, $\cos(0) = 1$ and $\cos(\arccos(f(\phi)))=f(\phi)$ in the integral
\begin{equation}
    \int\int_{S} \sin(\theta) d\theta d\phi = \int_{0}^{\pi/2} \Big[1- f^{(+)}(\phi)\Big] d\phi
    +\int_{\pi/2}^{\pi} \Big[1 - f^{(-)}(\phi)\Big] d\phi + 
    \int_{\pi}^{3\pi/2} \Big[1-f^{(+)}(\phi)\Big] d\phi
    + \int_{3\pi/2}^{2\pi} \Big[1-f^{(-)}(\phi)\Big] d\phi
    \label{eq:op_intd}
\end{equation}
Since both $f^{(+)}$ and $f^{(-)}$ are periodic with period $\pi$ and $\tan^2(x)=\cot^2(x-\pi/2)$, the following equalities hold
\begin{equation}
    \int_{0}^{\pi/2} f^{(+)}(\phi) d\phi = 
    \int_{\pi/2}^{\pi} f^{(-)}(\phi) d\phi =  
    \int_{\pi}^{3\pi/2} f^{(+)}(\phi) d\phi =
    \int_{3\pi/2}^{2\pi} f^{(-)}(\phi) d\phi
    \label{eq:op_equal}
\end{equation}
Hence, integral in Eq.~\ref{eq:op_intd} simplifies
\begin{gather*}
    \int\int_{S} \sin(\theta) d\theta d\phi = 2\pi - 4\int_{0}^{\pi/2} f^{(+)} d\phi
\end{gather*}
It can be proved that
\begin{gather*}
    \int_{0}^{\pi/2} f^{(+)} d\phi = \int_{-\pi/4}^{\pi/4}    \frac{\cos(\phi)}{\sqrt{2-\sin^2(\phi)}} d\phi = \frac{\pi}{3}
\end{gather*}
The solution of the integral is
\begin{equation}
    \int\int_{S} \sin(\theta) d\theta d\phi = 2\pi - 4\frac{\pi}{3} = \frac{2\pi}{3}
    \label{eq:op_resu_den}
\end{equation}

This is what we should expect for this geometry: the surface we integrated over is exactly one sixth of the surface of a unit sphere (since we have integrated the surface of a unit sphere only for those angles of $\theta$ and $\phi$ for which the intersection of the prolongation of $\vers{n}$ with the surface of the cube is contained in one face of the cube).

In order to solve the integral in the numerator of Eq.~\ref{eq:op_defi}, we use the equality $\int \cos^2(\theta)\sin(\theta) dx = -\cos^3(\theta)/3$ and we obtain an equation similar to Eq.~\ref{eq:op_intd}, with $[f^{(+)}]^3$ instead of $f^{(+)}$, and $[f^{(-)}]^3$ instead of $f^{(-)}$. Since the equalities in Eq.~\ref{eq:op_equal} hold also in this case, the desired integral can be simplified as follows
\begin{gather*}
    \int\int_{S} \cos^2(\theta) \sin(\theta) d\theta d\phi =
    \frac{2\pi}{3}-\frac{4}{3}\int_{0}^{\pi/2} \Big[f^{(+)}(\phi)\Big]^3d\phi
\end{gather*}
It can be proved that
\begin{equation}
    \int_0^{\pi/2} \Big[f^{(+)}(\phi)\Big]^{3} d\phi =
    \int_{-\pi/4}^{\pi/4}
    \frac{1-\sin^2(\phi)}{\big[2-\sin^2(\phi)\big]^{3/2}}\cos(\phi) d\phi = \frac{\pi}{3}-\frac{1}{\sqrt{3}}
    \label{eq:op_solun}
\end{equation}
In conclusion, the integral in the numerator of Eq.~\ref{eq:op_defi} is
\begin{equation}
    \int\int_{S} \cos^2(\theta) \sin(\theta) d\theta d\phi =
    \frac{2\pi}{3}-\frac{4}{3}\Bigg(\frac{\pi}{3}-\frac{1}{\sqrt{3}}\Bigg) = \frac{2\pi+4\sqrt{3}}{9}
    \label{eq:op_resu_num}
\end{equation}
Finally, we substitute Eq.~\ref{eq:op_resu_den} and Eq.~\ref{eq:op_resu_num} into Eq.\ref{eq:op_defi} to recover the expected value of $\max((\vers{n} \cdot \vers{e}_{k})^2)$ for a uniform distribution of orientations of $\vers{n}$
\begin{equation}
    \Big\langle \max \Big((\vers{n} \cdot \vers{e}_k)^2\Big) \Big\rangle = \dfrac{\dfrac{2\pi+4\sqrt{3}}{9}}{\dfrac{2\pi}{3}} = \dfrac{\pi+2\sqrt{3}}{3\pi}
    \label{eq:op_solu1}
\end{equation}
As a proof that this value is indeed correct, we remark that Batten \textit{et al.} noticed that, for an isotropic phase of superballs, $\langle P_2(\max(\vers{n} \cdot \vers{e}_k))\rangle \approx 0.55$, which is exactly what we obtain:
\begin{gather*}
    \Big\langle P_2\Big(\max\big(|\vers{n} \cdot \vers{e}_k|\big)\Big) \Big\rangle = \frac{1}{2}\Big\langle 3\max \Big((\vers{n} \cdot \vers{e}_k)^2\Big) - 1\Big\rangle = \frac{1}{2}\Bigg(3\frac{\pi+2\sqrt{3}}{3\pi}-1\Bigg) = \frac{\sqrt{3}}{\pi} \approx 0.55
\end{gather*}
Normalising $\max((\vers{n}\cdot\vers{e}_{k})^2)$ with the result in Eq.~\ref{eq:op_solu1} and imposing that the parameter goes to 1 when $\max((\vers{n}\cdot\vers{e}_{k})^2)=1$, \textit{i.e.} when the cube is perfectly aligned with $\vers{n}$, we can finally obtain a renormalised uniaxial order parameter for particles with cubic symmetry
\begin{equation}
    S = \dfrac{1}{N} \sum\limits_{i=1}^{N} S_i = \dfrac{1}{N} \sum\limits_{i=1}^{N} \frac{\pi\Big( 3\max \big((\vers{n} \cdot \vers{e}_{k,i})^2,k=1,2,3\big)-1\Big) -2\sqrt{3}}{2\pi-2\sqrt{3}}
    \label{eq:op_fine}
\end{equation}

\clearpage

%%%%%%%%%%%%%%%%%%%%%%%%%%%%%%%%%%%%%%%%%%%%
%%%%%%%%%%%%%%%%%%%%%%%%%%%%%%%%%%%%%%%%%%%%
%%%%% SECTION 3: MSD TRACER IN 0.20
%%%%%%%%%%%%%%%%%%%%%%%%%%%%%%%%%%%%%%%%%%%%
%%%%%%%%%%%%%%%%%%%%%%%%%%%%%%%%%%%%%%%%%%%%

\secti{S3. Mean square displacement of a spherical tracer immersed in a bath of nanocubes}

\begin{figure}[ht!]
    \centering
    \includegraphics[width=0.5\textwidth]{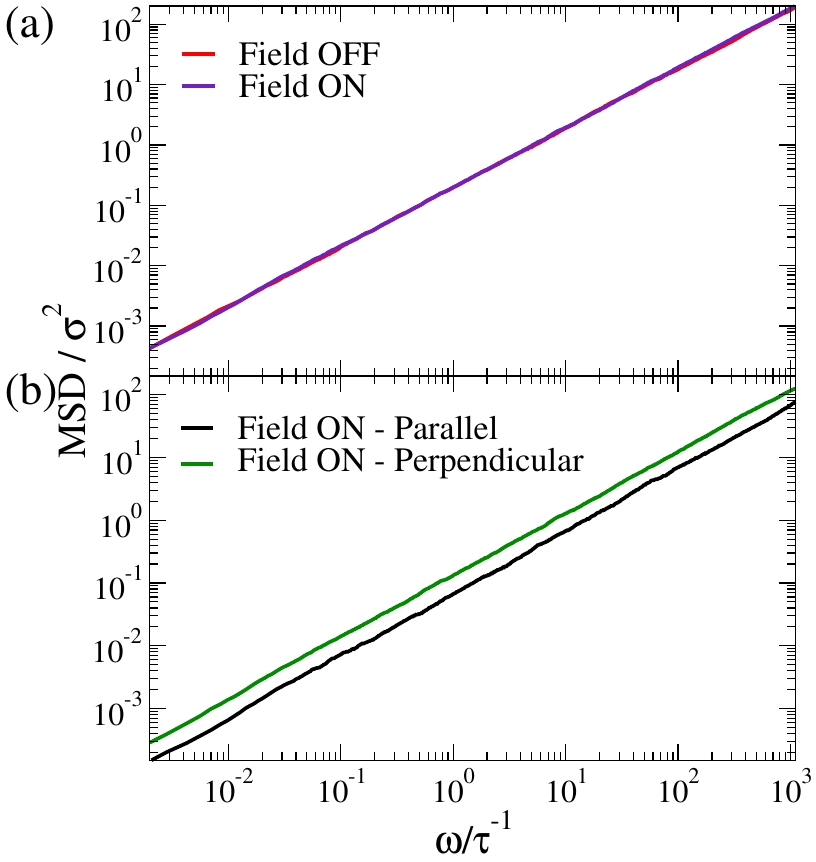}
    \caption{(Color on-line) Mean square displacement of a spherical tracer of size $3\sigma$ diffusing across a bath of nanocubes of side $\sigma$ at a packing fraction $\eta = 0.02$. Top panel: MSD calculated in the three spatial coordinates when the external field is on (blue line) and off (red line). Bottom panel: MSD parallel (black line) and perpendicular (green line) to the external field, when the field is on.}
    \label{fig:fig5MR_1}
\end{figure}

\begin{figure}[t!]
    \centering
    \includegraphics[width=0.5\textwidth]{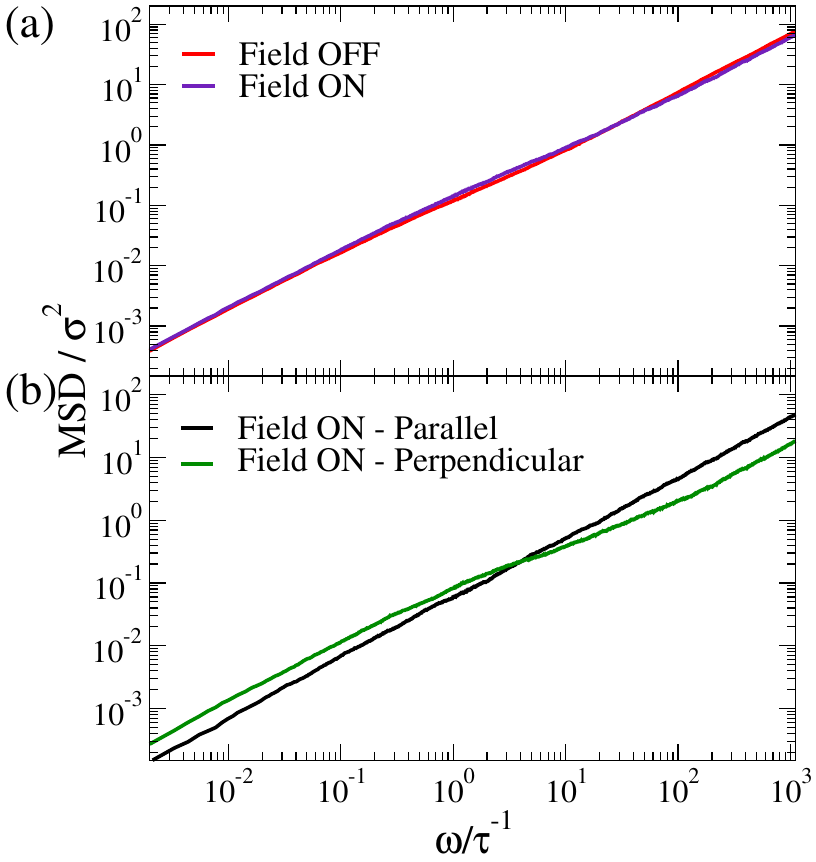}
    \caption{(Color on-line) Mean square displacement of a spherical tracer of size $3\sigma$ diffusing across a bath of nanocubes of side $\sigma$ at a packing fraction $\eta = 0.2$. Top panel: MSD calculated in the three spatial coordinates when the external field is on (blue line) and off (red line). Bottom panel: MSD parallel (black line) and perpendicular (green line) to the external field, when the field is on.}
    \label{fig:fig5MR}
\end{figure}

Figures \ref{fig:fig5MR_1} and \ref{fig:fig5MR} depict the mean square displacement (MSD) of a spherical tracer immersed in a bath of nanocubes with side $\sigma$ at packing fractions $\eta = 0.02$ and $\eta = 0.2$, respectively. The MSD is estimated when the external field is on and off. The MSD of the tracer particle can be defined from the tracer's position $\textbf{r}_d(t)$ at time $t$ as follows
\begin{equation}
    \langle \Delta \textbf{r}^2_d(t)\rangle = \langle(\textbf{r}^2_d(t)-\textbf{r}^2_d(0))^2\rangle 
    \label{eq:MR1}
\end{equation}
where the brackets refer to the averages over independent trajectories, and $d$ is the dimensionality of the displacements of the tracer. While the value $d=3$ indicates the total MSD of the tracer particle, $d=2$, and $1$ refer, respectively, to the MSDs perpendicular and parallel to the external field.
To calculate the MSDs we have run 1000 uncorrelated trajectories to simulate the motion of the tracer and bath particles.

\clearpage

%%%%%%%%%%%%%%%%%%%%%%%%%%%%%%%%%%%%%%%%%%%%
%%%%%%%%%%%%%%%%%%%%%%%%%%%%%%%%%%%%%%%%%%%%
%%%%% SECTION 4: BENCHMARK TRANSFORM
%%%%%%%%%%%%%%%%%%%%%%%%%%%%%%%%%%%%%%%%%%%%
%%%%%%%%%%%%%%%%%%%%%%%%%%%%%%%%%%%%%%%%%%%%

\secti{S4. Benchmarking between the Fourier and compliance-based methods to calculate the elastic and viscous moduli of a bath of nanocubes}

The compliance methods \cite{Evans2009, Nishi2018} intend to transform the time dependent compliance ($J(t)$) of the material to estimate its corresponding complex modulus ($G^{*}(\omega)$) in the frequency domain. The compliance can be expressed as
\begin{equation}
    J(t) = \left(\frac{\pi a_{sph}}{k_{\rm{B}}T}\right)\langle\Delta r^2(t)\rangle,
    \label{eqMR1}
\end{equation}
where $a_{sph}$ indicates the radius of the tracer, $k_{\rm{B}}$ the Boltzmann's constant, $T$ the temperature, and $\langle\Delta r^2(t)\rangle$ the tracer's mean square displacement (MSD). According to Evans \textit{et al.} \cite{Evans2009}, $J(t)$ can be related to $G^{*}(\omega)$ by
\begin{equation}
\frac{i\omega}{G^*(\omega)} = \left(1-e^{-1\omega t_1}\right)\frac{J(t_1)}{t_1} + 6De^{-i\omega t_{N_t}} + \sum_{k=2}^{N_t}\frac{J_k-J_{k-1}}{t_k-t_{k-1}}\left(e^{-i\omega t_{k-1}}-e^{-i\omega t_{k}}\right),
\label{eqMR2}
\end{equation}
where $N_t$ denotes the number of points within the time window where the MSD was computed, $D$ is linked to the inverse of the system's viscosity, and $J_k$ refers to $J(t)$ at time $t_k$. The solution of Eq.~\ref{eqMR2} results in the viscous ($G''(\omega)$) and elastic ($G'(\omega)$) moduli since $G^{*}(\omega) = G'(\omega) + iG''(\omega)$.   

\begin{figure}[ht!]
    \centering
    \includegraphics[width=0.45\textwidth]{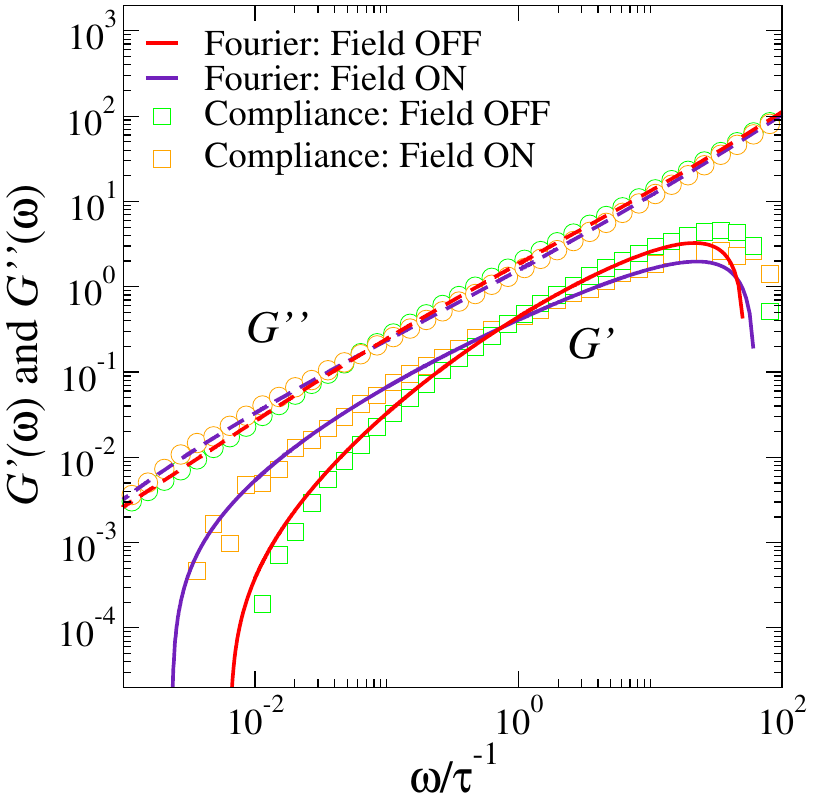}
    \caption{(Color on-line) Viscous $G''(\omega)$ (empty circles, dashed lines) and elastic $G'(\omega)$ (empty squares, solid lines) moduli obtained by the Fourier approach \cite{mason2000} (lines) and the compliance-based method \cite{Evans2009} (symbols) for an bath of nanocubes containing a spherical tracer of size $3\sigma$. Two scenarios were explored: when the external field is on (orange symbols, blue curves) and off (green symbols, red curves).}
    \label{fig:fig6MR}
\end{figure}

The comparison between $G''(\omega)$ and $G'(\omega)$ calculated by the Fourier approach by Mason \cite{mason2000} and the compliance method proposed by Evans \textit{et al.} \cite{Evans2009} is presented in Fig.~\ref{fig:fig6MR}. In both cases, the MSD of a spherical tracer with size $3\sigma$ diffusing in a bath of nanocubes is calculated when the external field is on and off.    

\bibliography{References_SI}